\documentclass{nature}

\newif\ifSRSUB \SRSUBfalse

\usepackage{graphicx}
\usepackage{booktabs,longtable}
\usepackage{color}
\usepackage{lscape}

\title{The interplay of mutations and electronic properties in disease-related genes}

\author{Chi-Tin Shih$^{1}$, Stephen A. Wells$^{2}$, Ching-Ling Hsu$^{3}$ Yun-Yin Cheng$^{1}$, \& Rudolf A. R\"{o}mer$^{2,*}$}

\begin{document}

\maketitle

\begin{affiliations}
\item Department of Physics, Tunghai University, 40704 Taichung, Taiwan and The National Center for Theoretical Sciences, 30013 Hsinchu, Taiwan
\item Department of Physics and Centre for Scientific Computing, University of Warwick, Gibbet Hill Road, Coventry, CV4 7AL, UK
\item Department of Physics, Chung-Yuan Christian University, 32023 Chung-Li, Taiwan \\
$^*$Correspondence: r.roemer@warwick.ac.uk
\end{affiliations}

\begin{abstract}
Electronic properties of DNA are believed to play a crucial role in many phenomena in living organisms, for example the location of DNA lesions by base excision repair (BER) glycosylases and the regulation of tumor-suppressor genes such as p53 by detection of oxidative damage. However, the reproducible measurement and modelling of charge migration through DNA molecules at the nanometer scale remains a challenging and controversial subject even after more than a decade of intense efforts. Here we show, by analysing $162$ disease-related genes from a variety of medical databases with a total of almost $20,000$ observed pathogenic mutations, a significant difference in the electronic properties of the population of observed mutations compared to the set of all possible mutations. Our results have implications for the role of the electronic properties of DNA in cellular processes, and hint at the possibility of prediction, early diagnosis and detection of mutation hotspots.
\end{abstract}


Cells tend to accumulate over time genetic changes such as nucleotide
substitutions, small insertions and deletions, rearrangements of the genetic
sequences and copy number changes.\cite{She03} These changes in turn affect
protein-coding or regulatory components and lead to health issues such as
cancer, immunodeficiency, ageing-related diseases and other disorders. A cell
responds to genetic damage by initiating a repair process or programmed cell
death.\cite{Fra07}
In recent years, a vast number of detailed databases have been assembled in
which rich information about the type, severity, frequency and diagnosis
of many thousand of such observed mutations has been
stored.\cite{OMIM,SteBM03,PetMKI07,LohG04} This abundance of data is based on
the now standard availability of massively parallel sequencing
technologies.\cite{Nag06}
Harvesting these genomic databases for new cancer genes and hence potential
therapeutic targets has already demonstrated its usefulness\cite{EnkMJY09} and
several recent international cancer genome projects continue the required
large-scale analysis of genes in tumours.\cite{ICGM10}

The possible relevance of charge transport in DNA damage has recently also
attracted considerable interest in the bio-chemical and bio-physical
literature.\cite{StaLT06,Cha07,BerC07,GenBB10} Direct measurement of charge transport
and/or transfer in DNA remains a highly controversial topic due to the very
challenging level of required manipulation at the nano-scale.\cite{GuoGH08}
Ab-initio modelling of long DNA strands is similarly demanding of
computational resources and so some of the most promising computational
approaches necessarily use much simplified models based on coarse-grained
DNA.\cite{Cha07} 
Here we compute and datamine the results of charge transport calculations based on two such effective models for each possible mutation in $162$ of the most important disease-associated genes from four
large gene data\-bases. The models are (i) the standard one-dimensional chain of coupled nucleic bases with onsite ionisation potentials\cite{Cha07,ShiRR08} as well as a novel 2-leg ladder model with diagonal couplings and explicit modelling of the sugar-phosphate backbone.\cite{WelSR09}

\section*{Results}

\subsection{Point Mutations and Electronic Properties}
We consider  native genetic sequences and mutations of
disease-associated genes as retrieved from the {\em Online Mendelian Inheritance
  in Man} (OMIM)\cite{OMIM} of NCBI, the {\em Human Gene Mutation Database}
(HGMD),\cite{SteBM03} {\em International Agency of Research on Cancer}
(IARC)\cite{PetMKI07} as well as {\em Retinoblastoma Genetics}.\cite{LohG04}
We have selected these genes such that (i) those from OMIM have a
well-known sequence with known phenotype as well as at least $10$ point
mutations, (ii) all other selected cancer-related genes have also at least
$10$ point mutations and (iii) all non-cancer related genes from HGMD have at
least $200$ point mutations (cp.\ Supplementary Table \ifSRSUB S1\else\ref{tab-S1}\fi).


Many different types of mutation are possible in a genetic sequence
including point mutations, deletion of single base pairs (producing a frame
shift), and large-scale deletion or duplication of multiple base pairs.  Here, we restrict our attention to point mutations as it allows us to
directly compare the sequence before and after the mutation.
We study the magnitude of the \emph{change} in charge
transport (CT) for pathogenic mutations when compared to all possible
mutations either {\em locally}, i.e.\ at the given hotspot site, or {\em
  globally} when ranked according to magnitude of CT change. We find that the
vast majority of mutations shows good agreement with a hypothesis where {\em
  smallest change in electronic properties} --- as measured by a change in CT
--- {\em corresponds to a mutation} that has appeared in one of the
aforementioned databases {\em of pathogenic genes}.

A gene with ${\cal N}$ base pairs (bps) has a native nucleotide sequence
$(s_1,s_2,\cdots,s_{\cal N})$ along the coding strand. The gene has a total of
$3\cal{N}$ possible point mutations, which we denote as the set $M_{\rm all}$,
of which a subset $M_{\rm pa}$ are known pathogenic mutations.
A point mutation is represented by the pair $(k,s)$, where $k$ is the
position of the point mutation in the genomic sequence and $s$ is the mutant
nucleotide which replaces the native nucleotide.
We shall write a mutation from a native base P to a mutant base Q as
``Pq''. We note that there are a total of twelve possible point mutations in a
DNA sequence (from any one of four bases to any one of three alternatives). Of
these twelve, four are {\em transitions}, in which a purine base replaces a
purine or a pyrimidine replaces a pyrimidine, and eight are {\em
  transversions} in which purine is replaced by pyrimidine or vice
versa. Biologically, transitions are in general much more common than
transversions.\cite{ColJ94}
Indeed, the set of observed pathogenic mutations for our $162$ genes contains
$10999$ transitions and $8883$ transversions, whereas in the set of all
mutations their ratio is by definition $1:2$. The observed pathogenic
mutations are thus already a biased selection from the set of possible
mutations, favouring transitions. However, this local onsite chemical
shift is not sufficient to fully explain our data as we will show later.

We compute and datamine the results of quantum mechanical transport
calculations based on two effective H\"{u}ckel models\cite{Pow10} for each
possible mutation in those $162$ genes. 
Both models assume $\pi$--$\pi$
orbital overlap in a well-stacked double helix.  The parameters are chosen to
represent hole transport. Using the transfer matrix
method\cite{KraM93,NdaRS04} we calculate the spatial extent of (hole)
wavefunctions of a given energy on a length of DNA with a given genetic
sequence.  Wavefunction localisation is directly related to
conductance\cite{KraM93} and we therefore find it convenient to report our
results in terms of conductance.

To determine the effect of a mutation, we consider sub-sequences of length $L$
bps; there are $L$ such sequences that include a given site $k$. For all $L$
sequences we calculate quantum-mechanical charge transmission coefficients
${T}$ (in units of ${e^2}/{\hbar}$, averaged across a range of
incident energies, as detailed in Methods) for the native and mutant
sequences. We describe the effect of the mutation on the electronic properties
of the DNA strand near to the mutation site using the mean square difference,
${\Gamma}=\left\langle |{T}_{\rm native}-{T}_{\rm mutant}|^2
\right\rangle$, averaged across all $L$ sequences. Larger values of
${\Gamma}$ therefore correspond to a greater difference in electronic
structure between the native and mutant sequences.
The length $L$ must be long enough to allow for substantial delocalisation
across multiple base pairs,\cite{KloRT05} but should remain below the typical
persistence length of $\sim 150$ bps\cite{Hag88} such that any overlap or
crossing by packing, e.g.\ by wrapping around histone complexes in chromatin,
can be ignored. In this study we have considered lengths of $20,40,60$
bps. This requires, for each of the $\cal{N}$ sites in a gene, $L$
calculations for each sequence of length $L$ and for each of $4$ possible bases
at that site; which, for the more than $11\times 10^6$ bases in our dataset of
$162$ genes, is more than $5\times 10^9$ quantum mechanical transport
calculations.

\subsection{Local and global ranking}

\begin{figure}[h]
\centering
\ifSRSUB\else
(a)\includegraphics[width=0.45\columnwidth]{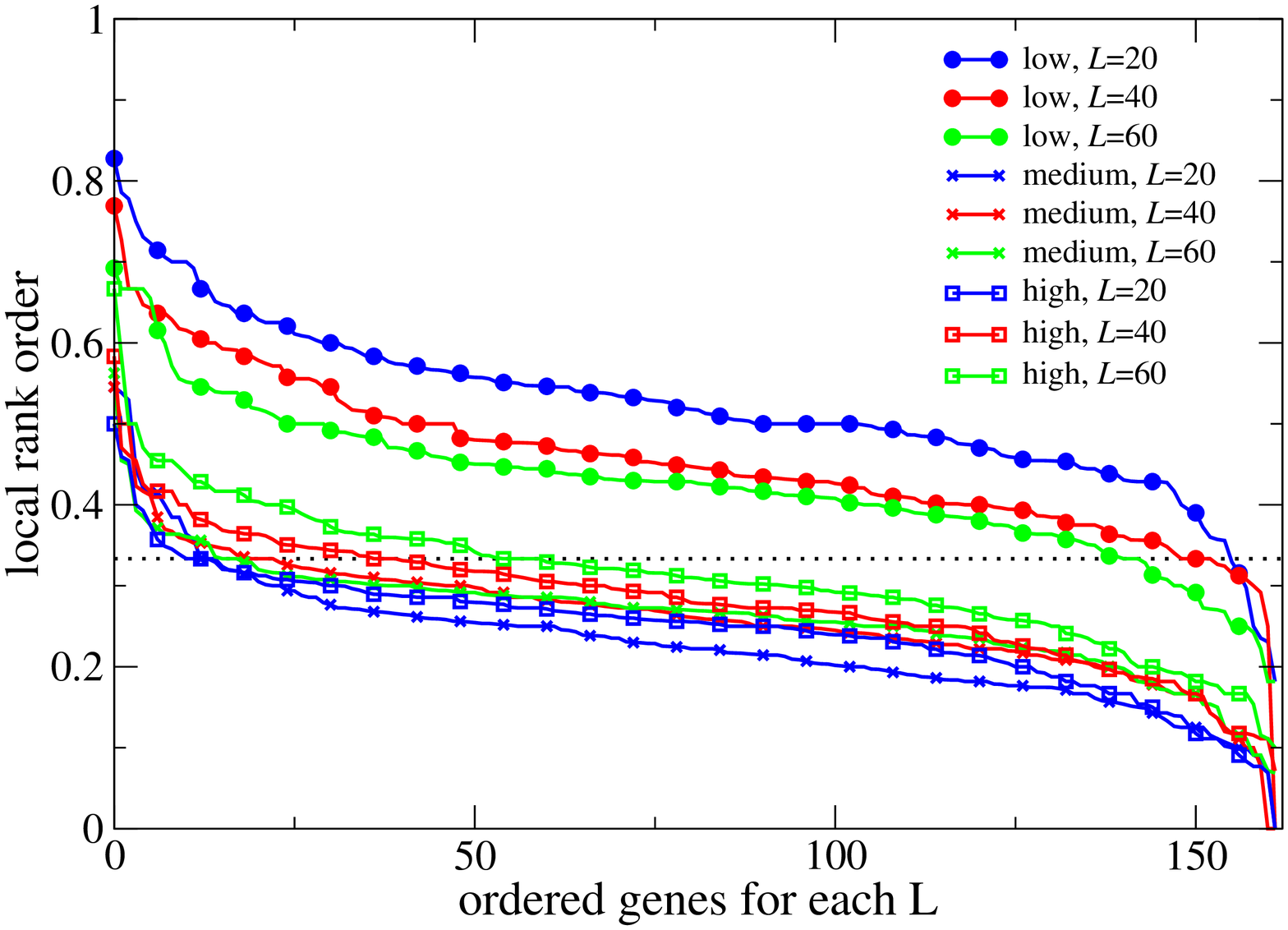}
(b)\includegraphics[width=0.45\columnwidth]{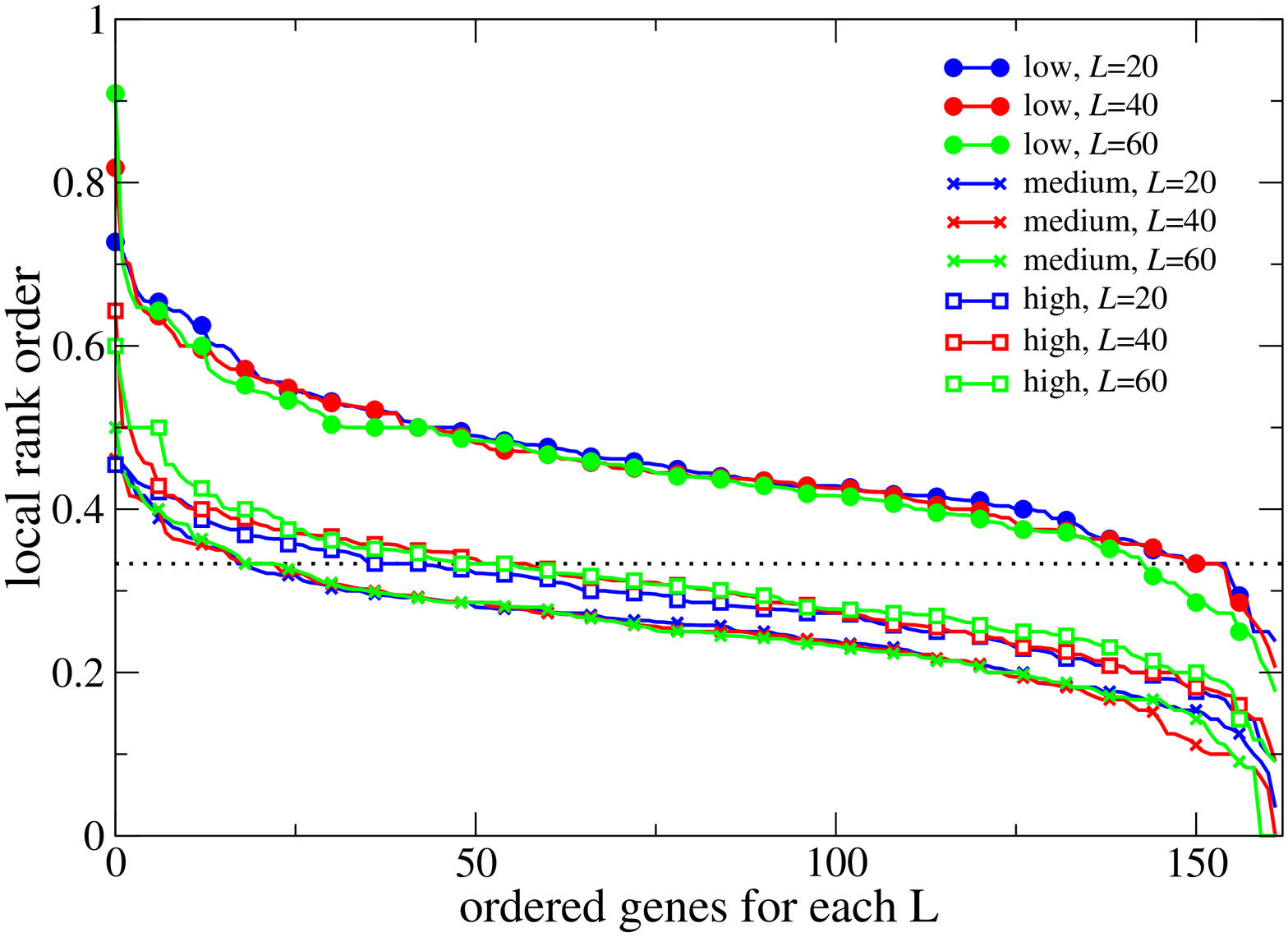}
(c)\includegraphics[width=0.45\columnwidth]{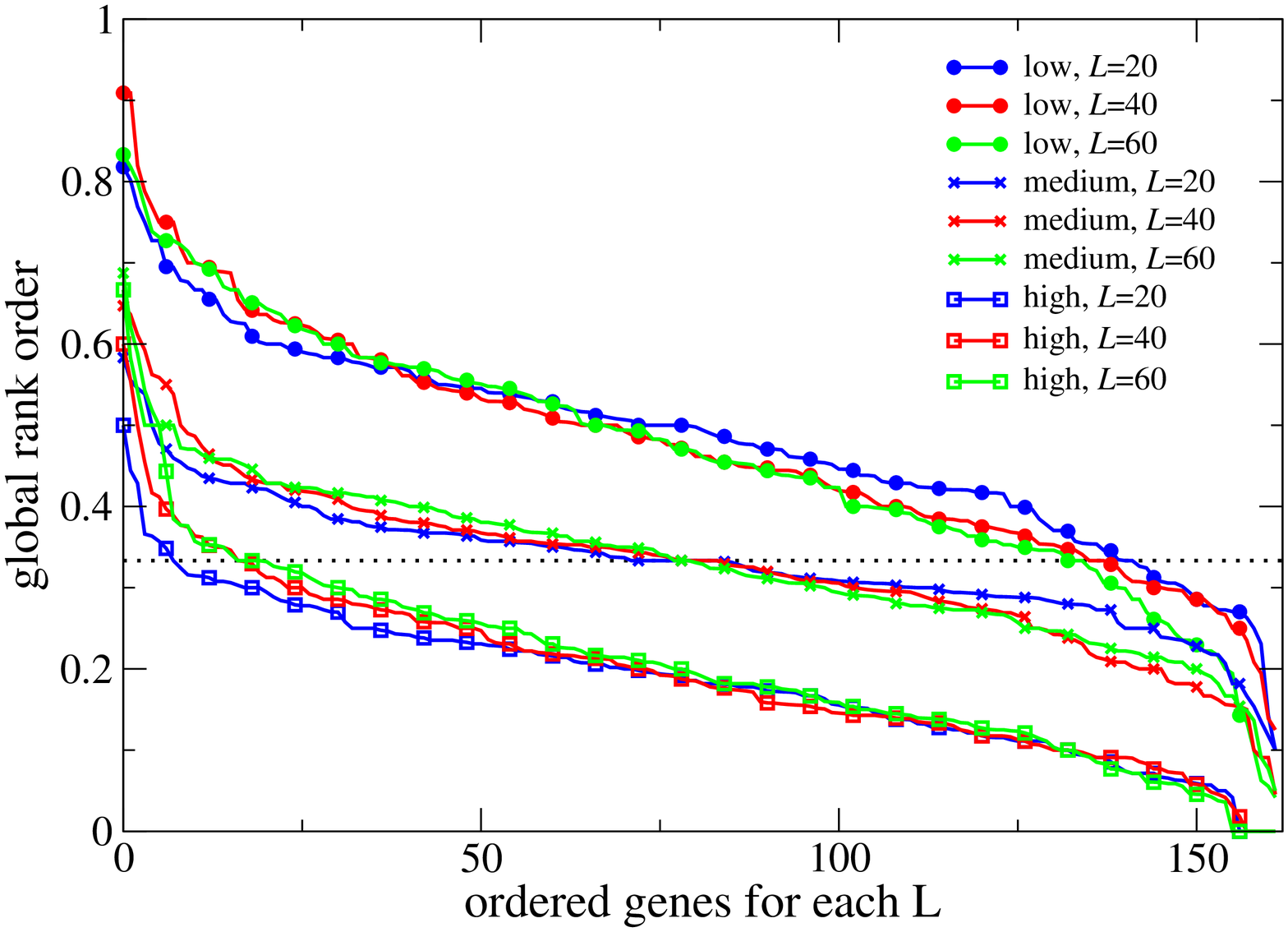}
(d)\includegraphics[width=0.45\columnwidth]{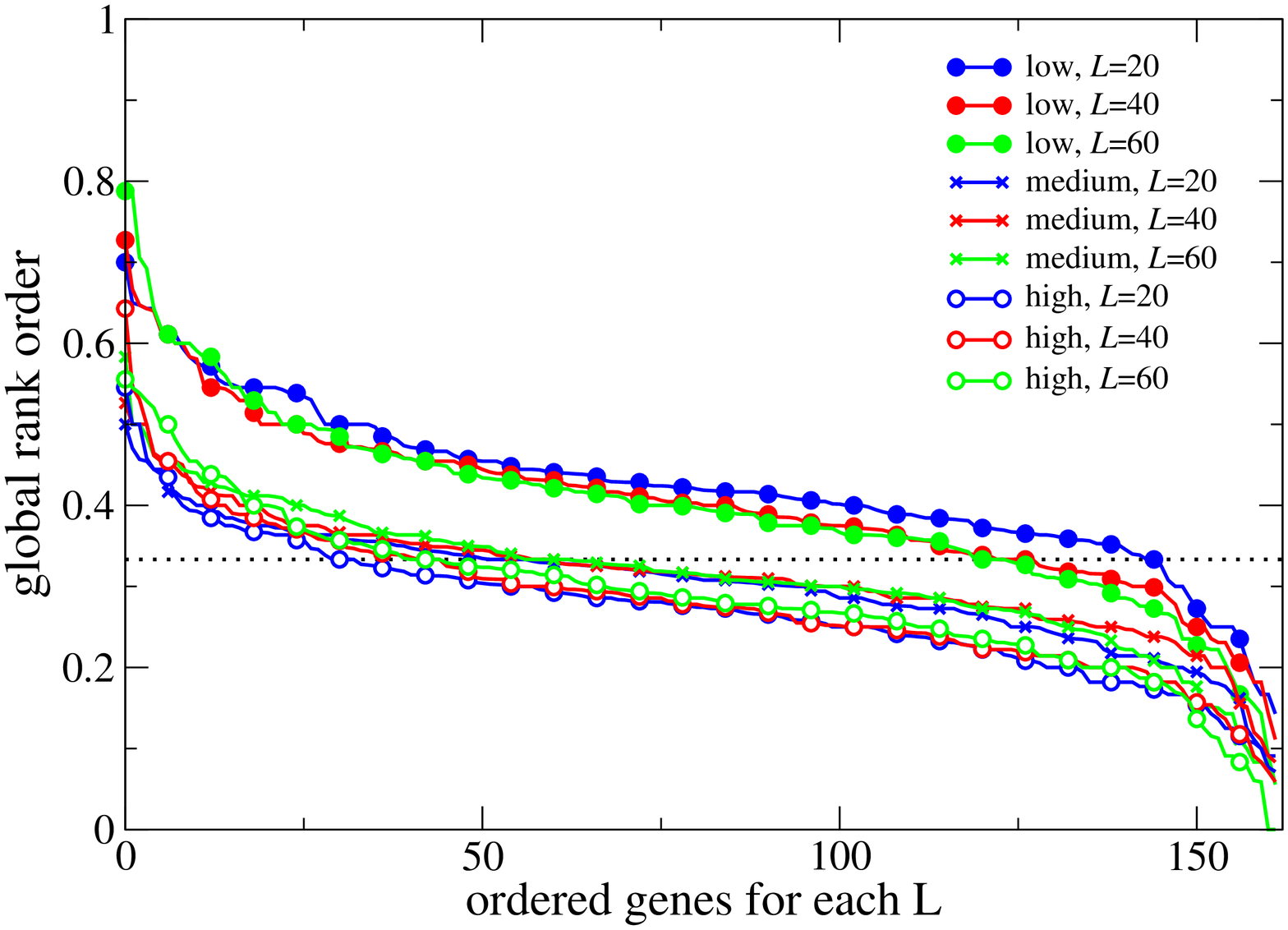}
\fi
\caption{Sorted prevalence of the low, medium and high CT change among
  {\em local} (a+b) and {\em global} (c+d) rankings for pathogenic mutations
  in $162$ genes using the 1D (a+c) and the 2-leg (b+d) models. Results are
  consistent for all three lengths $L=20,40,60$. The $1/3$ value expected by
  chance is shown as a dashed horizontal line. Low rankings are dramatically
  more prevalent locally and globally than chance would suggest.  }
\label{fig-LG_ranking-all}
\end{figure}

We first compare $\Gamma$ of each observed pathogenic mutation with the other two
non-pathogenic ones at the same position and determine a {\em local ranking}
(LR) of CT change. There are three possibilities of LR, namely {\em low}, {\em
  medium} and {\em high}. Note that those hotspots with more than one
pathogenic mutations are excluded in the LR analysis.
We have also sorted the LR ranking for each gene
according to prevalence in Fig.~\ref{fig-LG_ranking-all}(a+b). We find that for
$L=20$, $40$ and $60$ the low CT
change corresponds to $155$ ($95\%$), $148$ ($91\%$) and $140$ ($86\%$) of all
$162$ genes with pathogenic mutations. Examples of LR for the
pathogenic mutations of {\it p16} and {\it CYP21A2} are shown in Supplementary
Fig.~\ifSRSUB S3\else\ref{fig-S3}\fi.
We graphically summarise the results for all $162$ disease-associated genes in
Fig.~\ifSRSUB S5\else\ref{fig-sub-all}\fi. For each gene, we have shown a positive deviation from
the $33\%$ line by orange ---supporting the scenario of small CT change for
pathogenic mutations --- and by blue when the results seem to show no or
negative indication with CT change. It is clear that the correlation between low
CT change and mutation hotspots is well pronounced.

We can also consider a {\em global} ranking (GR) by sorting CT change
$\Gamma$ for {\em all} possible $3{\cal N}$ mutations of a gene with ${\cal
  N}$ bps in order to get a ranking of {\em every} observed pathogenic mutation. By
dividing each ranking by $3{\cal N}$ we compute the normalised GR $\gamma$ of
the mutation, with values between $0$ and $1$. Smaller values of $\gamma$ mean
smaller CT change. By analogy to the local ranking, we divide the $\gamma$ of
the pathogenic mutations into three groups as before, i.e.\ low ($\gamma <
33.3\%$), medium ($33.3\% \leq \gamma < 66.7\%$), and high ($\gamma \geq
66.7\%$) CT change. 
The results of the GR for the $162$ genes are shown in the bottom row (c)
and (d) of Fig.~\ref{fig-LG_ranking-all}.
As for the LR results, we observe many
$\gamma$ values with low CT change (cp.\ Supplementary
Figs.\ \ifSRSUB S3\else\ref{fig-sub-L_ranking-p16}\fi and \ifSRSUB S4\else\ref{fig-sub-L_ranking-all}\fi).
Hence the LR and GR results consistently show that observed pathogenic mutations
are generally biased towards smaller change in CT than the set of all possible
mutations (cp.\ Supplementary Figs.\ \ifSRSUB S5\else\ref{fig-sub-all}\fi and \ifSRSUB S6\else\ref{fig-sub-all-avg}\fi).

\subsection{Distributions of change in charge transport}

\begin{figure}
\centering
\ifSRSUB\else
(a)\includegraphics[width=0.45\columnwidth]{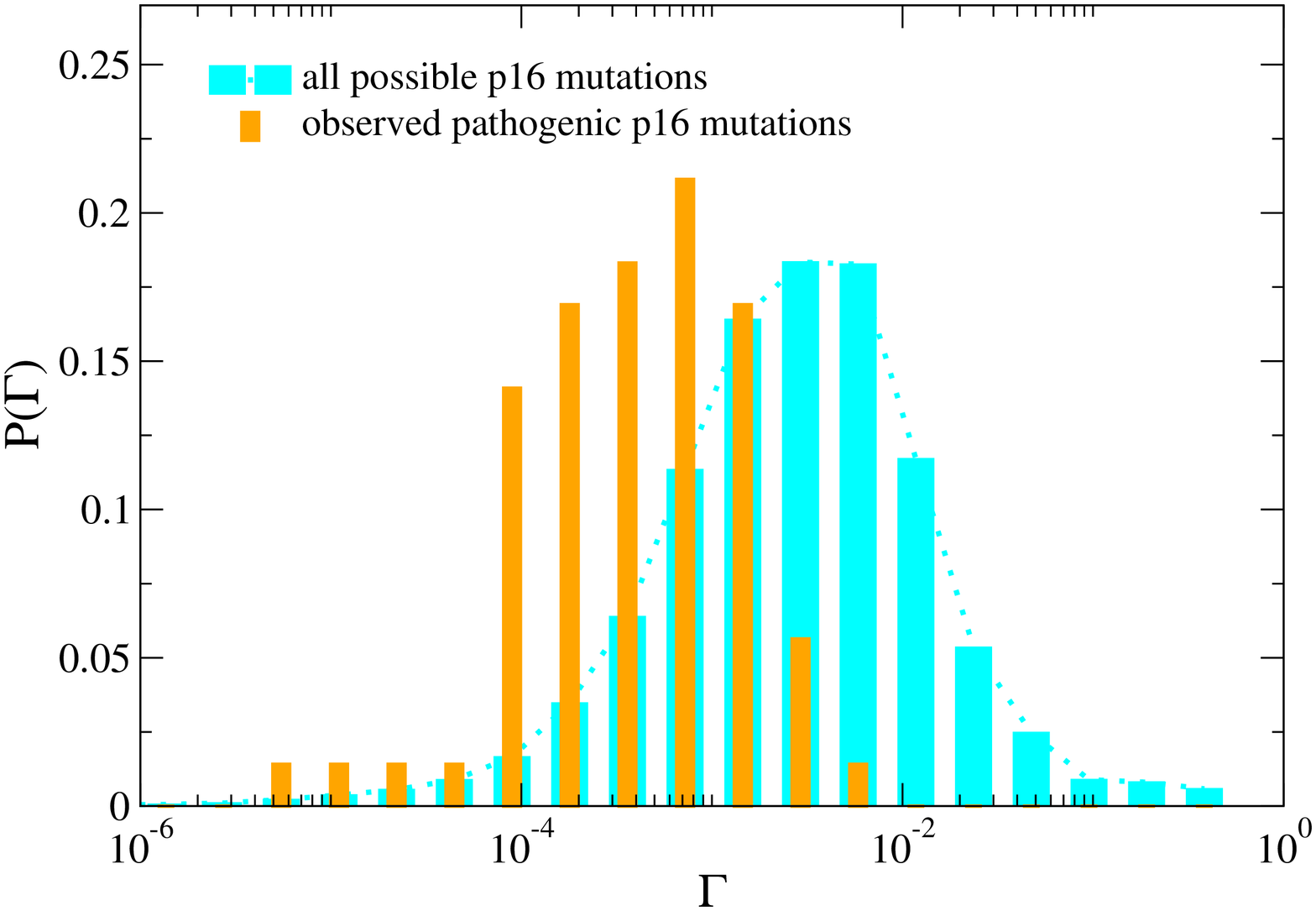}
(b)\includegraphics[width=0.45\columnwidth]{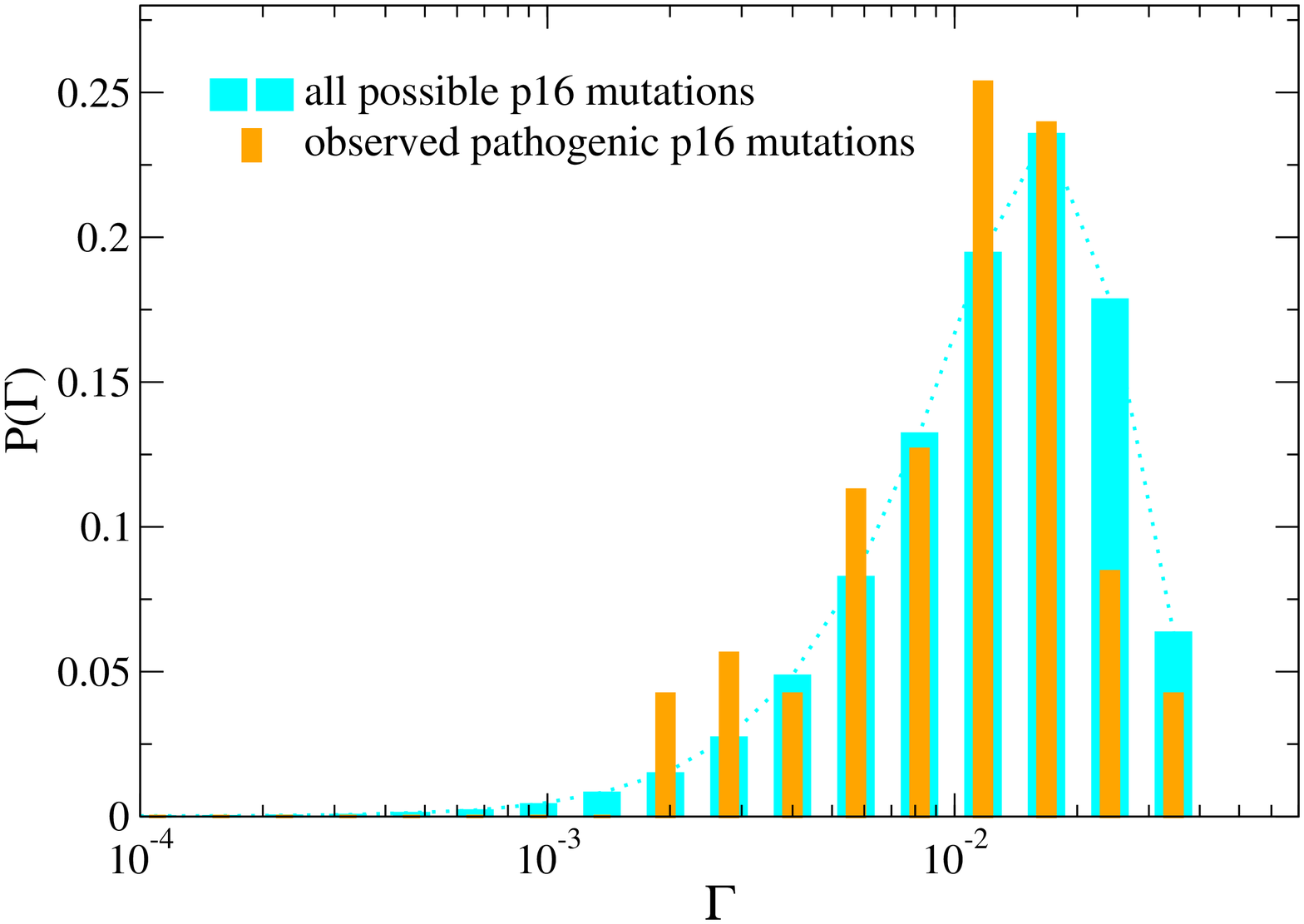}
(c)\includegraphics[width=0.45\columnwidth]{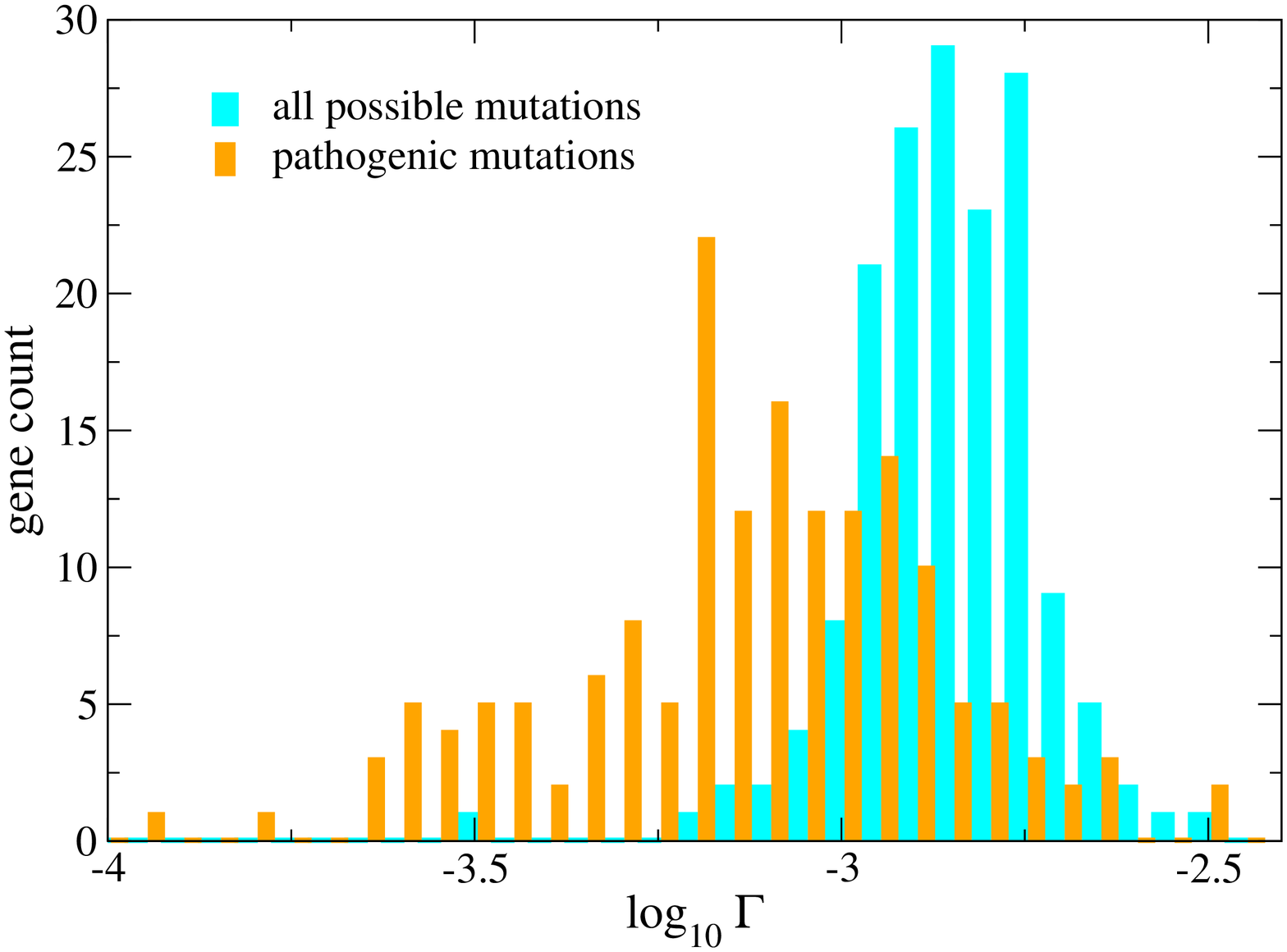}
(d)\includegraphics[width=0.45\columnwidth]{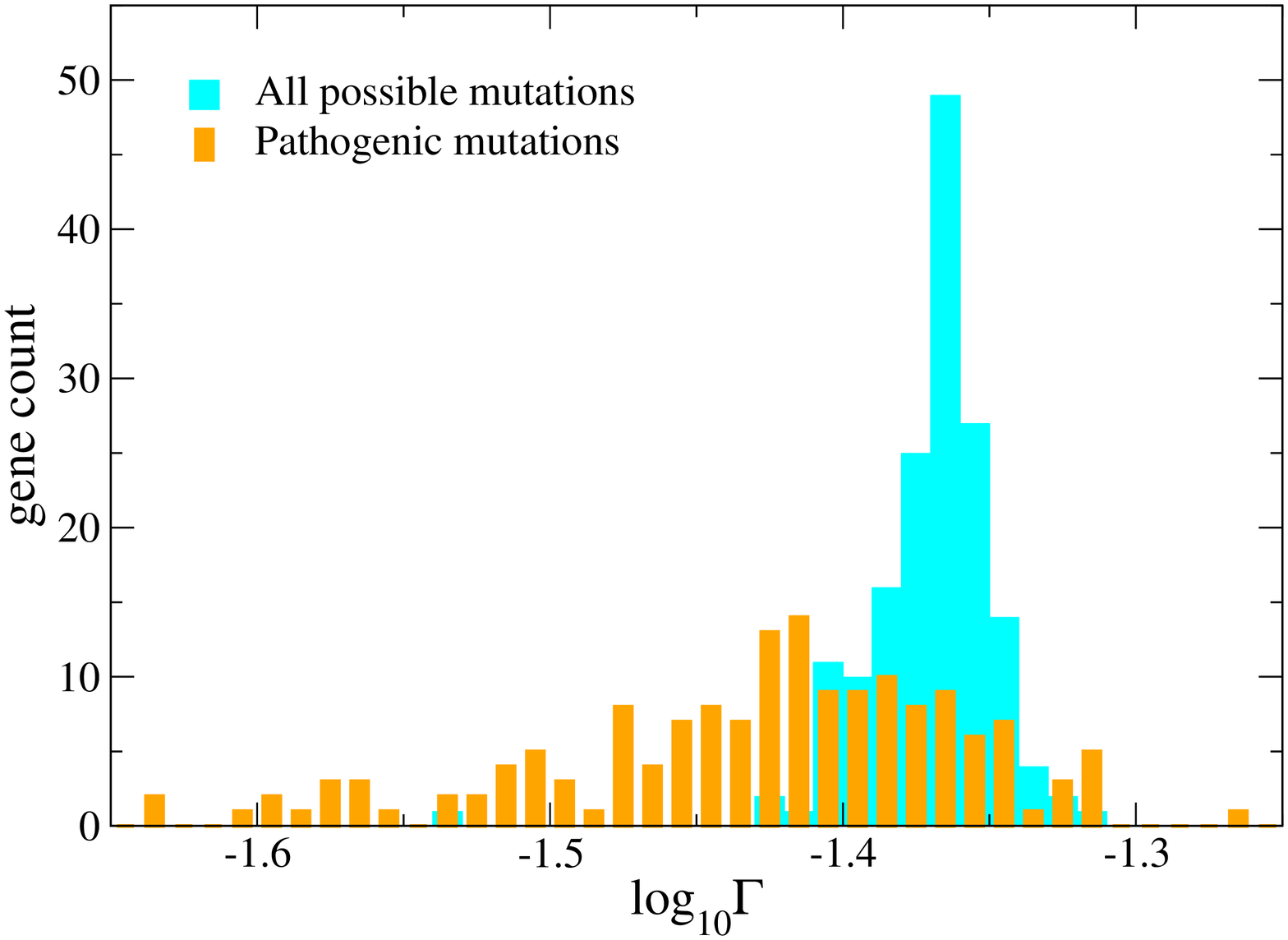}
(e)\includegraphics[width=0.45\columnwidth]{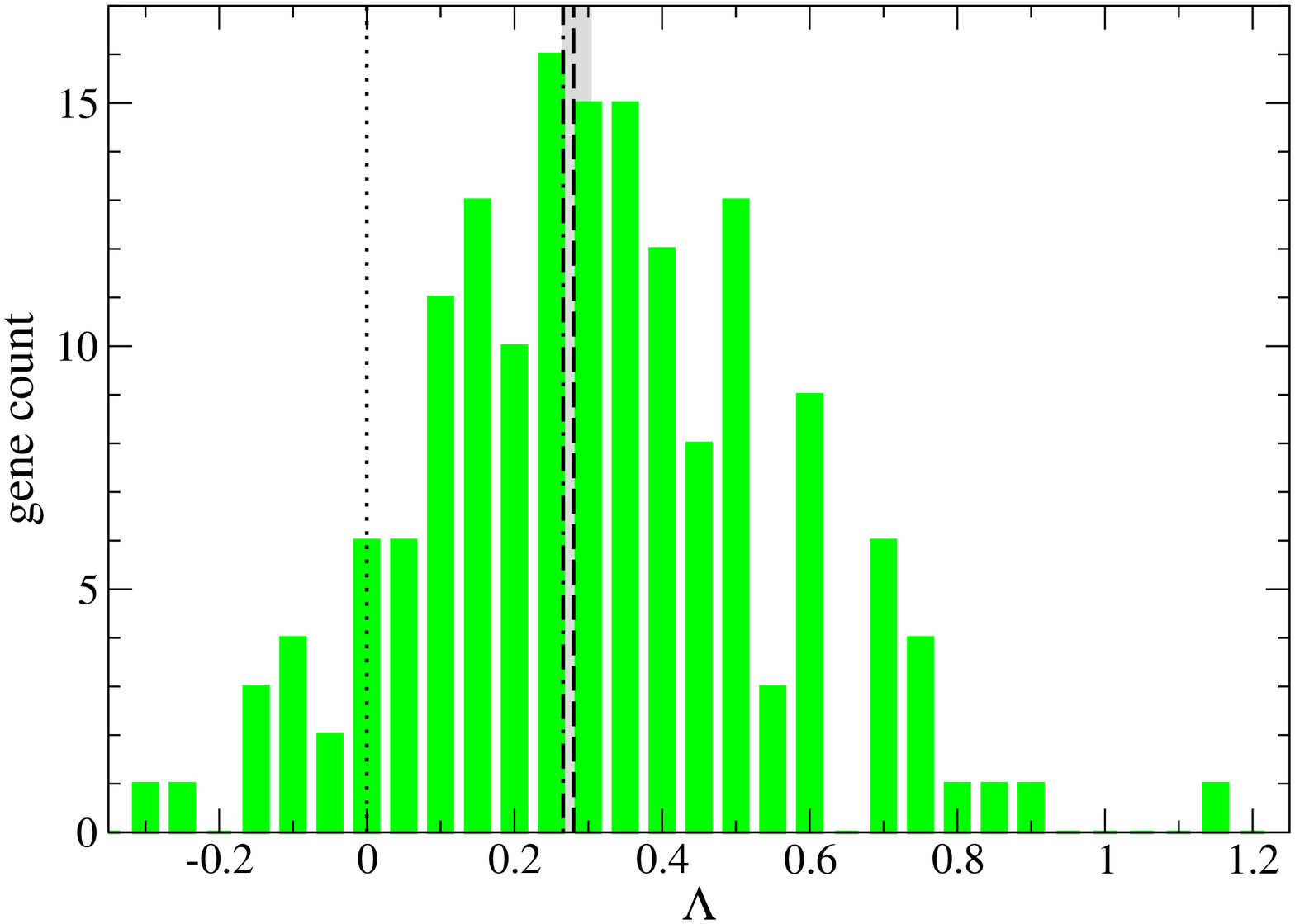}
(f)\includegraphics[width=0.45\columnwidth]{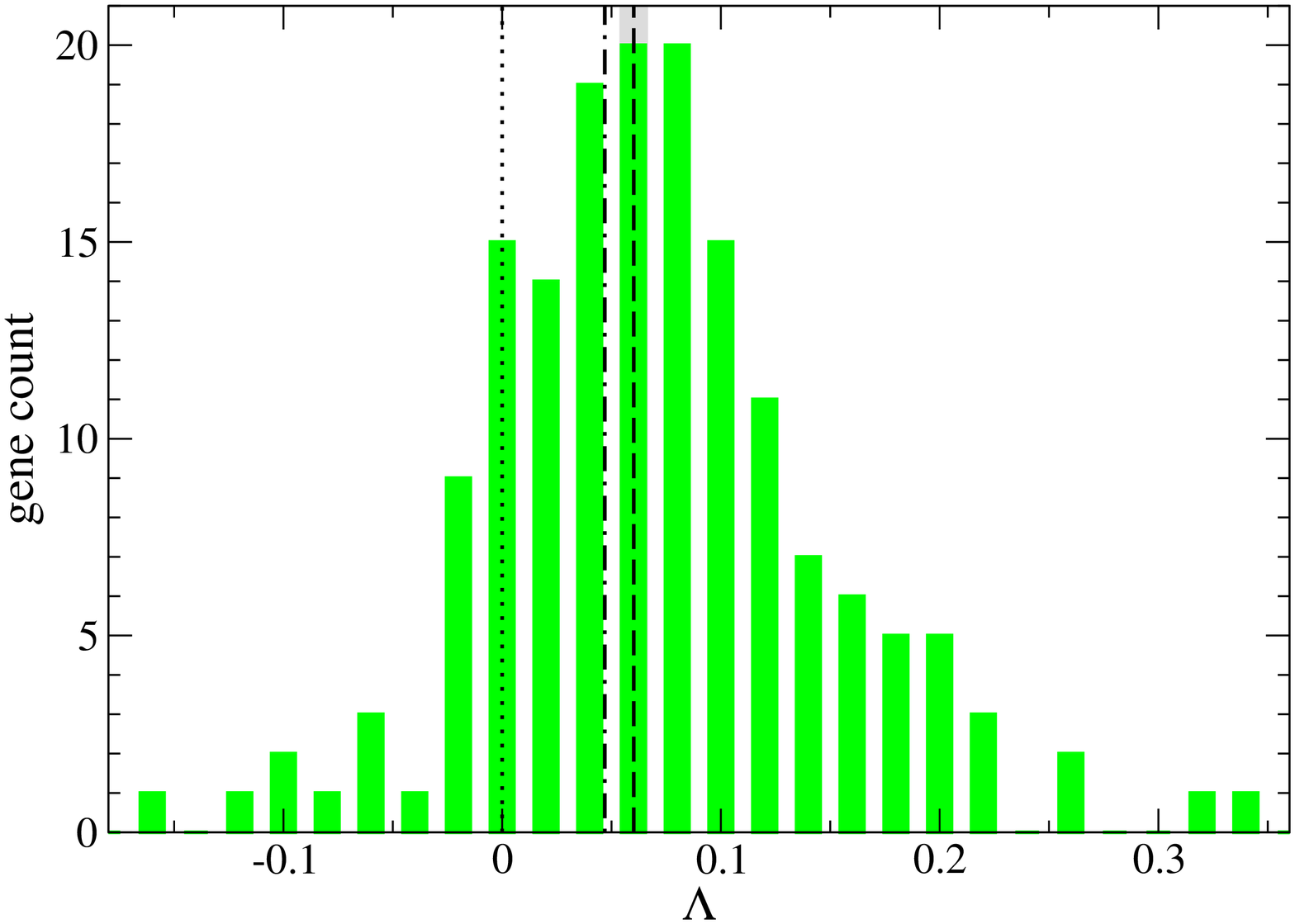}
\fi
\caption{(a+b) Distribution of the change in charge transport
  ${\Gamma}$ for pathogenic (orange bars) and all possible (cyan bars)
  mutations for the $p16$ (CDKN2A) gene with $26740 $ base pairs and $111$
  known pathogenic mutations.  (c+d): Distribution of the average
  (logarithmic) change in charge transport $\left\langle \log {{\Gamma}}
  \right\rangle$ for all pathogenic (orange bars) and all possible (cyan bars)
  mutations for all $162$ genes.  (e+f): Distribution of the global shift
  $\Lambda$ values for all genes, showing a consistent tendency to positive
  values. The average $\bar{\Lambda}$ (dashed) and weighted average
  $\tilde{\Lambda}$ (dash-dotted) values are indicated by vertical lines
  similarly to the $0$ line (dotted). The grey bars denote the error of mean
  for $\left\langle \bar{\Lambda} \right\rangle$.  The results for the 1D and
  2-leg models are displayed in panels (a,c,e) and (b,d,f), respectively.  All
  results shown are for $L=40$, data for $L=20$ and $60$ are similar.  }
  \label{fig-Gamma_percent-p16}
  \label{fig-Gamma_percent-all}
  \label{fig-GS-all}
\end{figure}

In Figure \ref{fig-Gamma_percent-p16} we show as an example results for the
distribution of ${\Gamma}$ for the $p16$ DNA strand for both 1D and 2-leg
models.
In panels (a+b), it is clear that the $111$ observed pathogenic mutations of $p16$
have on average {\em smaller changes} in the CT properties as compared to all
possible $80220$ mutations, for both the 1D and 2-leg models. We find that
results for the vast majority of the other $161$ genes are quite similar.
The distributions of ${\Gamma}$ values in
Fig.~\ref{fig-Gamma_percent-p16}(a+b) are approximately log-normal. We
therefore calculate, for each of the 162 genes in our dataset, an average
$\log {\Gamma}$ value for the distributions of all and pathogenic
mutations. Histograms of the distributions of these $\left\langle \log
{\Gamma} \right\rangle$ values are shown in
Fig.~\ref{fig-Gamma_percent-all}(c+d). It is once again clear that the
distributions for observed pathogenic mutations are shifted towards lower
${\Gamma}$ values in both the 1D and the 2-leg models.

We next define a {\em global CT shift} for a gene $g$ as $\Lambda_{g} =
\left\langle \log {\Gamma}_{g,{\rm all}} \right\rangle - \left\langle \log
{\Gamma}_{g,{\rm pa}} \right\rangle$. Positive values of $\Lambda_{g}$
indicate that the observed pathogenic mutations of gene $g$ have a lower average
${\Gamma}$. For each of our 162 genes we obtain the distribution of
$\Lambda_{g}$ for the 1D and 2-leg models as shown in
Figs.~\ref{fig-GS-all}(e+f).
We can define, for the whole set of $162$ genes, an average global shift
$\bar{\Lambda}=\sum_{g} \Lambda_{g}/162$, weighting all genes equally; we can
also weight the results by the number of observed pathogenic mutations for each gene
$|M_{\rm pa}|_{g}$ for a {\em weighted} average global shift
$\tilde{\Lambda}=\frac{1}{\sum_{g} |M_{\rm pa}|_{g}} \sum_{g} |M_{\rm pa}|_{g}
\Lambda_{g}$.  These values are also indicated in Figs.~\ref{fig-GS-all}(e+f)
and in both models there is a tendency towards lower average
$\bar{\Lambda}_{g}$ for observed pathogenic mutations.

\subsection{Transitions and transversions}

In our models we would expect transitions to cause, in general, a smaller
change in CT than transversions, as the change in onsite energy and in
transfer coefficients is smaller for a transition than a transversion.
However, as we will demonstrate here, the increased proportion of transitions
among the observed pathogenic mutations is {\em not} sufficient to account for the
distributions seen in Fig.~\ref{fig-GS-all}.

\begin{figure}[h]
\centering
\ifSRSUB\else
(a)\includegraphics[width=0.45\columnwidth]{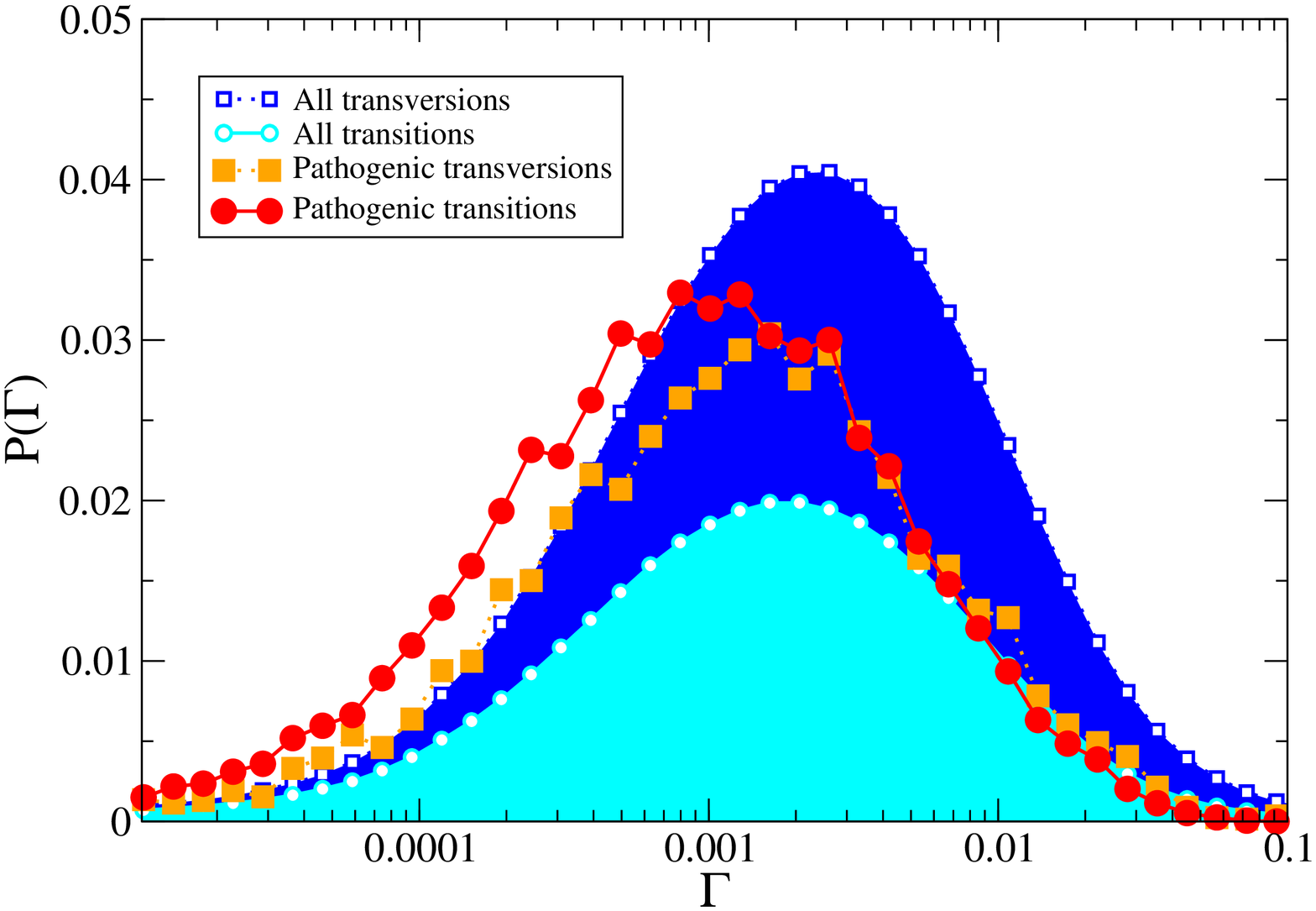}
(b)\includegraphics[width=0.45\columnwidth]{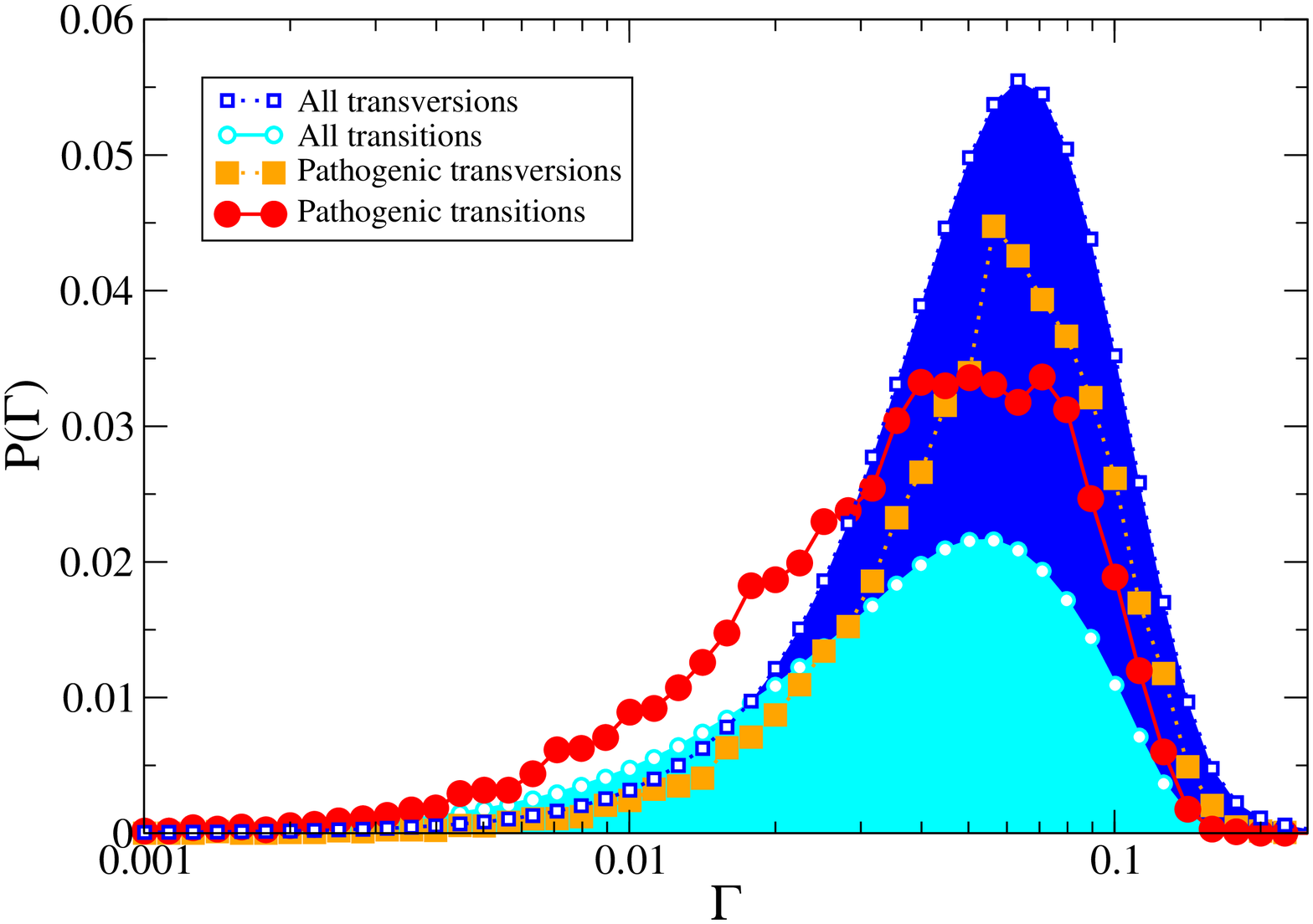}\\
(c)\includegraphics[width=0.45\columnwidth]{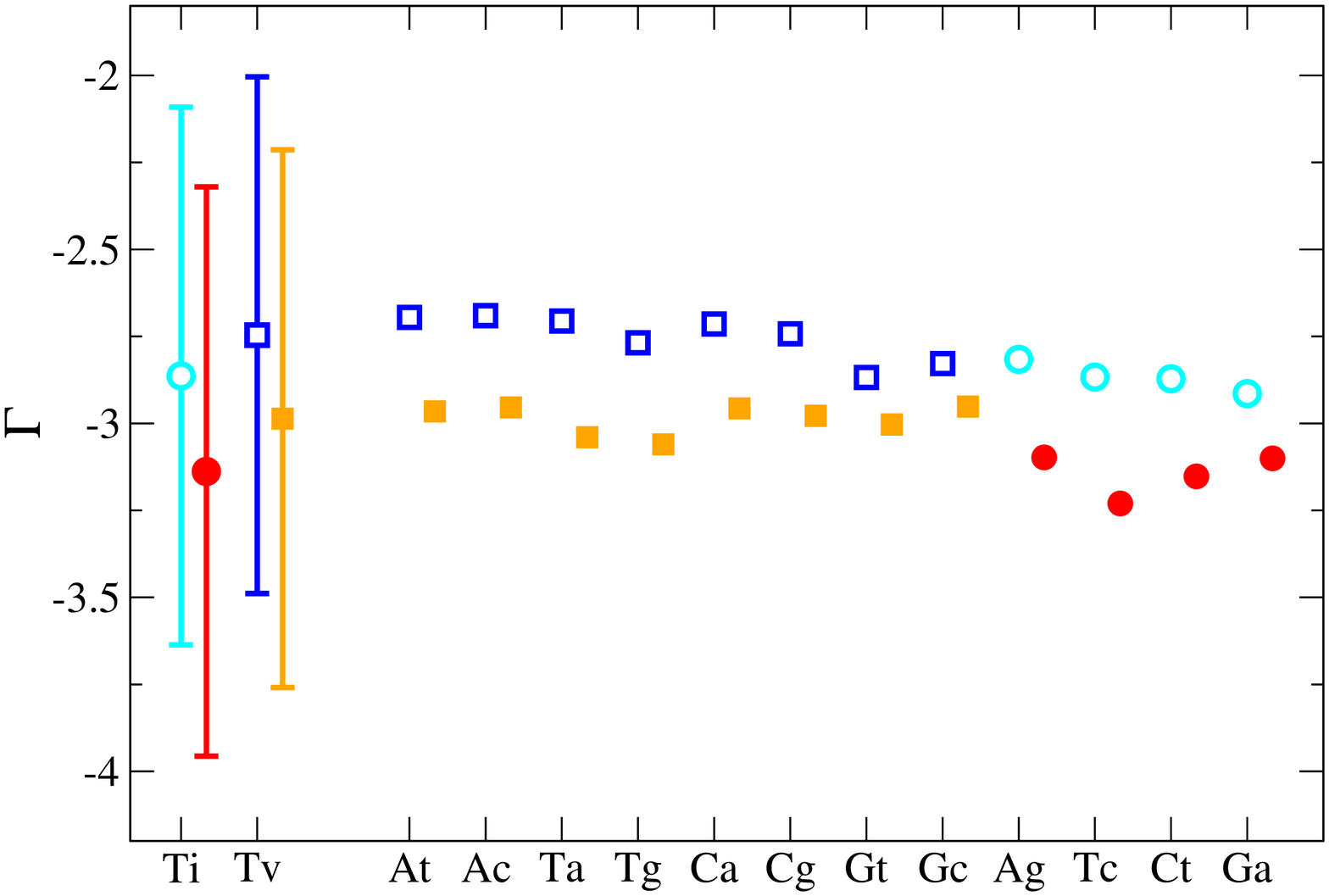}
(d)\includegraphics[width=0.45\columnwidth]{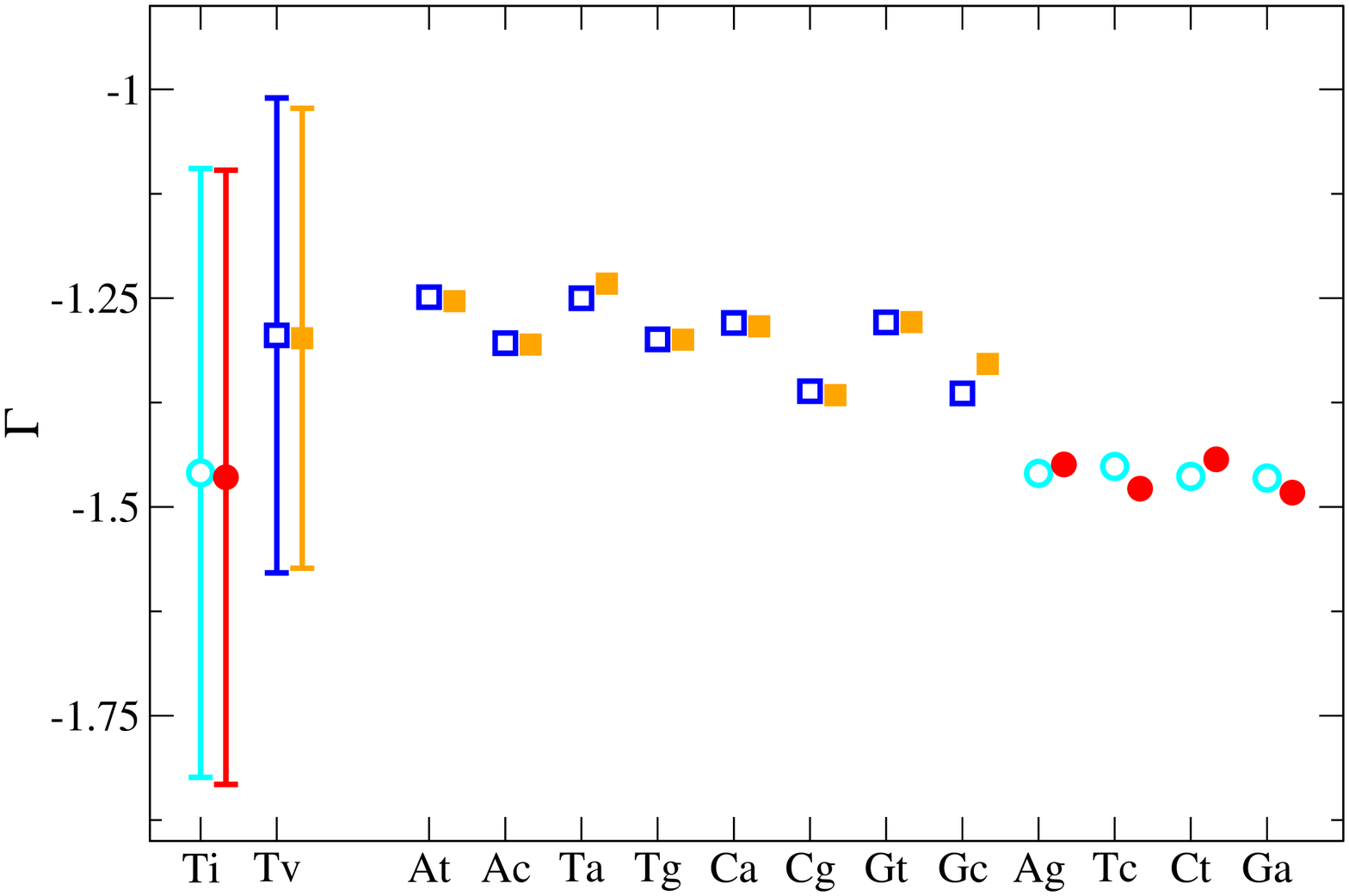}
\fi
\caption{Distributions of $\Gamma$ for the 1D (a) and 2-leg (b) models
  for all genes, with mutations divided into transitions and
  transversions. The distributions are normalised by the size of the mutation
  dataset.  Lines are guides to the eye only.  The means (symbols) and
  standard deviations (error bars) of the distributions of $\log {\Gamma}$
  are shown in panels (c) and (d) for the 1D and 2-leg models.  {\em Estimated
    errors of the means are smaller than the symbols.} Distributions are shown
  for transition (Ti) and transversion (Tv) mutations, and for the twelve
  types of point mutation individually. Open symbols (blue, cyan) are for the
  set of all mutations, filled symbols (orange, red) for the set of
  pathogenic mutations.
}
\label{fig-gamma-162}
\end{figure}

In Fig.~\ref{fig-gamma-162}(a+b) we show the distribution of ${\Gamma}$
values for our entire dataset of all $\simeq 34 \times 10^6$ possible
mutations and $19882$ known pathogenic mutations, dividing the datasets into
transitions and transversions. For both models, the transitions are shifted to
slightly lower ${\Gamma}$ values than the transversions. However, in the
2-leg model, the distribution for observed pathogenic transitions appears co-located
with the distribution for all transitions, and likewise for transversions. In
the 1D model, by contrast, the observed pathogenic transitions are visibly shifted to
lower ${\Gamma}$ values than the set of all transitions, and the same is
true for transversions.

In Fig.~\ref{fig-gamma-162}(c+d) we represent the distributions of
${\Gamma}$ values for each of the twelve types of point mutation by points
for the mean values of $\log {\Gamma}$ and bars indicating the standard
deviation of the distribution of $\log {\Gamma}$. In the 2-leg model, the
distributions for observed pathogenic mutations are essentially coincident with the
distributions for all mutations for each type Pq. The positive $\bar{\Lambda}$
and $\tilde{\Lambda}$ shift results in the 2-leg model are thus accounted for
by the set of observed pathogenic mutations being biased towards transitions.  The 1D
model displays a quite different behaviour; in each case the mean of the
distribution for the observed pathogenic mutations of any type Pq, lies from $7.5$ to
$20$ standard errors {\em below} the mean for all possible mutations of type
Pq.
Hence the probability that the observed pathogenic mutations are a random subset of
all mutations, with respect to their electronic properties in the 1D model, is
comparable to the probability of drawing twelve values more than $7.5$
standard deviations below the mean from a normal distribution, which is less
than $10^{-168}$. The observed difference between CT change between
observed pathogenic and all possible mutations is thus statistically highly
significant irrespective of whether transitions or transversions are involved.
In the 2D model, by contrast, the means of the $\log {\Gamma}$
distributions for observed pathogenic mutations can lie either above or below those
for all mutations for different types Pq, and the difference in the means ---
between $0.03$ and $5.5$ standard errors --- is much smaller.

\begin{figure}[h]
\centering
\ifSRSUB\else
\includegraphics[width=0.9\columnwidth,angle=-270,scale=0.7]{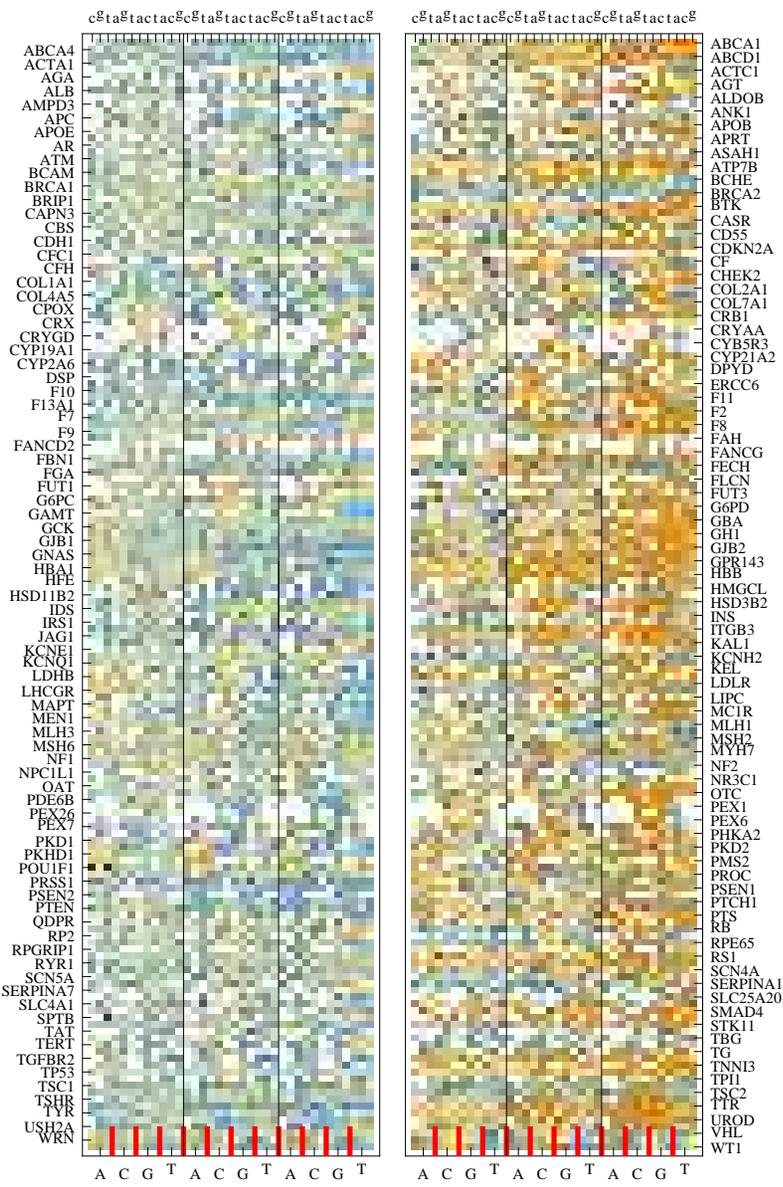}
\fi
\caption{Distribution of subset shifts $\lambda$ for the 1D (top) and
  2L (bottom) models over all $162$ genes split into the $12$ possible
  mutations (Ac, Ag, At, Ca, \ldots, Tc, Tg). The capital letters on the right
  axes denote the original base pairs, whereas the lowercase letters in the
  left axes show the mutant base. The short red tick marks on the right axes
  distinguish different original bases.  The system sizes $L=20$, $40$ and
  $60$ are shown in the bottom, centre and top row for each model. The orange
  shading corresponds to positive $\lambda$ and blue to negative.  The white
  squares correspond to cases for which either no corresponding pathogenic
  mutations are known ($1029$ cases) or for which the subset shift is
  inconclusive ($3$ cases for the 2-leg model).  }
\label{fig-162by36}
\end{figure}

Let us also consider, for each gene $g$, simulation length $L$ and each
mutation type Pq whether the {\em subset shift} $\lambda = \left\langle \log
{\Gamma}_{\rm all} \right\rangle - \left\langle \log {\Gamma}_{\rm pa}
\right\rangle_{g, L, {\rm Pq}}$ is positive or negative. This gives us, for
each model, $162 \times 3 \times 12 = 5832$ data points, less $1029$ cases
where no calculation is possible as no pathogenic mutations of type Pq are
known for gene $g$. These $\lambda$ data are presented in
Fig.~\ref{fig-162by36}.
In the 2-leg model there are approximately equal numbers of negative and
positive $\lambda$ values. This is consistent with a null hypothesis where the
observed pathogenic mutations of a type Pq have the same distribution of
${\Gamma}$ vales as for all mutations of that type. In the 1D model, by
contrast, such a null hypothesis is decisively rejected: there is a
preponderance of positive $\lambda$ values by almost $2:1$ ($3326$ positive to
$1513$ negative) and the binomial probability of obtaining such a result at
random would be approximately $10^{-153}$. The two analyses agree that
observed pathogenic mutations display a significant bias towards smaller changes in
electronic properties in the 1D model.

\section*{Discussion}

Our CT models act as probes of the statistics of the DNA sequence. It is
possible that we are merely observing a correlation; i.e.\ that
mutations are more likely to occur in areas of the genome with certain
statistical properties, for reasons not causally related to charge transport,
and these properties correlate with biased CT properties in our 1D model. Such
a correlation between quantum transport and mutation hotspots would in itself
be a valuable and novel observation in bioinformatics. There are known
chemical biases in the occurence of mutations, such as the enhanced transition
rate in C-G doublets,\cite{BlakeHT92} the bias towards GC base pairs rather
than AT pairs in biased gene conversion\cite{GalD07,Mar03} and the tendency of
holes to localise on GG and GGG sequences and there cause oxidative
damage.\cite{NunHB01} However, since our observed bias is consistent across
all twelve types of point mutation, these known biases cannot fully account
for our data.

There are also plausible causal connections between our data and cellular
genetic processes where the electronic properties of DNA may be
significant. One such process is gene regulation, where charge transport along
the DNA strand can couple to redox processes in DNA-bound proteins, inducing
protein conformational change and unbinding.\cite{AugMB07} Similarly, it has
been proposed that DNA repair glycosylases containing redox-active [4Fe-4S]
clusters\cite{BoaYB07} may localise to the site of DNA lesions through a
DNA-mediated charge transport mechanism.\cite{YavSOD06} The recognition of
specific areas in the DNA sequence by DNA-binding proteins generally may
involve electrostatic recognition of the target DNA sequence.\cite{CheKK08}
Furthermore, homologous recombination\cite{FerA01} --- a process which is
vital to the repair of double-strand breaks, a most serious DNA
lesion,\cite{Jac02,KhaJ01} and also to genetic recombination --- relies on the
mutual recognition of homologous chromosomes before strand invasion can
occur. Homologous double-stranded DNA sequences are capable of mutual
recognition even in a protein-free environment,\cite{BalBRW08} presumably via
electronic or electrostatic interactions.\cite{KorL01}

All the above processes, especially those involving protein--DNA or DNA--DNA
recognition, would be less disrupted by a smaller change in the electronic
environment along the coding strand. From this point of view, the observed
mutations are biased to cause {\em less} disruption to gene regulation and DNA
damage repair in the cell. This may seem counterintuitive at first. However,
in order for a mutation to appear in our dataset of pathogenic mutations, the
cell and the organism must develop viably for long enough for a mutant
phenotype to be observed. Mutations which cause large disruptions to DNA regulation and repair are more likely to be lethal to the cell at an early stage and will thus be absent from disease databases. Similarly, mutations which are more visible to DNA repair mechanisms are less likely to persist and to appear in databases.

Genetic repair and regulation mechanisms cannot know whether the consequences
of a mutation are beneficial, neutral or harmful.  We would therefore predict
that neutral mutations should display the same bias, towards smaller change in
electronic structure, as we observe in the pathogenic mutations. As a first
test of this prediction, we have considered the case of the TP53 gene, with
$20303$ base pairs and for which there are known $2003$ pathogenic mutations,
$366$ silent mutations and $113$ intronic mutations.\cite{PetMKI07} We have
simulated these silent and intronic mutations using the 1D model. Histograms
of the distribution of $\Gamma$ values for these mutations are given in
supplementary material, see Fig.\ \ifSRSUB S7\else\ref{fig-neutral}\fi. In Table
\ref{tab-neutral} we analyze the statistical properties for the resulting $\Gamma$
distributions; our results demonstrate that, for both transitions and
transversions, the silent and intronic mutations are similar to the pathogenic
mutations and significantly {\em disimilar} to the population of all possible mutations, as predicted.

\section*{Methods}

\subsection{Models of charge transport in DNA.}

\begin{figure}[h]
\centering
\ifSRSUB\else
(a)\includegraphics[width=0.45\columnwidth]{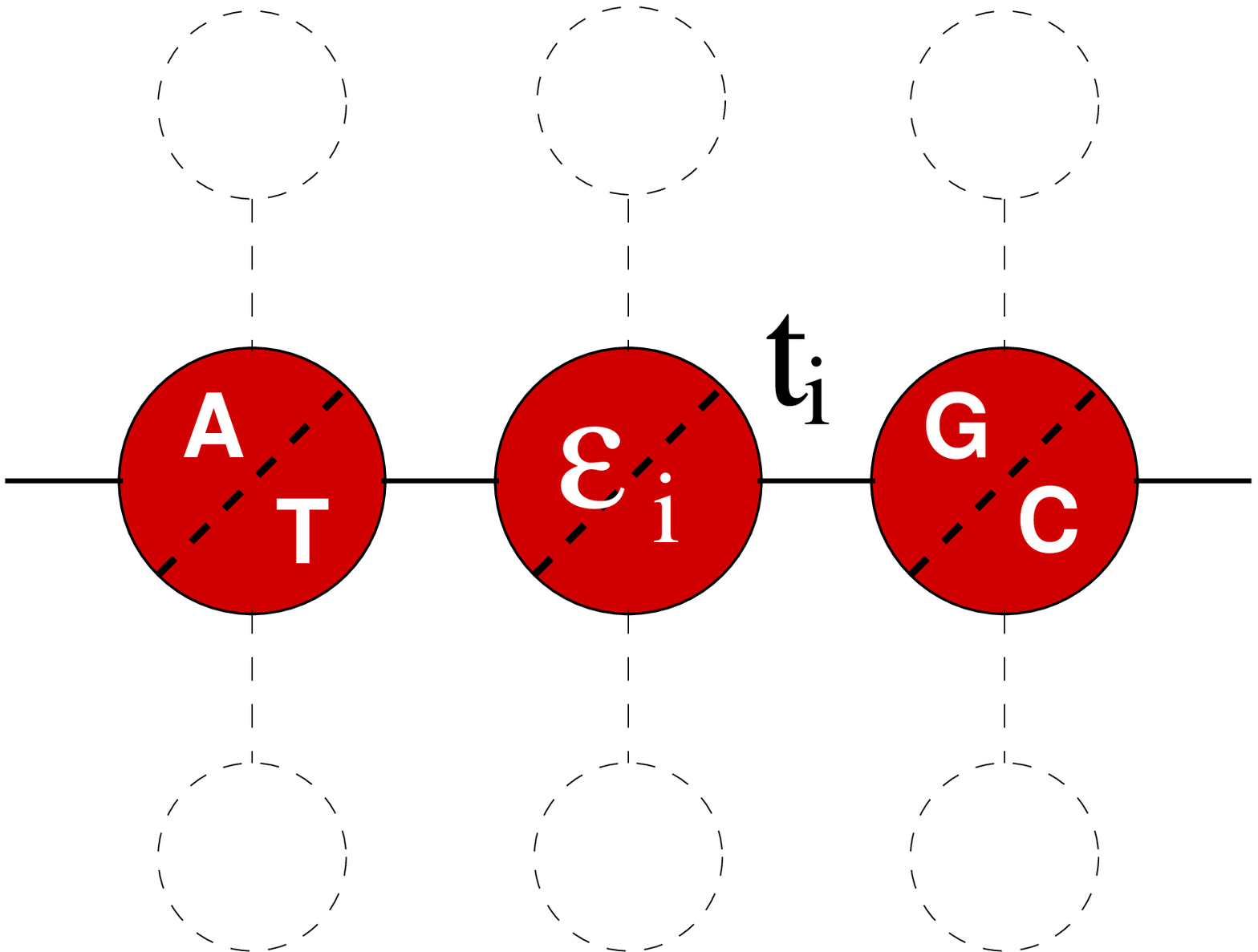}
(b)\includegraphics[width=0.45\columnwidth]{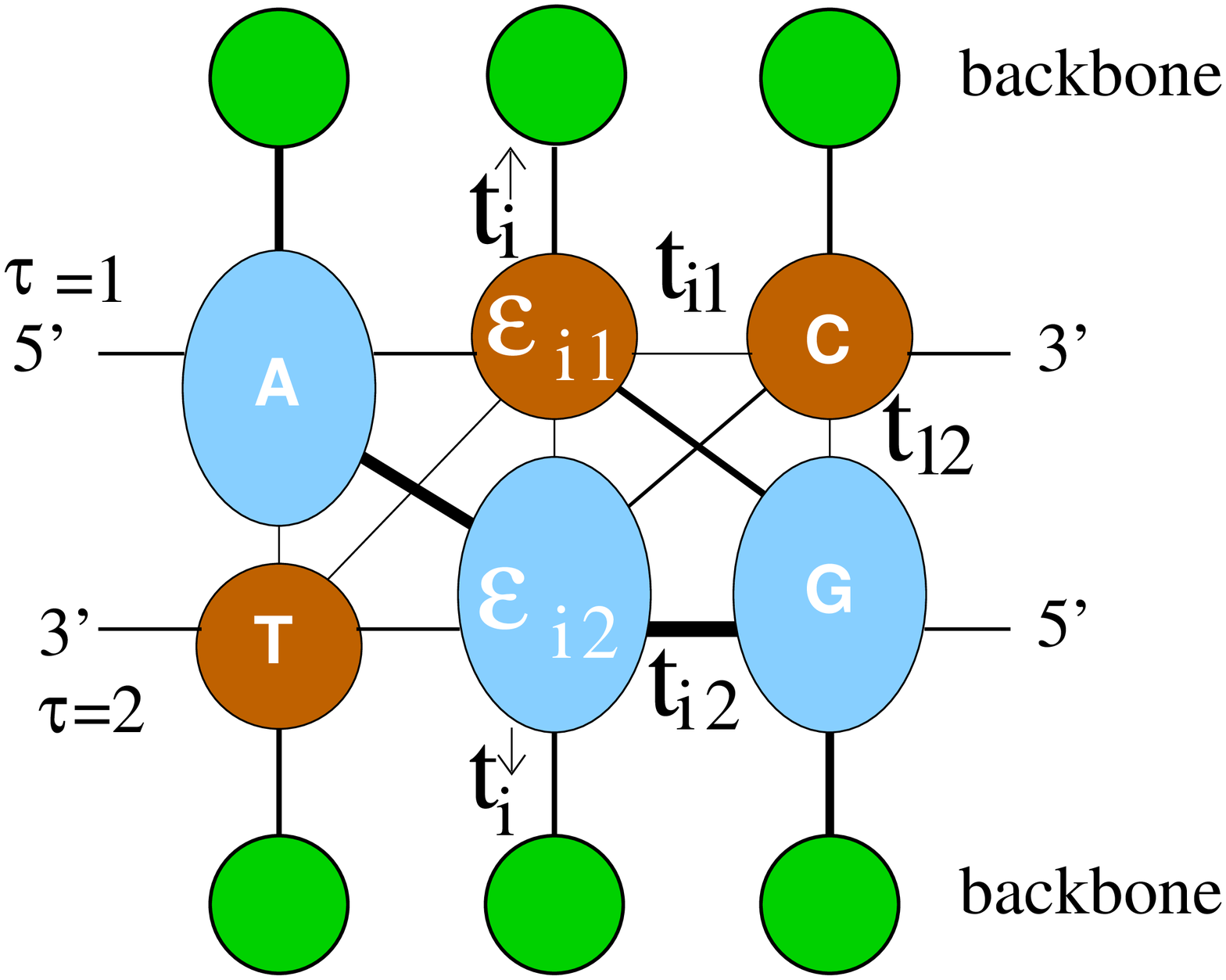}
\fi
\caption{Schematic models for charge transport in DNA. The nucleobases
  are given as circles (red, denoting pairs) and ellipses (blue, brown for
  single nucleotides).  Electronic pathways are shown as solid lines of
  varying thickness to indicate variation in strength. Model (a) indicates the
  1D model where the sugar-phosphate backbone is ignored. In model (b), brown
  circles denote the smaller pyrimidines, blue ellipses are the large purines
  and green circles denote the sugar-phosphate backbone sites. Note that
  diagonal hopping between purines is favoured, and between pyrimidines
  disfavoured, by the larger size of the purines.}
  \label{fig-models}
\end{figure}

The simplest model of coherent hole transport in DNA is given by an effective
one-dimensional H\"{u}ckel-Hamiltonian for CT through nucleotide HOMO
states,\cite{Cha07} where each lattice point represents a nucleotide base
(A,T,C,G) of the chain for $n=1, \ldots, N$.  In this tight-binding formalism,
the on-site potentials $\epsilon_n$ are given by the ionisation potentials
$\epsilon_{\rm G}=7.75 e$V, $\epsilon_{\rm C}=8.87 e$V, $\epsilon_{\rm A}=8.24
e$V and $\epsilon_{\rm T}=9.14 e$V, at the $n$th site,
cp.\ Fig.~\ref{fig-models}; the hopping integrals $t_{n,n+1}$ are assumed to be
nucleotide-independent with $t_{n,n+1}=0.4e$V.\cite{Cha07}
A model which is less coarse-grained is provided by the diagonal, 2-leg ladder
model shown in Fig.~\ref{fig-models}. Both strands of DNA and the backbone are
modelled explicitly and the different diagonal overlaps of the larger purines
(A,G) and the smaller pyrimidines (C,T) are taken into account by suitable
inter-strand couplings.\cite{RakVMR02,WelSR09} The intra-strand couplings are
$0.35 e$V between identical bases and $0.17 e$V between different bases; the
diagonal inter-strand couplings are $0.1 e$V for purine-purine, $0.01 e$V for
purine-pyrimidine and $0.001 e$V for pyrimidine-pyrimidine. Perpendicular
couplings to the backbone sites are $0.7 e$V, and perpendicular hopping across
the hydrogen bond in a base pair is reduced to $0.005 e$V.

The 2-leg model\cite{WelSR09} allows inter-strand coupling between the purine
bases in successive base pairs, in accordance with electronic structure
calculations \cite{RakVMR02}, and should therefore be a better model for bulk
charge transport along the DNA double helix; the 1D model, by contrast, makes
use of the site energies of only the bases on the coding strand, \cite{ShiRR08}
and so is most representative of the electronic environment along that strand.
We also find that the 2-leg model recovers some of the coding strand
dependence of the 1D model upon decreasing the diagonal hoppings. For $28$
genes, we find that reducing only the diagonal hopping elements by two leads
to a much greater agreement with the 1D results similar to
Fig.\ \ref{fig-gamma-162}(c).

\subsection{Calculation of quantum transmission coefficients.}
\label{sec-calc-quantum}

The quantum transmission coefficient $T(E)$ for a DNA sequence with length $N$
bps for different injection energy $E$ can be calculated for both models by
using the transfer matrix method.\cite{Roc03,NdaRS04} Let us define
$T_{j,L}(E)$ as the transmission coefficient for a part of a given DNA
sequence which starts at base pair position $j$ and is $L$ base pairs long.
The {\em position-dependent averaged transmission coefficient} at the $k-$th
base pair for transmission length $L$ bps is defined as
\begin{equation}
  {T}^{(k)}_{L}=\frac{1}{L}\sum_{j=k-L+1}^{k}
  \int^{E_1}_{E_0} \frac{T_{j,L}(E)}{E_1-E_0} dE\quad .
\label{eq:ave_te}
\end{equation}
Here $j$ ranges from $k-L+1$ to $k$ such that each subsequence of length $L$
contains the $k$th base pair. $E_0$ and $E_1$ are the lower and upper bounds
of the incident energy of the carriers, e.g.\ for the $1D$ model used here,
the values are $5.75$ and $9.75 e$V, respectively; for the 2-leg model the
bounds are $7$ and $11 e$V. We have used an energy resolution of $\Delta E =
0.005 e$V.
Then we examine the difference between transmission coefficients of the
normal and mutated genomic sequence of a point mutation\cite{ShiRR08} and
hence denote by $T_{j,L}^{(k,s)}$ the transmission coefficient of the same
segment of DNA as $T_{j,L}^{(k)}$ but with the point mutation
$(k,s)$. ${\Gamma}^{(k,s)}_{L}$ is the averaged effect of the point
mutation $(k,s)$ on CT properties for all subsequences of length $L$
containing the mutation,
\begin{equation}
 {\Gamma}^{(k,s)}_{L}=\frac{1}{L}\sum_{j=k-L+1}^{k} 
 \int^{E_1}_{E_0}
  \frac{|T_{j,L}(E)-T^{(k,s)}_{j,L}(E)|^2}{E_1-E_0} dE \quad .
\label{eq:gamma}
\end{equation}





\begin{addendum}
 \item[Supplementary Information] available
 \item This work was supported by the National Science Council in Taiwan (CTS,
   Grant No.\ 97-2112-M-029-002-MY3) and the UK Leverhulme Trust (RAR, SAW,
   Grant No.\ F/00215/AH). Part of the calculations were performed at the
   National Center for High-Performance Computing in Taiwan. We are grateful
   for their help.
 \item[Author Contributions] CTS and RAR coordinated the international collaboration and wrote the main manuscript text. CTS, RAR and SAW wrote the programs and performed the main computation. YYC and CLH analyzed the source databases and performed the data preprocessing. All authors analyzed the data and reviewed the manuscript.
 \item[Competing Interests] The authors declare that they have no competing
   financial interests.
 \item[Correspondence] Correspondence and requests for materials should be
   addressed to CTS~(email: ctshih@thu.edu.tw) or RAR~(email:
   r.roemer@warwick.ac.uk).
\end{addendum}

\clearpage
\newpage


\begin{table}\centering
{\small
\begin{tabular}{ p{2in}|c|c|c|c|c }
 & $\overline{\log_{10} \Gamma}$ & SEM & $\sigma$ & $p_{\rm all}$ & $p_{\rm pa}$ \\
\hline \hline
All transitions & $-1.753$ & $0.003$ & $0.427$ & -& - \\
\hline
Pathological transitions & $-1.840$ & $0.015$ & $0.431$ & $1.01\times^{-8}$& - \\
 \hline
Silent transitions & $-1.868$ & $0.029$ & $0.440$ & $6.62\times 10^{-5}$ & $0.391$ \\
 \hline
Intron transitions & $-1.805$ & $0.048$& $0.391$& $0.320$ & $0.526$ \\
 \hline
\hline
All transversions & $-1.605$ & $0.002$ & $0.422$ & -& - \\
 \hline
Pathological transversions &  $-1.710$& $0.012$ & $0.4190$ & $<10^{-10}$ & - \\
 \hline
Silent transversions & $-1.691$& $0.036$& $0.432$ & $0.016$& $0.610$ \\
 \hline
Intron transversions & $-1.739$ & $0.054$ & $0.337$& $0.032$& $0.636$ \\
 \hline
\end{tabular}
\caption{ {\small\rm Mean logarithm of CT change $\Gamma$ for gene TP53 using the 1D model with $L=20$.  Data are divided into transition and transversions. We give standard errors of the mean (SEM) and standard deviations ($\sigma$) for each distribution. From these we estimate the probability of each distribution being a random sample from the set of all mutations, $p_{\rm all}$, or being a sample from a population similar to the pathogenic mutations, $p_{\rm pa}$ (cp.\ Fig.\ \ifSRSUB S7\else\ref{fig-neutral}\fi). There are $224$ silent transitions and $142$ silent transversions; $67$ intronic transitions and $46$ intronic transversions. The pathogenic mutations and all possible mutations outnumber the silent and intronic populations by factors of $10$--$1000$ and so it is the SEM for the smaller populations that is significant.  It is clear that the mean CT change $\overline{\log_{10} \Gamma}$ for the silent and intronic populations is far more similar to the pathogenic populations than to the entire population of all possible mutations. This is true for both transitions and transversions, although the $p$-value for the intronic transitions is not statistically significant (i.e.\ $\geq 0.05$) which we attribute to the small number of available intronic data.} }
\label{tab-neutral}
}
\end{table}

\clearpage
\newpage
%
%
\clearpage
\clearpage
\clearpage
\clearpage
\clearpage

\ifSRSUB\end{document}\else

\clearpage\newpage
\setcounter{figure}{0}
\setcounter{table}{0}
\def\thefigure{S\arabic{figure}}
\def\thetable{S\arabic{table}}
\setcounter{page}{1}

\section*{Supplementary Material}

\section*{Comparing the Averaged Electronic Properties for the Pathogenic and Non-pathogenic Mutations for Each Gene}

We denote the genomic sequence of a gene with length ${\cal N}$ base pairs
(bps) as $(s_1,s_2,\cdots,s_{\cal N})$. Each point mutation of a given gene is
characterized by the set $(k,s)$, where $k$ and $s$ are the position of the
point mutation in the genomic sequence and the mutant nucleotide which
replaces the nucleotide $s_k$ of normal DNA, respectively.
There are totally $3\cal{N}$ possible point mutations of a gene with
$\cal{N}$ bps. The sets of these $3\cal{N}$ mutations and the pathogenic
mutations for the gene are denoted as $M_{\rm all}$ and $M_{\rm pa}$,
respectively. $M_{\rm pa}$ is a subset of $M_{\rm all}$.
For every possible point mutation, we compute the {\em mean} quantum
mechanical transmission coefficient ${T}^{(k)}_{L}$ of a subsequence with
length $L$ of the {\em wild-type} gene. Here the mean is determined by
averaging over all individual transmission coefficients ${T}_{j,L}$ with
$j=k-L+1, k-L+2, \ldots, k$. In this way, the influence of the full
neighborhood of hotspot $k$ is taken into account and not just the mutation
itself.
The results of ${T}^{(k)}_{L}$ for $k \in M_{\rm pa}$ already show some
signatures of atypical CT reponse for the 1D model.\cite{Shi06a} However, the
signal is much less pronounced in the 2-leg model.
Hence we study the {\em difference} in CT between a healthy DNA base and the
$3$ possible mutations. For example the hotspot $14585$ of $p53$ contains the
correct $C/G$ base pair in the wild but of the three possible mutations $C/G
\rightarrow G/C$, $C/G \rightarrow A/T$ and $C/G \rightarrow T/A$ only the
last one is know to lead to cancer.\cite{PetMKI07} Averaging again over all
incident energies and subsequences of length $L$ containing the hotspot
$(k,s)$, we can characterize the {\em average change} in CT as
\begin{equation}
 {\Gamma}^{(k,s)}_{L,q}=\frac{1}{L}\sum_{j=k-L+1}^{k} 
 \int^{E_1}_{E_0}
  \frac{|T_{j,L}(E)-T^{(k,s)}_{j,L}(E)|^q}{E_1-E_0} dE \quad .
\label{eq:gammaq}
\end{equation}
with $q=1$ or $2$. We find that results for $q=1$
and $2$ are similar. Hence in the manuscript we restrict our discussion to
$q=2$. We calculate such $\Gamma$ estimates for all possible $3\cal{N}$
mutations of each gene and compare the probability distribution of CT change
${\Gamma}^{(k,s)}_{L,q}$ for $(k,s)\in M_{\rm all}$ and $(k,s)\in M_{\rm
  pa}$ for each gene.
The result for the $p16$ gene was shown in Fig.~\ref{fig-Gamma_percent-p16}(a) as an example. As a control group, we also shuffled the $p16$ sequence randomly under the conditions that (1) the contents of the $4$ bases are not changed, and (2) the
positions of the mutations can be moved but the numbers of the $12$ types of
mutations are not changed. The distributions of the averaged $\Gamma$ for 1D
and 2-leg models with $L=40$ of the $20$ shuffled sequences are shown in
Fig.\ \ref{fig-sub-Gamma_shuffled-p16}. It is clear that the distributions of
$\Gamma$ for the $M_{all}$ and $M_{pa}$ are almost identical.

\section*{CT Change for the 12 Type of Mutations}

The comparison of $\Gamma$ between the pathogenic and all possible mutations
for the $12$ types of point mutations is shown in
Fig.\ \ref{fig-sub-gamma-12type}. It is clear for the 1D model (a--l)
$\Gamma$ tends to be smaller for the pathogenic mutations. However, the
difference is not visible for the 2L model (m--x).

\section*{Local ranking of point mutations at hotspot sites}

In order to study the local effects of pathogenic mutations on CT, we compare
$\Gamma^{(k,s)}_{L,2}$ of each pathogenic mutation $(k,s)$ with the other two
non-pathogenic ones at the same position $k$ and determine the {\em local
  ranking} (LR) of CT change for $(k,s)$. There are three possibilities of LR,
namely {\em low}, {\em medium} and {\em high}. Note that those hotspots $k$
with more than one pathogenic mutations are excluded in the LR analysis. As an
example, percentages of the three LR for the pathogenic mutations of $p16$ are
shown in the left panels of Fig.~\ref{fig-sub-L_ranking-p16}.
of pathogenic mutations with low CT change are evidently larger than the
medium and high ones for all $L$.
Let us again ask how significant this tendency is across all $162$
genes. Figure \ref{fig-sub-L_ranking-all} shows similar ranking analysis
results as in Fig.~\ref{fig-sub-L_ranking-p16} but now for {\em all} $M_{\rm
  pa}$.
We see that the
tendency towards low CT change in the pathogenic mutations is quite strong
overall.
In Fig.~\ref{fig-LG_ranking-all} we have sorted the LR ranking for each gene
according to prevalence. We find that for $L=20$, $40$ and $60$ the low CT
change corresponds to $155$ ($95\%$), $148$ ($91\%$) and $140$ ($86\%$) of all
$162$ genes with pathogenic mutations.  Note that similarly consistent is the
result for large CT with only about $30$ of all genes having high CT change.

\section*{Global CT rankings at hotspot sites}

Another way to compare the CT change is a {\em global} ranking (GR).  We have
sorted the CT change $\Gamma^{(k,s)}_{L,2}$ for {\em all} possible $3{\cal N}$
mutations of a gene with ${\cal N}$ bps in order to get a ranking of {\em
  every} pathogenic mutation $(k,s)$. By dividing each ranking by $3{\cal N}$
we compute the normalised GR $\gamma^{(k,s)}_{L,2}$ of the mutation with
values between $0$ and $1$. As before for $\Gamma^{(k,s)}_{L,q}$, smaller
values of $\gamma^{(k,s)}_{L,q}$ mean smaller CT change. To characterise the
CT change in a quantitative way, we divide the $\gamma^{(k,s)}_{L,2}$ of the
pathogenic mutations into again three groups as before, i.e.\ low ($\gamma <
33.3\%$), medium ($33.3\% \leq \gamma < 66.7\%$), and high ($\gamma \geq
66.7\%$) CT change.  The distributions of the GR for the complete set of
pathogenic mutations of $p16$ is shown in Fig.~\ref{fig-sub-L_ranking-p16} as
an example. As for the LR results, the pathogenic genes lead to many
$\gamma^{(k,s)}_{L,2}$ values with low CT change. This is most pronounced in
the 1D model as shown in Fig.~\ref{fig-sub-L_ranking-p16}(c).
The results of the GR for the $162$ genes are shown in the bottom row (c)
and (d) of Figs.~\ref{fig-sub-L_ranking-all} and \ref{fig-LG_ranking-all}. We
see that the GR results are fully consistent with the LR rankings.

\section*{Consistency of CT rankings for all DNA sequences}

The prevalence ordering as shown in Fig.~\ref{fig-LG_ranking-all} does not
imply that the order of the genes themselves is the same in all parts (a),
(b), (c) and (d) of the figure. Therefore we have calculated the correlations
in the ordering and found that in both models and across models and for all
$L=20$, $40$ and $60$, we find positive correlation coefficients. Hence genes
which have a low change in CT for, e.g., the local ranking at $L=20$, also
retain this low rank for the other $L$ values as well as the global
ranking. Similarly, this positive correlations implies that in those few case
where the mutations in a gene lead to high CT change, they do so across all
local as well as global rankings. This confirms that our results are
internally consistent.

We graphically summarise the results for all $162$ disease-related genes in
Fig.~\ref{fig-sub-all}. For each gene, we have shown a positive deviation from
the $0.33$ line by orange ---supporting the scenario of small CT change for
pathogenic mutations --- and by blue when the results seem to show no or
negative indication with CT change. The criteria corresponds to local and
global ranking results for $L=20$, $40$ and $60$ for the 1D and the 2-leg
models.
Similarly, in
Fig.~\ref{fig-sub-all-avg}, we average of all $12$ criteria and show the
resulting, overall agreement with the CT hypothesis: $161$ of $162$ genes are
above the $33\%$ line and hence show that for both 1D and 2-leg model and
averaged over lengths $20$, $40$ and $60$, a small CT change correlates with
the existence and position of pathogenic mutations.
Only for STK11 do we see that there is no overall agreement.

\section*{Difference and similarities in the two models}

The 2-leg model\cite{WelSR09} allows inter-strand coupling between the purine
bases in successive base pairs, in accordance with electronic structure
calculations,\cite{RakVMR02} and should therefore be a better model for bulk
charge transport along the DNA double helix; the 1D model, by contrast, makes
use of the site energies of only the bases on the coding strand,\cite{ShiRR08}
and so is most representative of the electronic environment along that strand.
We also find that the 2-leg model recovers some of the coding strand dependence of the 1D model upon decreasing the diagonal hoppings. For $28$ genes, we find that reducing only the diagonal hopping elements by $1/2$ leads to a much greater agreement with the 1D results similar to Fig.\ \ref{fig-gamma-162}(c).

\clearpage
\newpage
%
%
\def\thefigure{S\arabic{figure}}
\def\thetable{S\arabic{table}}

\clearpage
\setcounter{figure}{0}
\begin{figure}
(a)\includegraphics[width=0.45\columnwidth]{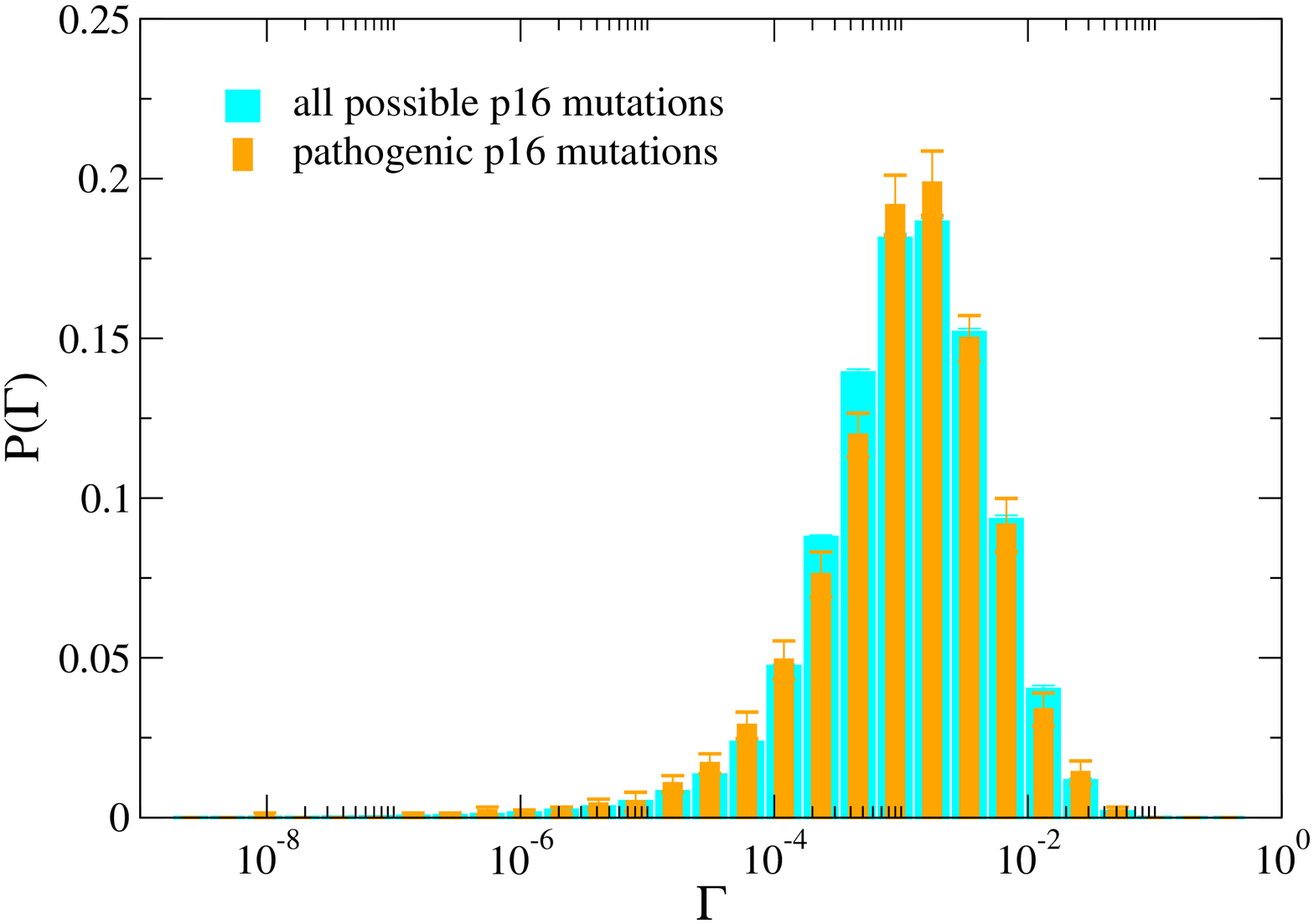}
(b)\includegraphics[width=0.45\columnwidth]{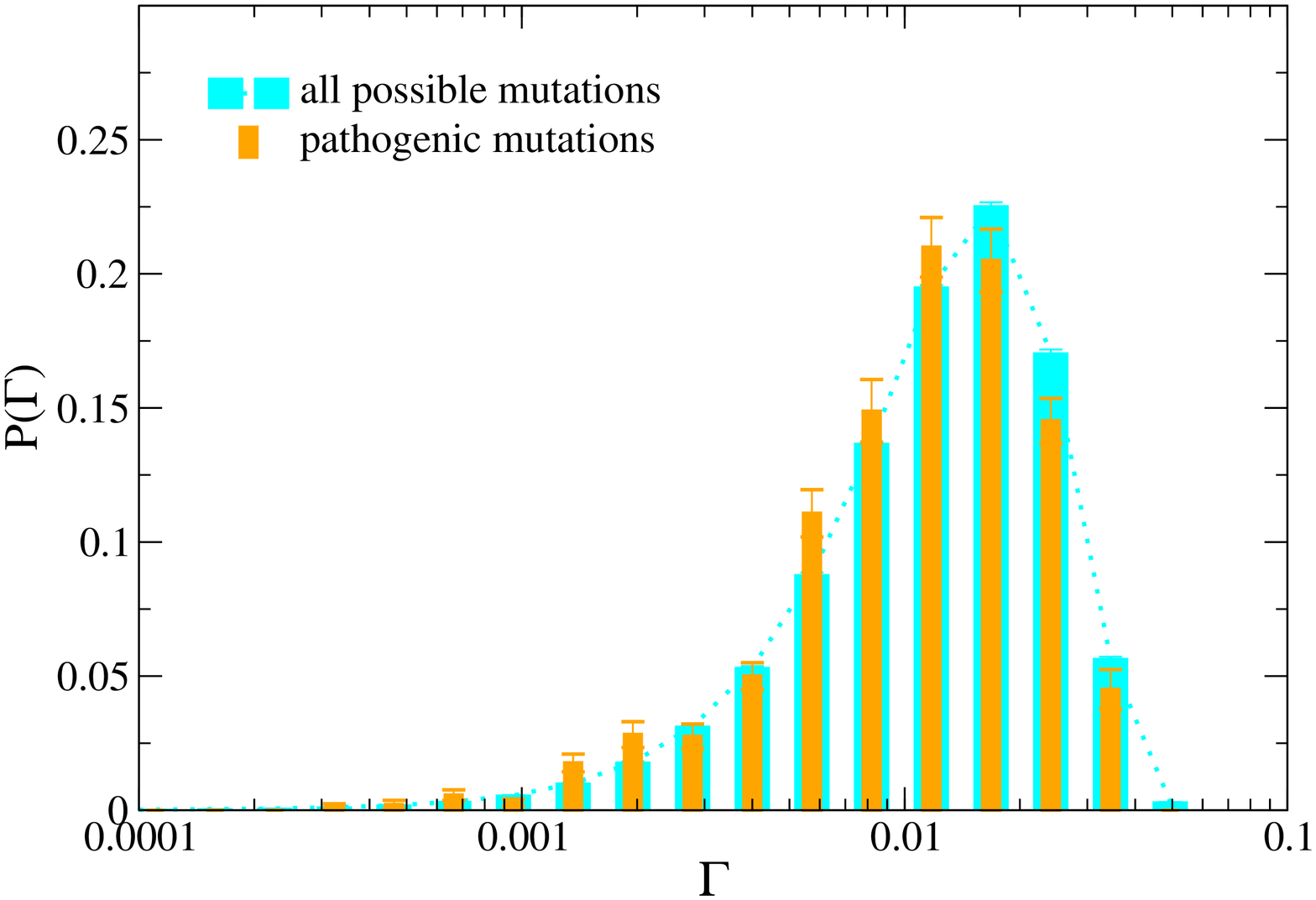}
\caption{(Supplementary) Distribution of the change in charge transport
  in (a) 1D and (b) 2L models $\Gamma$ for pathogenic (orange bars) and all
  possible (cyan bars) mutations averaged for the $20$ shuffled $p16$ (CDKN2A)
  DNA strands with $26740 $ base pairs. All results shown are for $L=40$, data
  for $L=20$ and $60$ are similar.  }
\label{fig-sub-Gamma_shuffled-p16}\label{fig-S1}
\end{figure}
\clearpage
\begin{figure}
\centering
(a) \includegraphics[width=0.2\columnwidth]{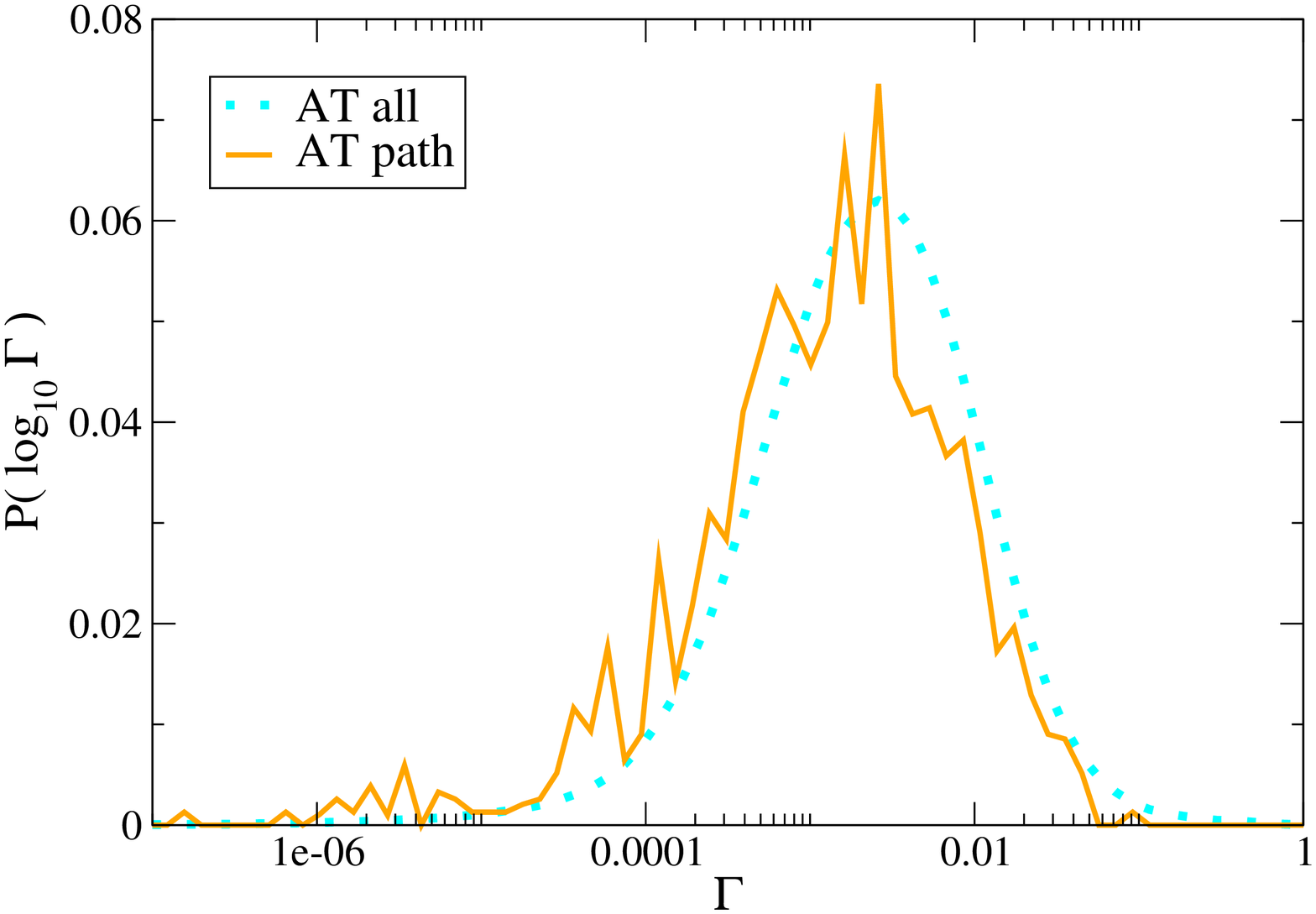}
(b) \includegraphics[width=0.2\columnwidth]{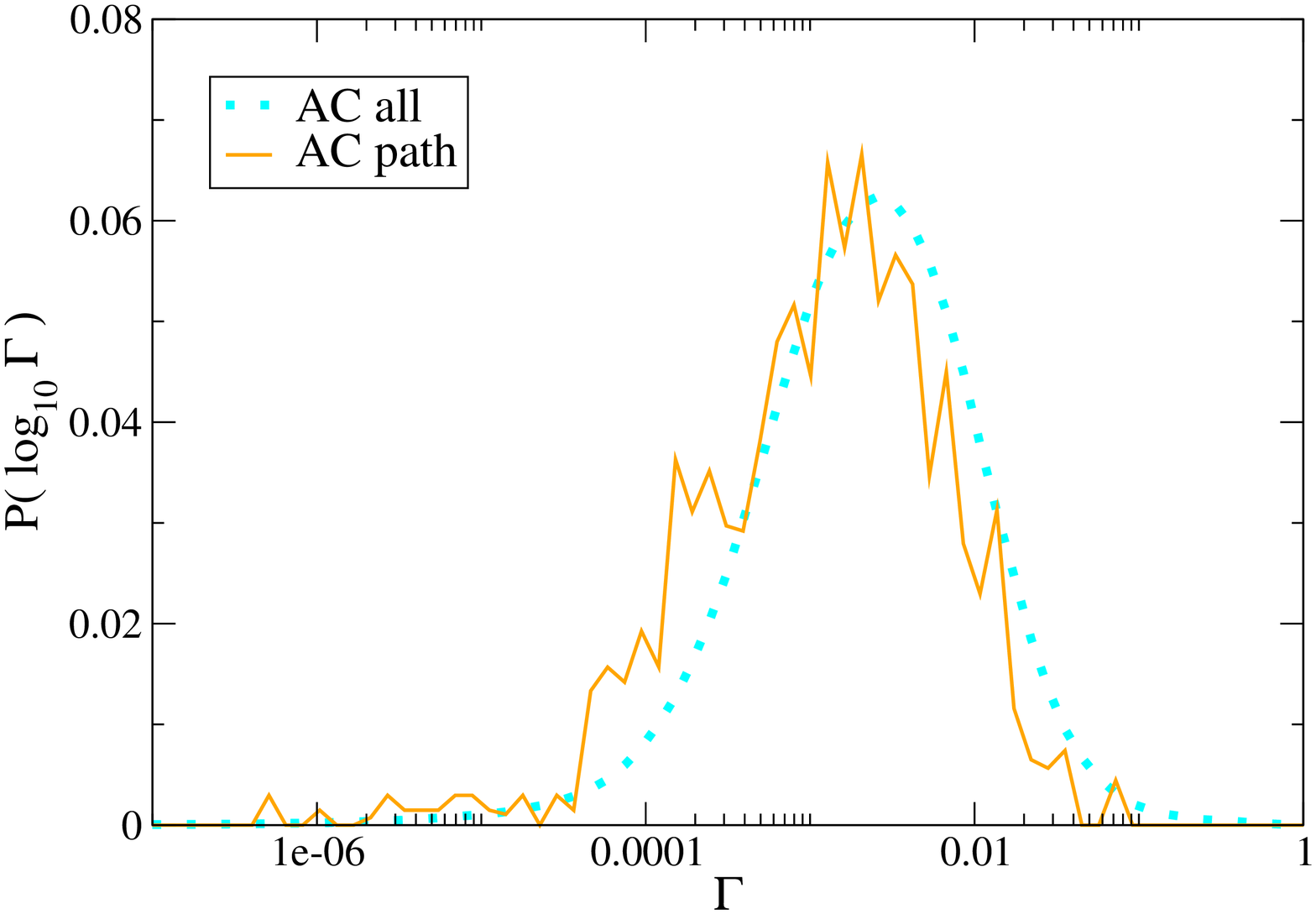}
(c) \includegraphics[width=0.2\columnwidth]{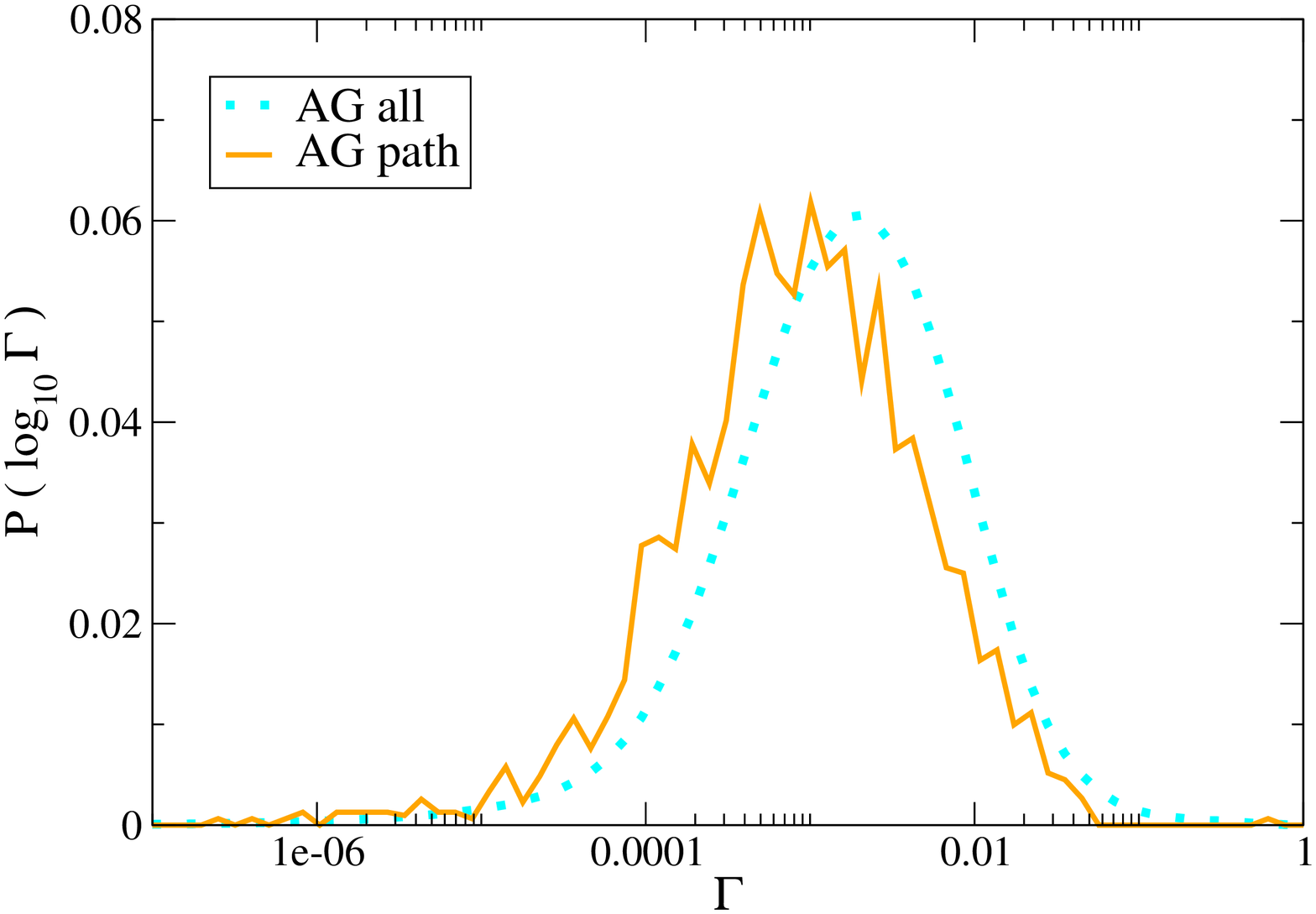}
(d) \includegraphics[width=0.2\columnwidth]{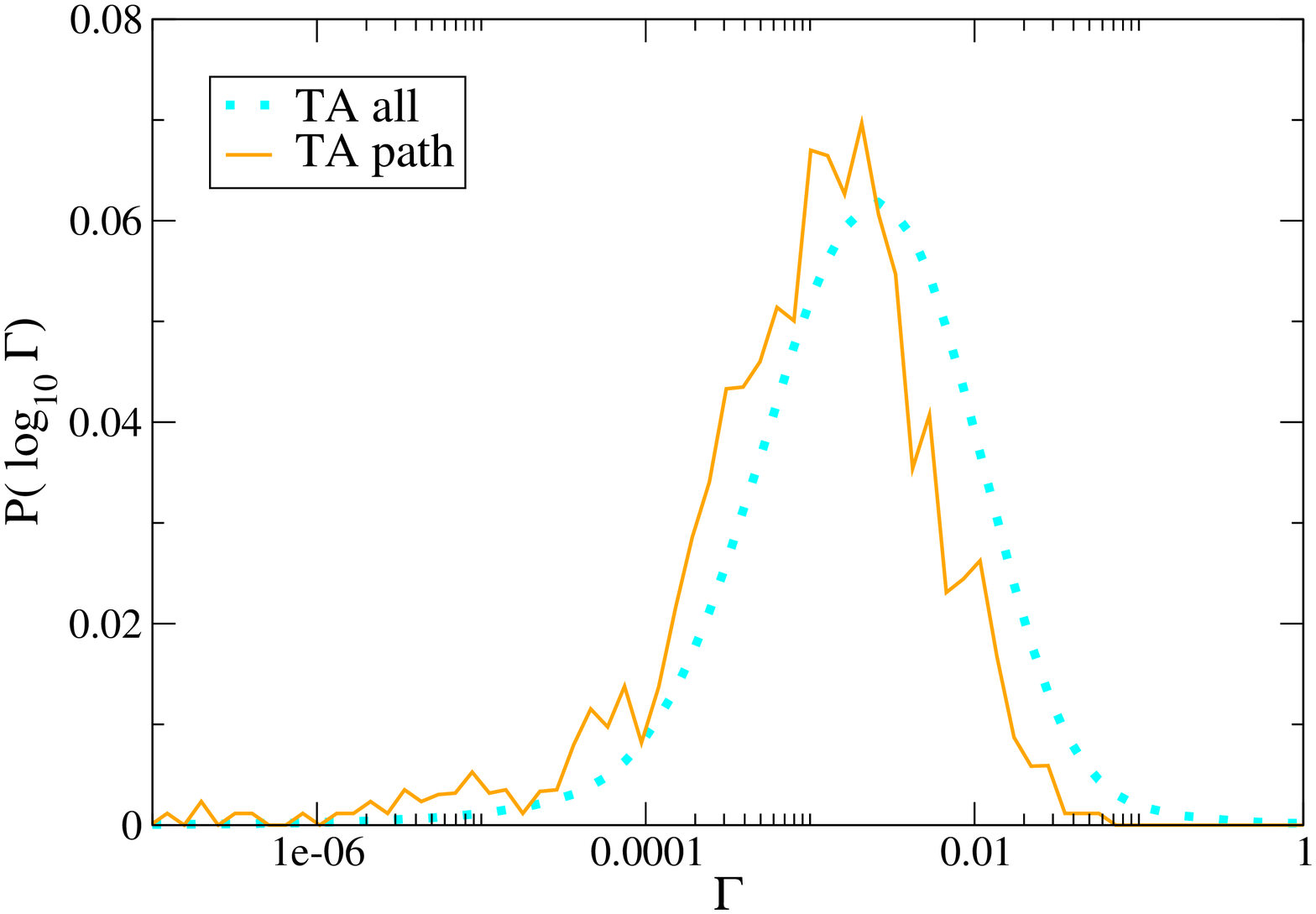}
(e) \includegraphics[width=0.2\columnwidth]{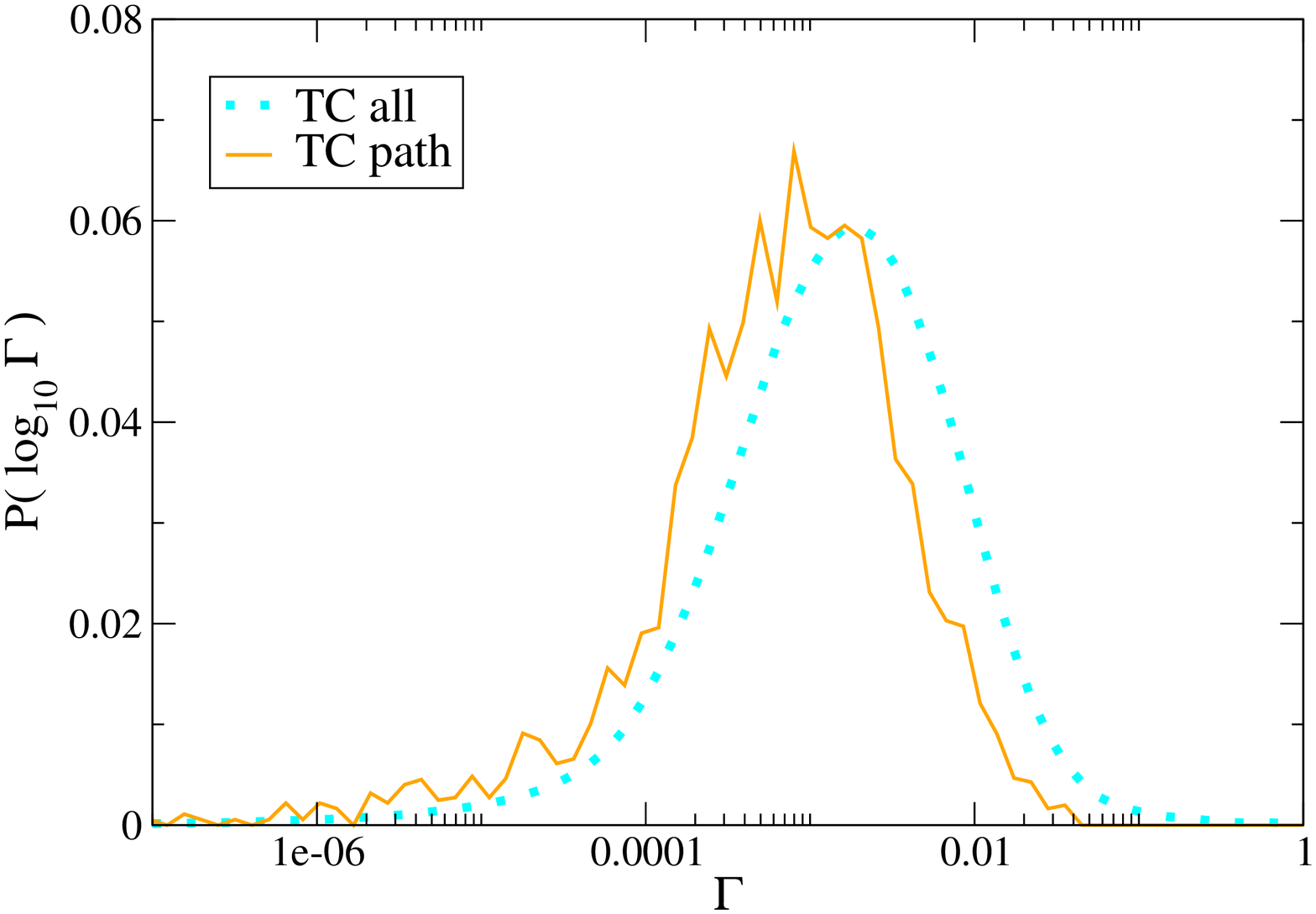}
(f) \includegraphics[width=0.2\columnwidth]{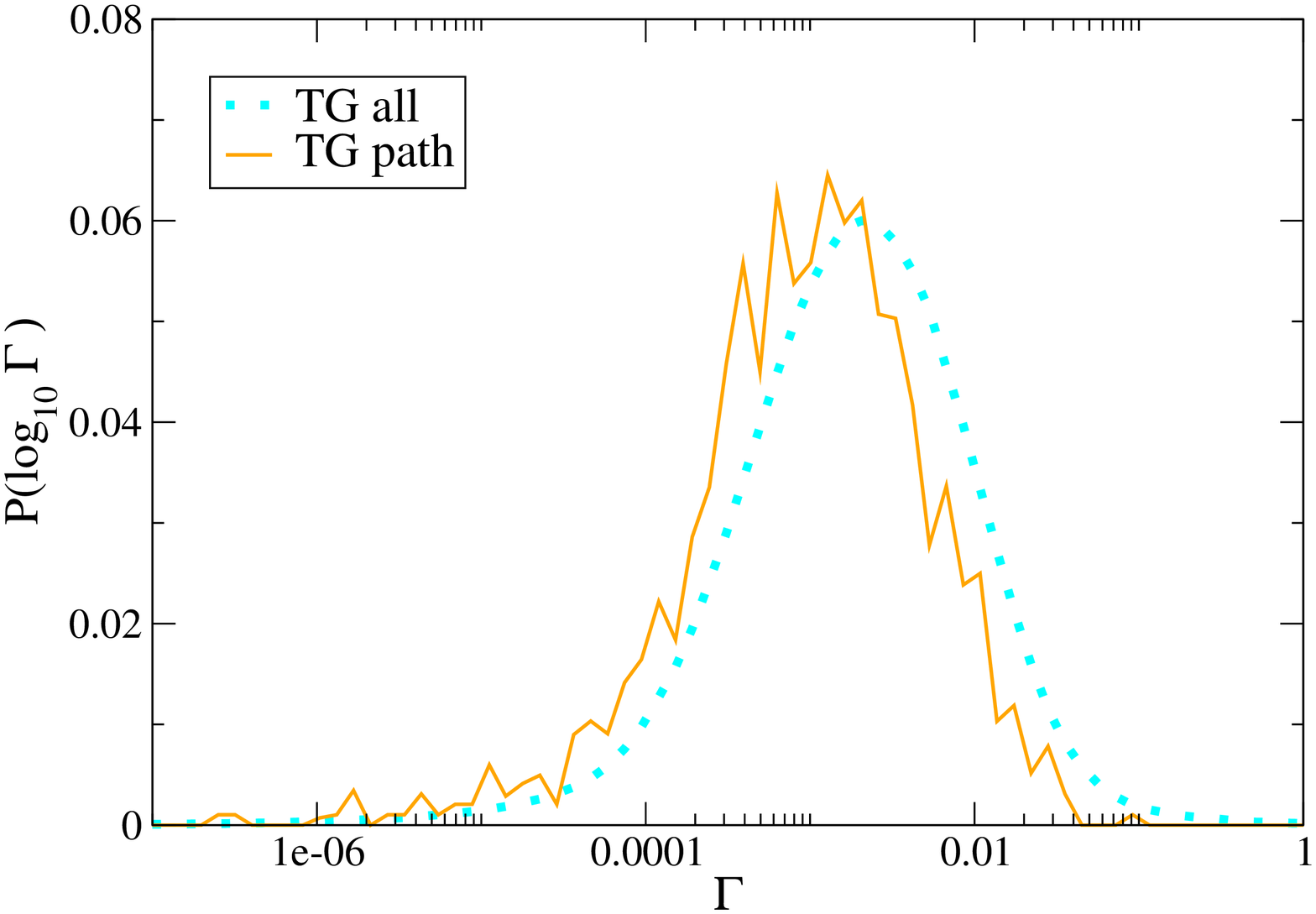}
(g) \includegraphics[width=0.2\columnwidth]{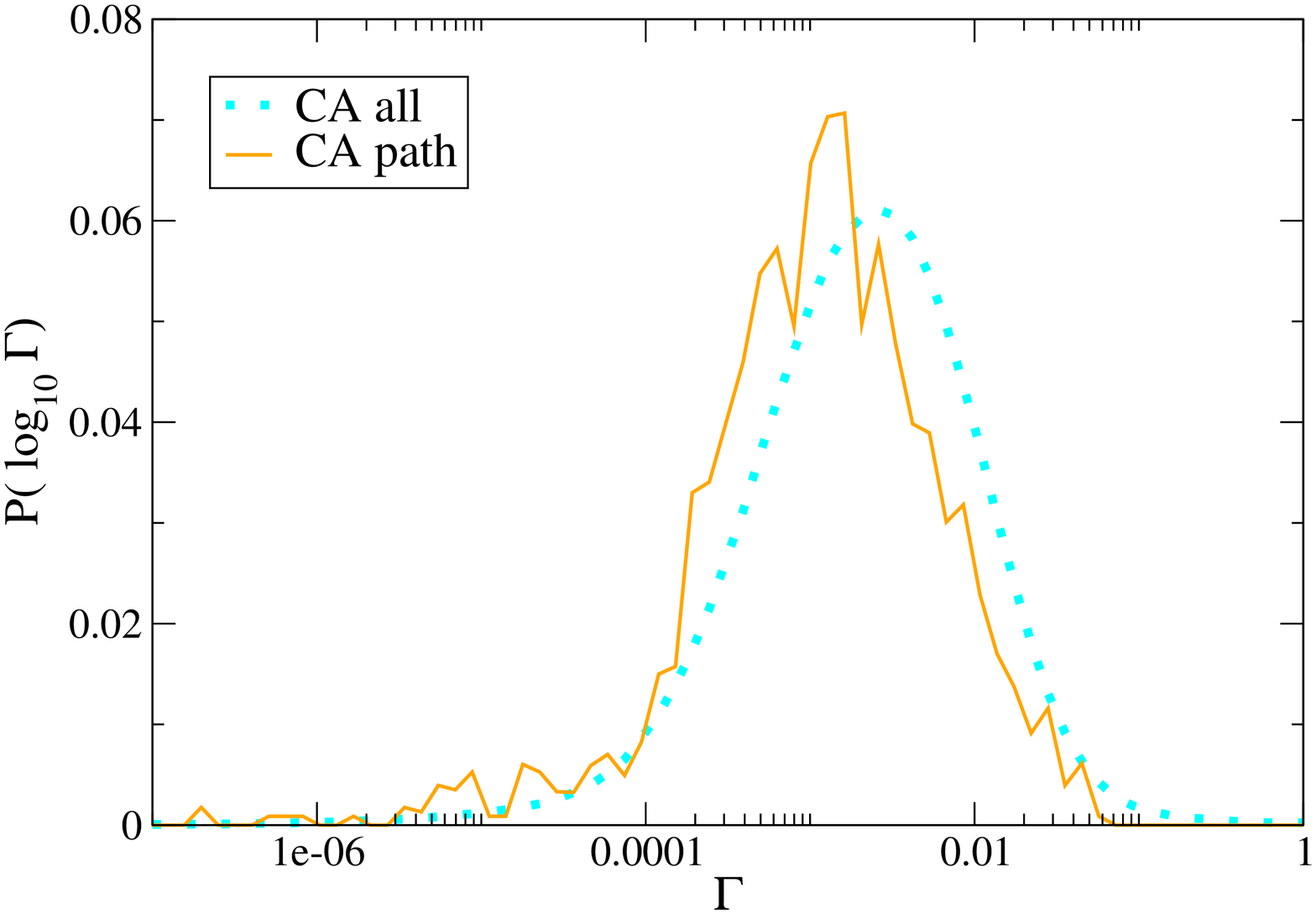}
(h) \includegraphics[width=0.2\columnwidth]{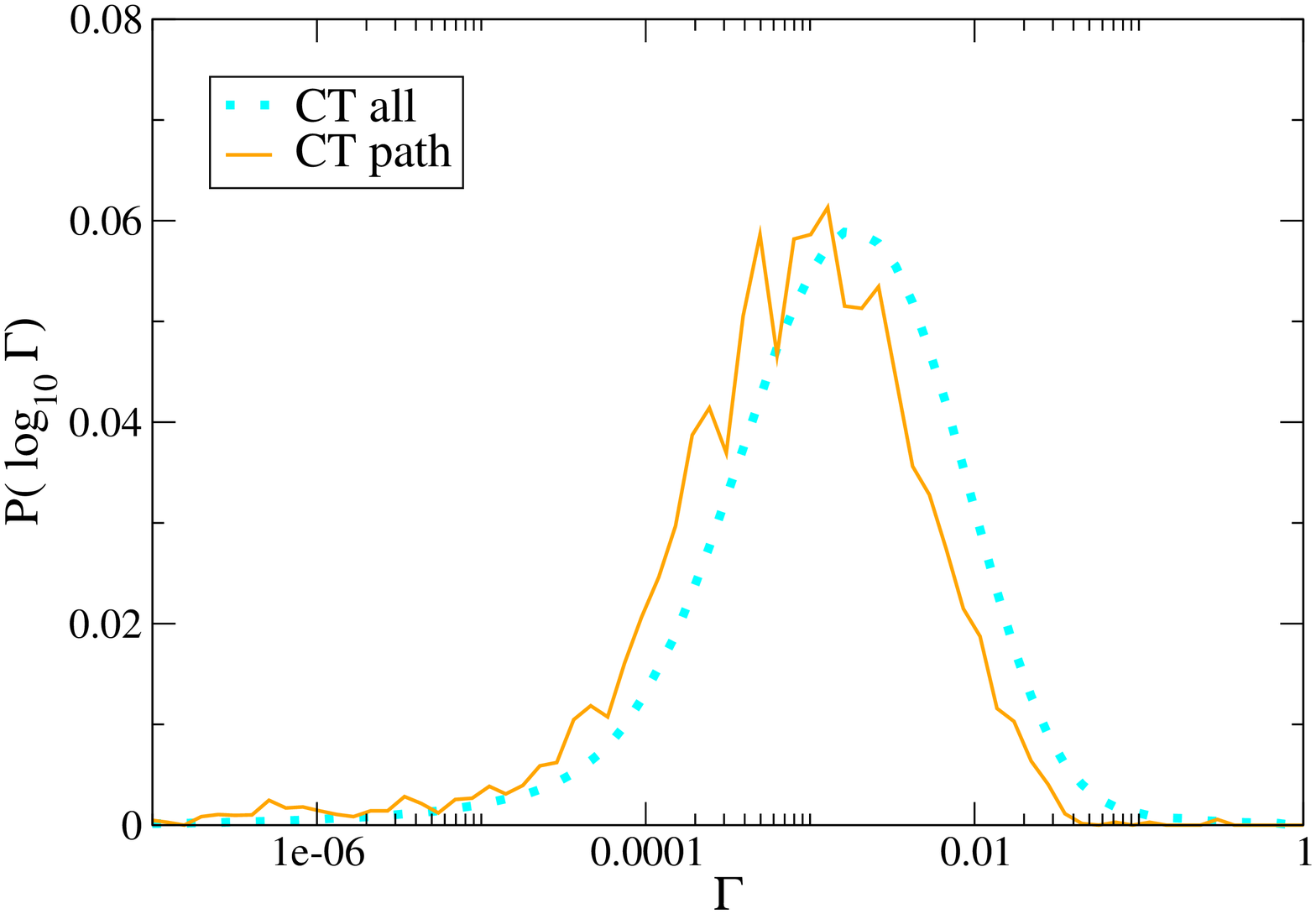}
(i) \includegraphics[width=0.2\columnwidth]{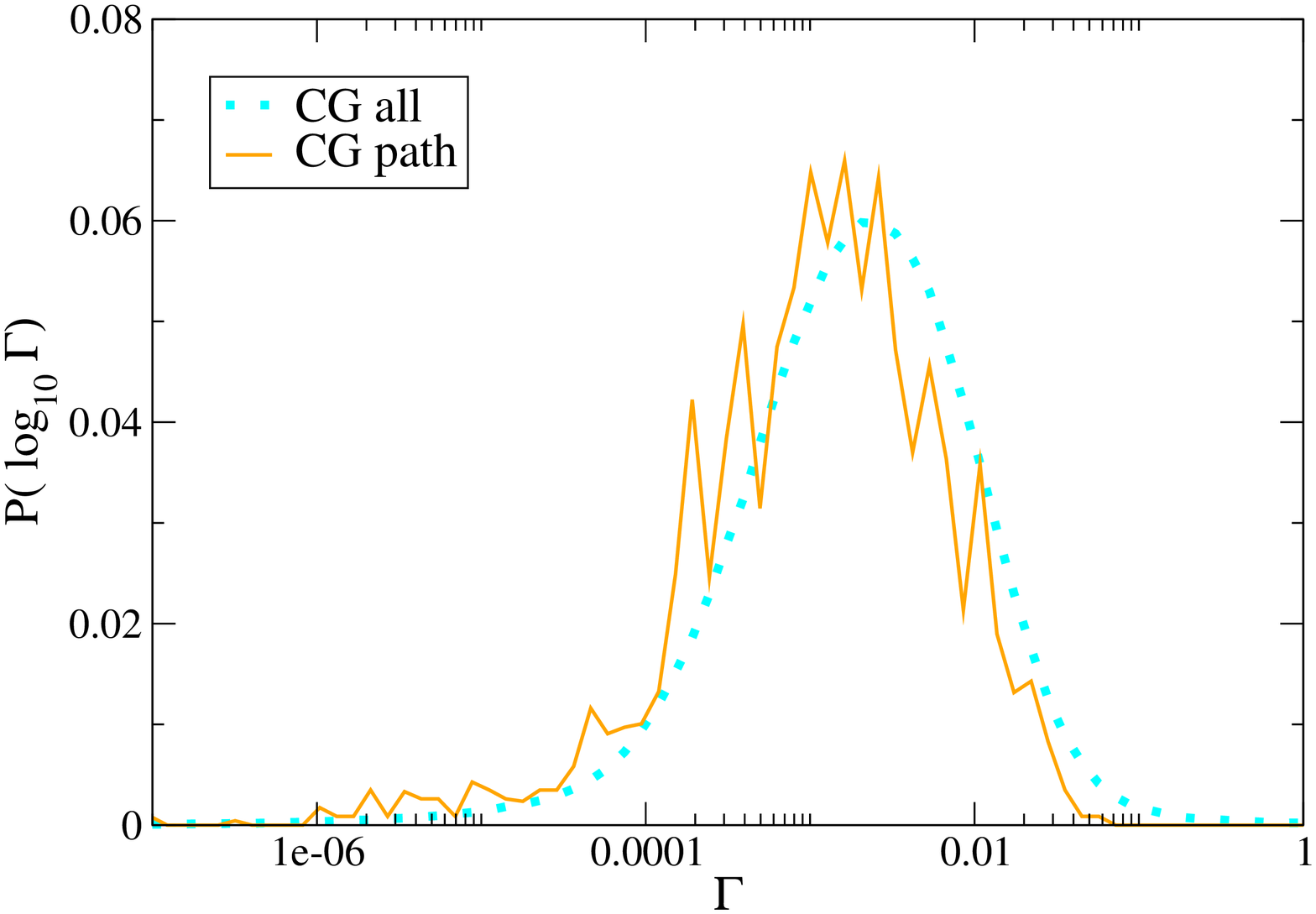}
(j) \includegraphics[width=0.2\columnwidth]{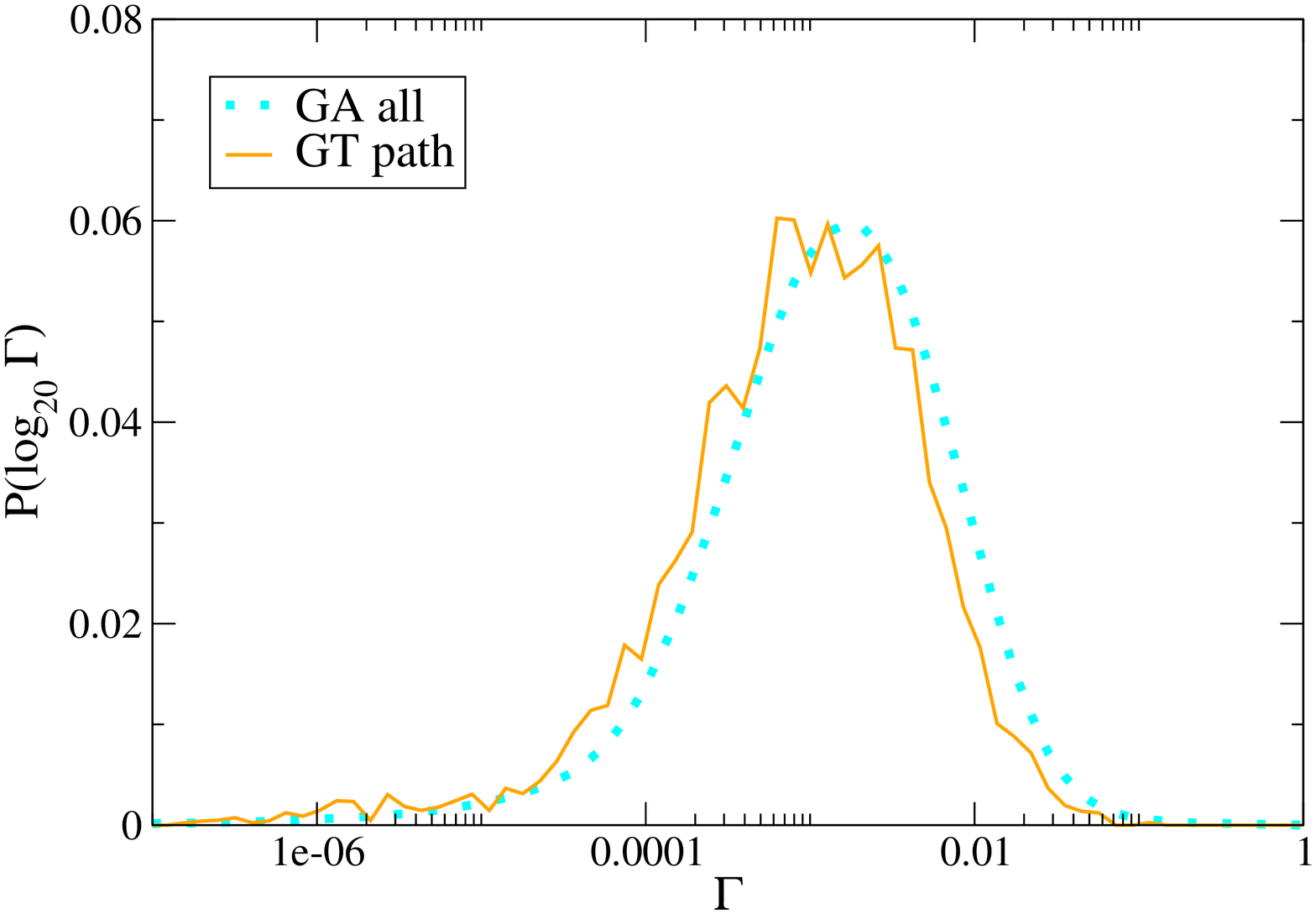}
(k) \includegraphics[width=0.2\columnwidth]{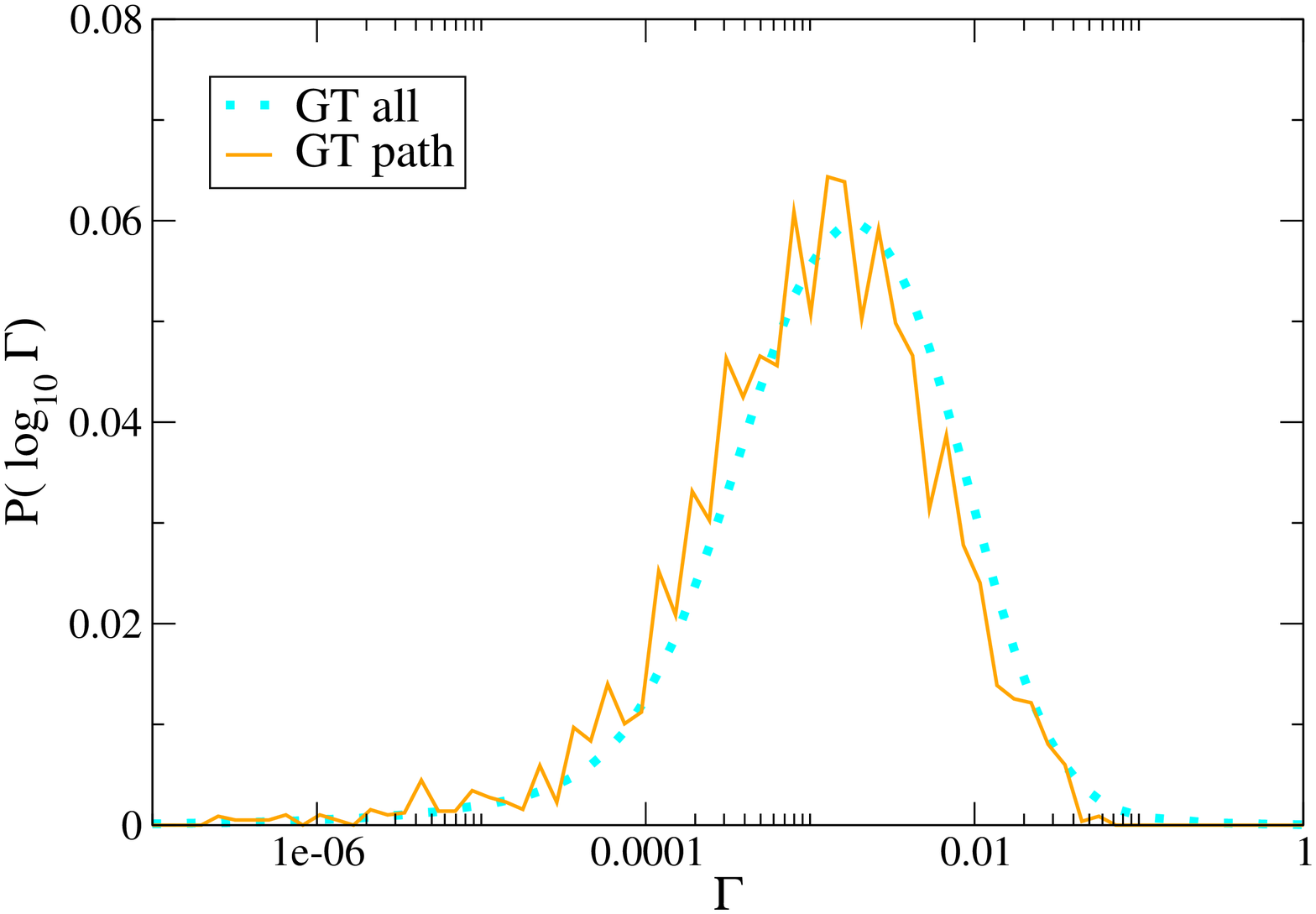}
(l) \includegraphics[width=0.2\columnwidth]{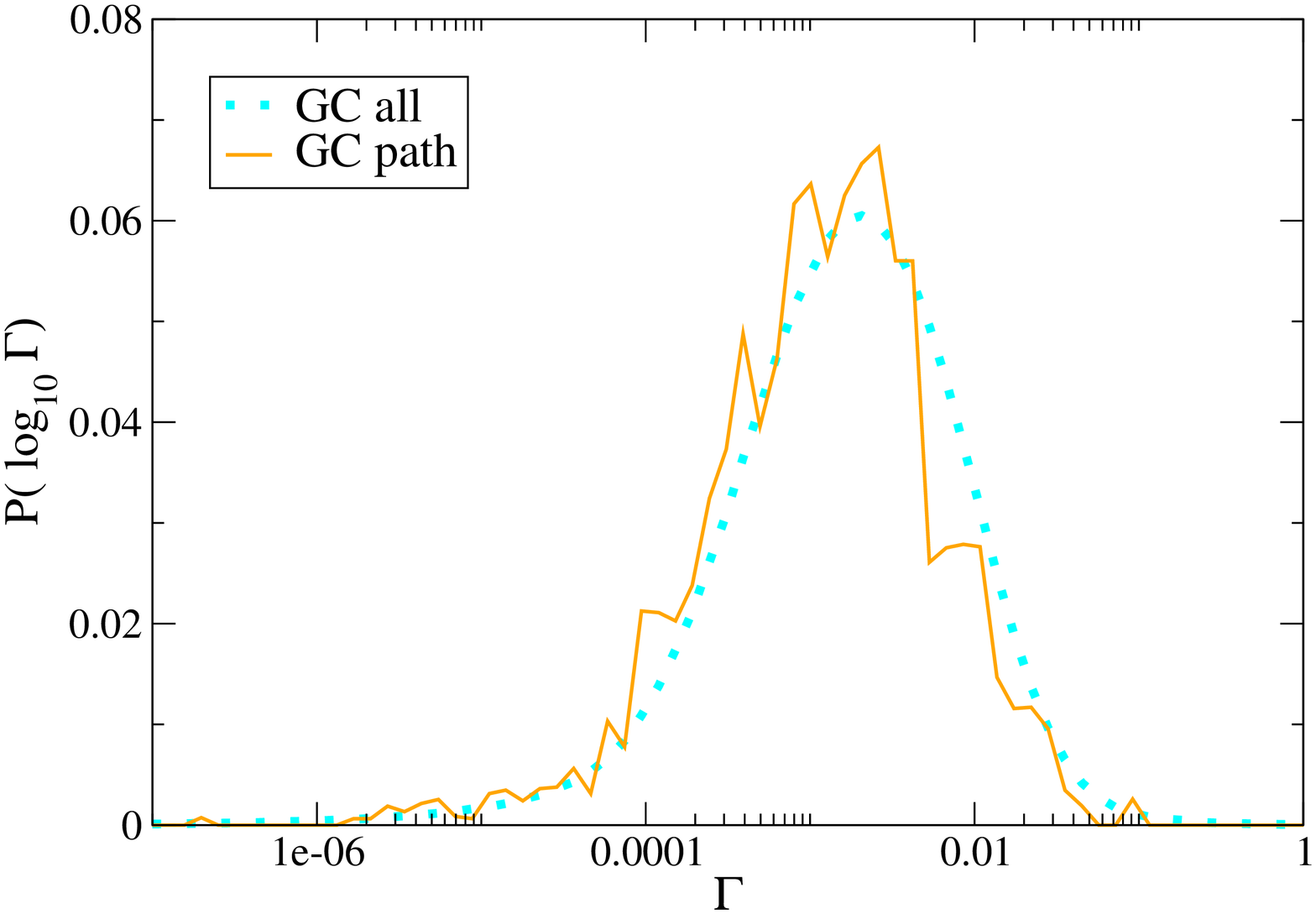}
(m) \includegraphics[width=0.2\columnwidth]{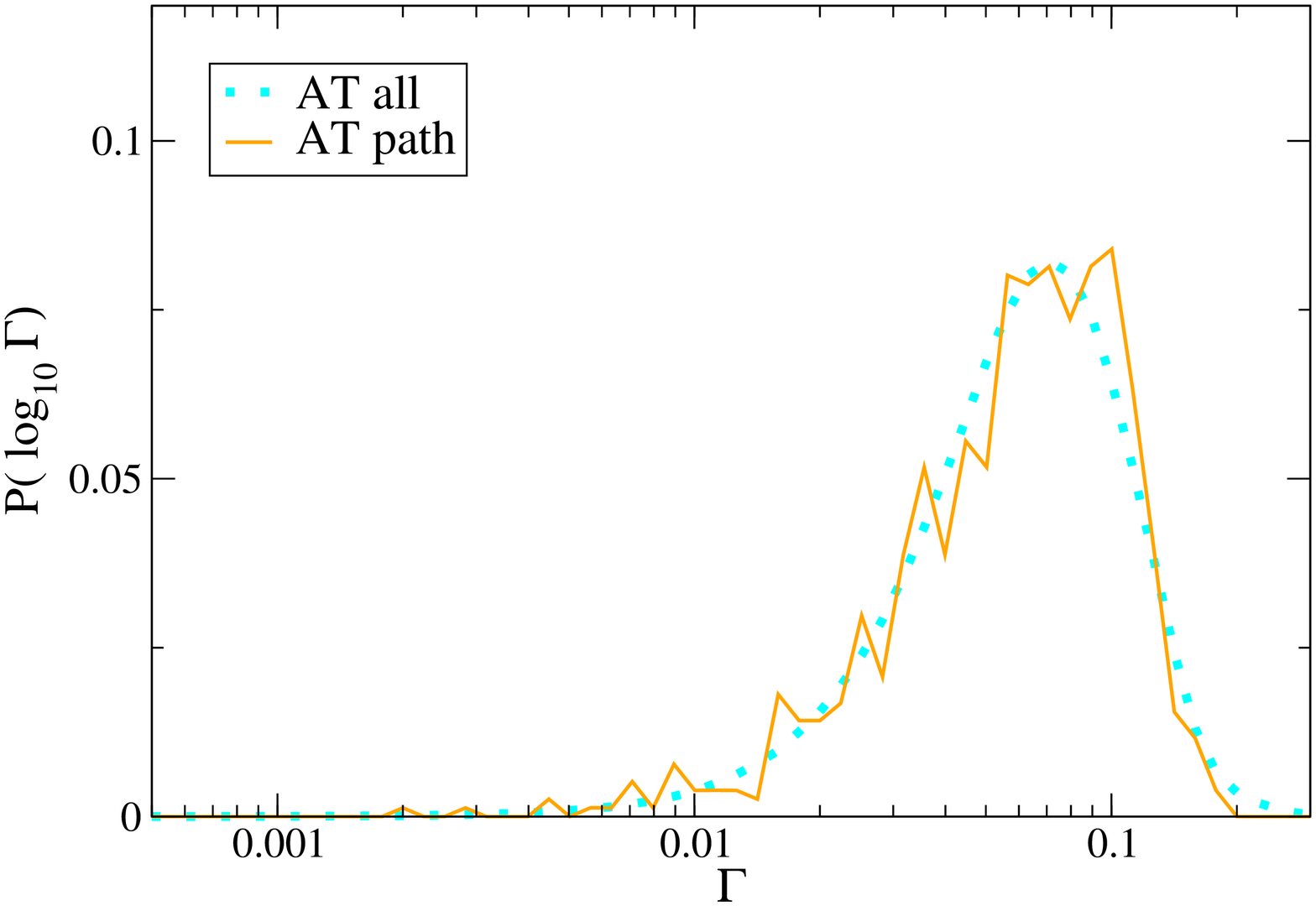}
(n) \includegraphics[width=0.2\columnwidth]{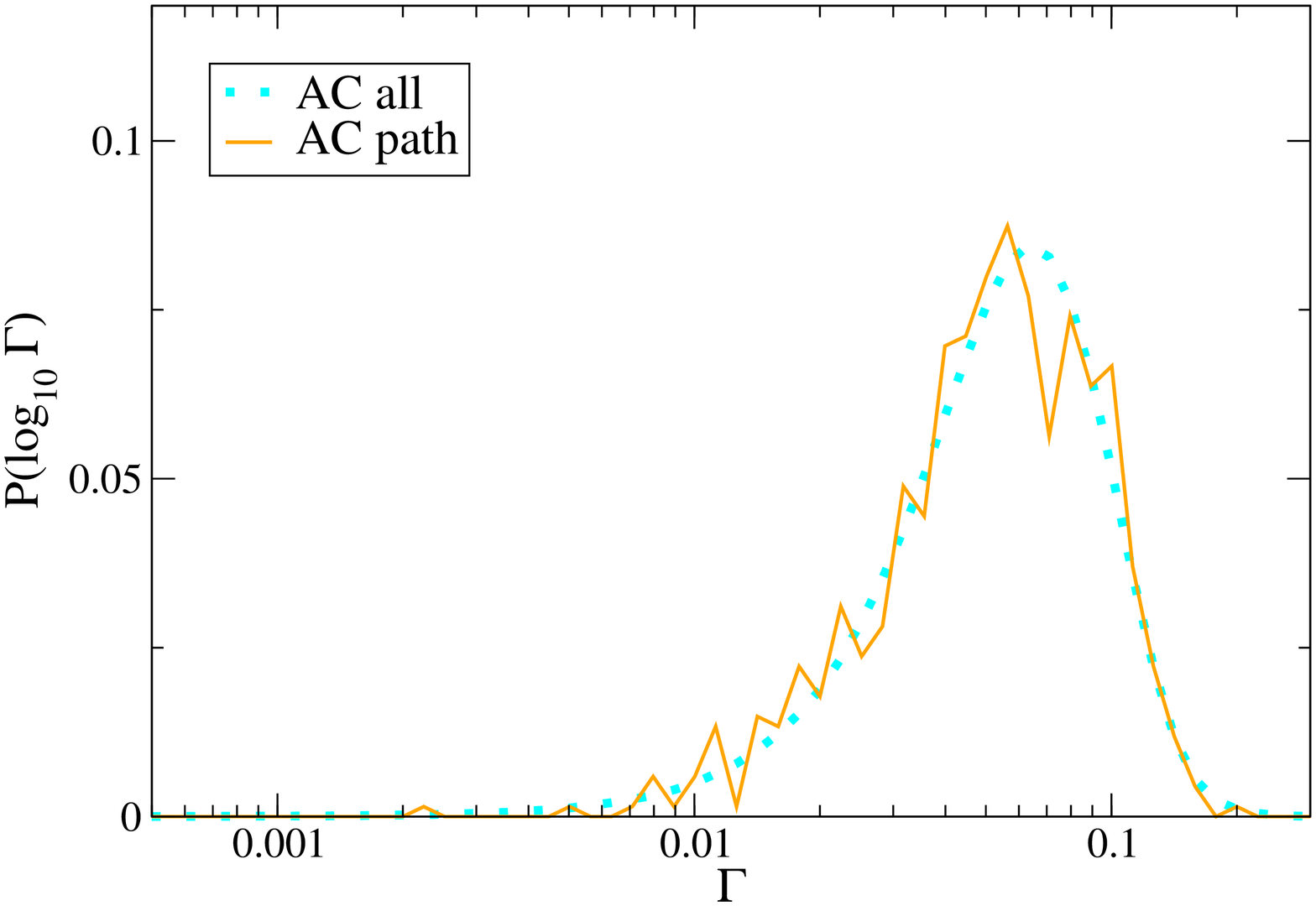}
(o) \includegraphics[width=0.2\columnwidth]{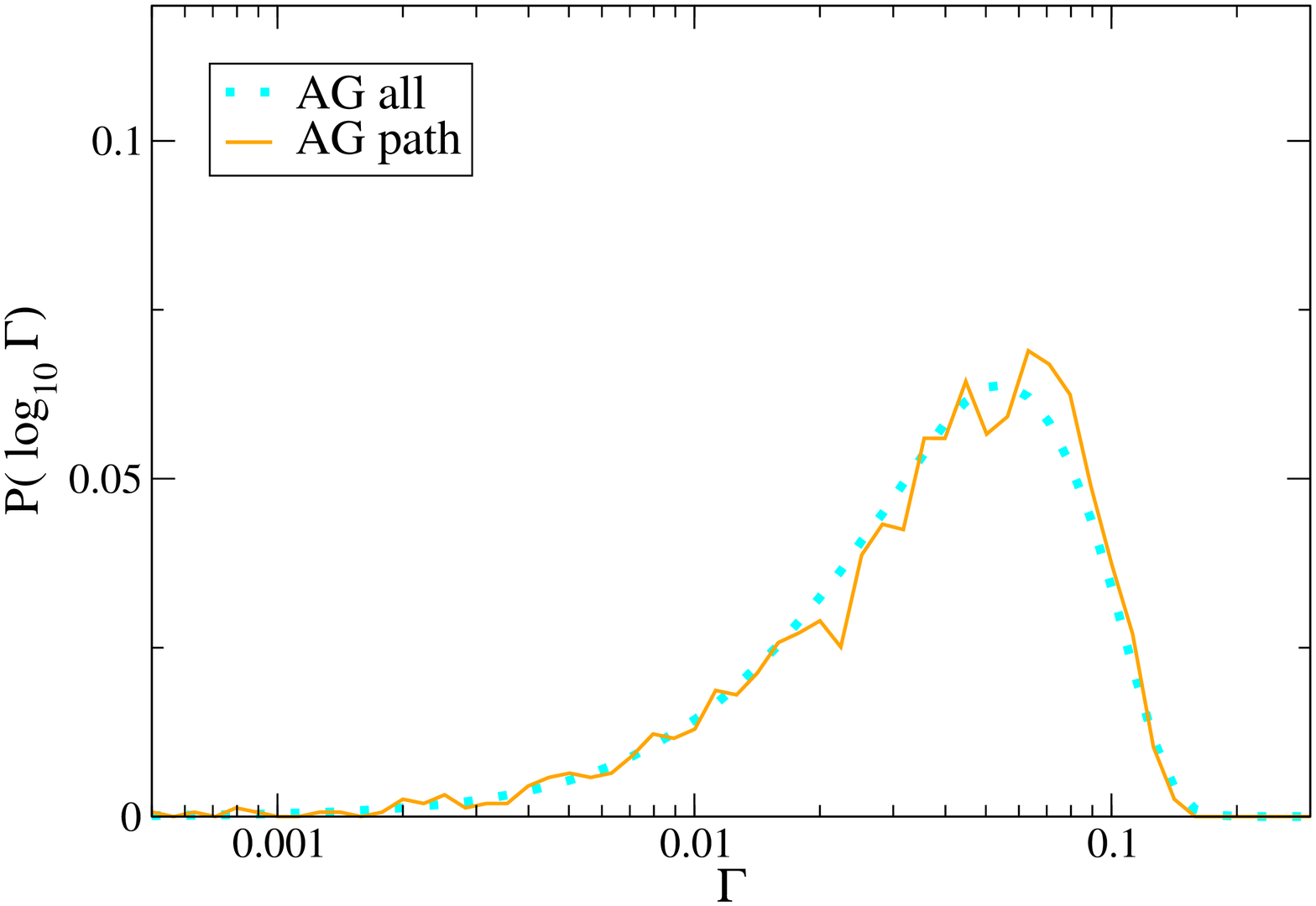}
(p) \includegraphics[width=0.2\columnwidth]{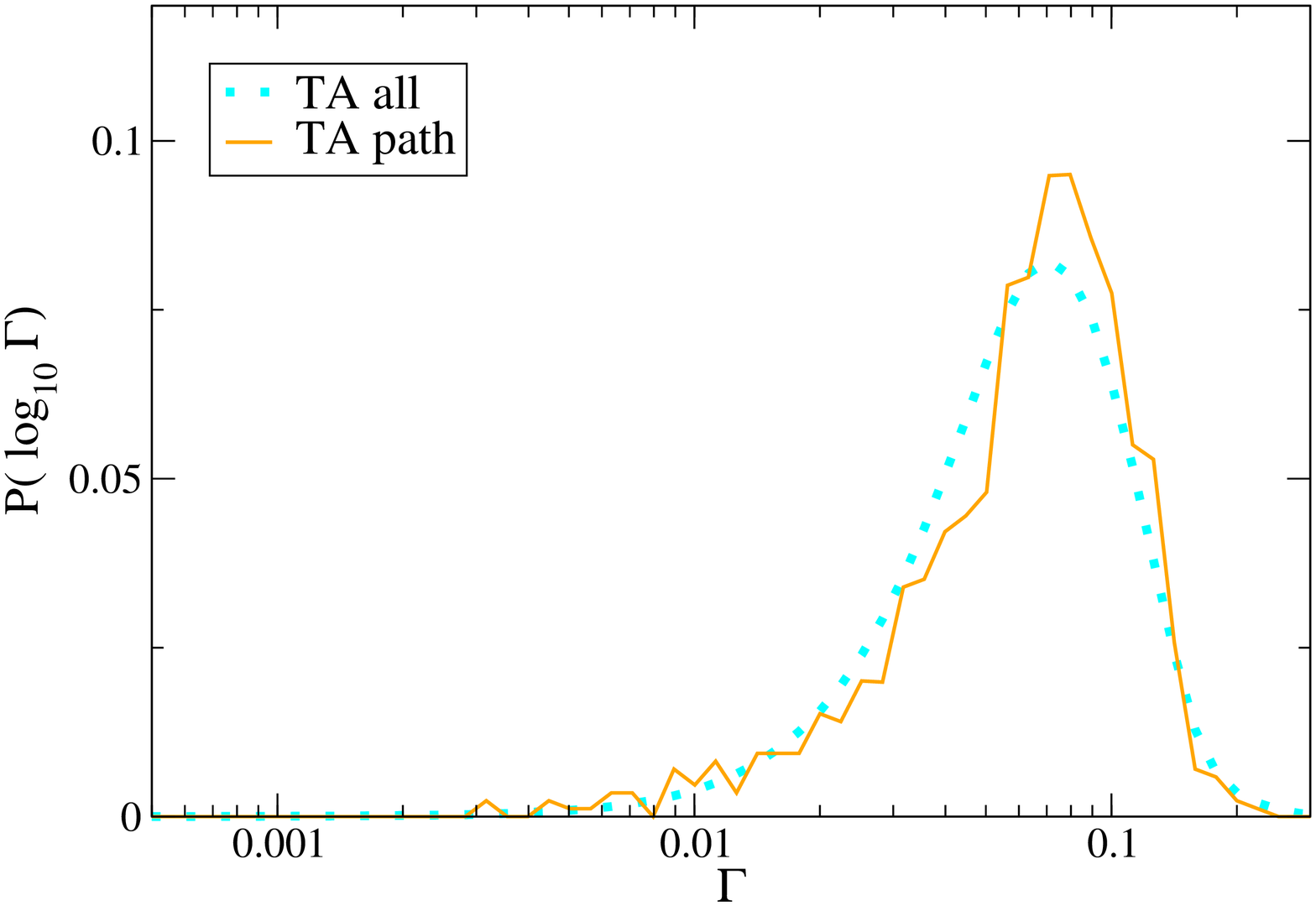}
(q) \includegraphics[width=0.2\columnwidth]{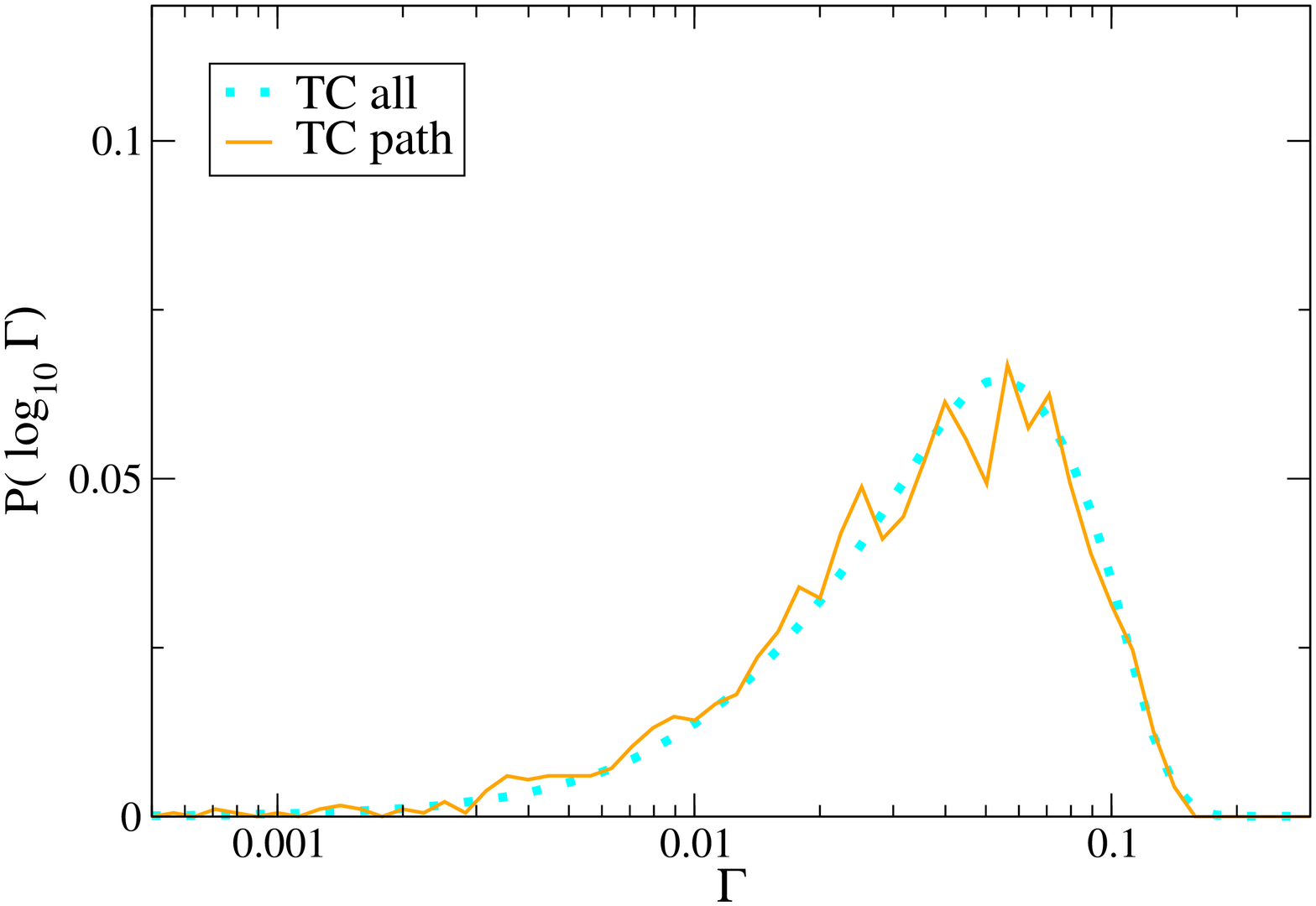}
(r) \includegraphics[width=0.2\columnwidth]{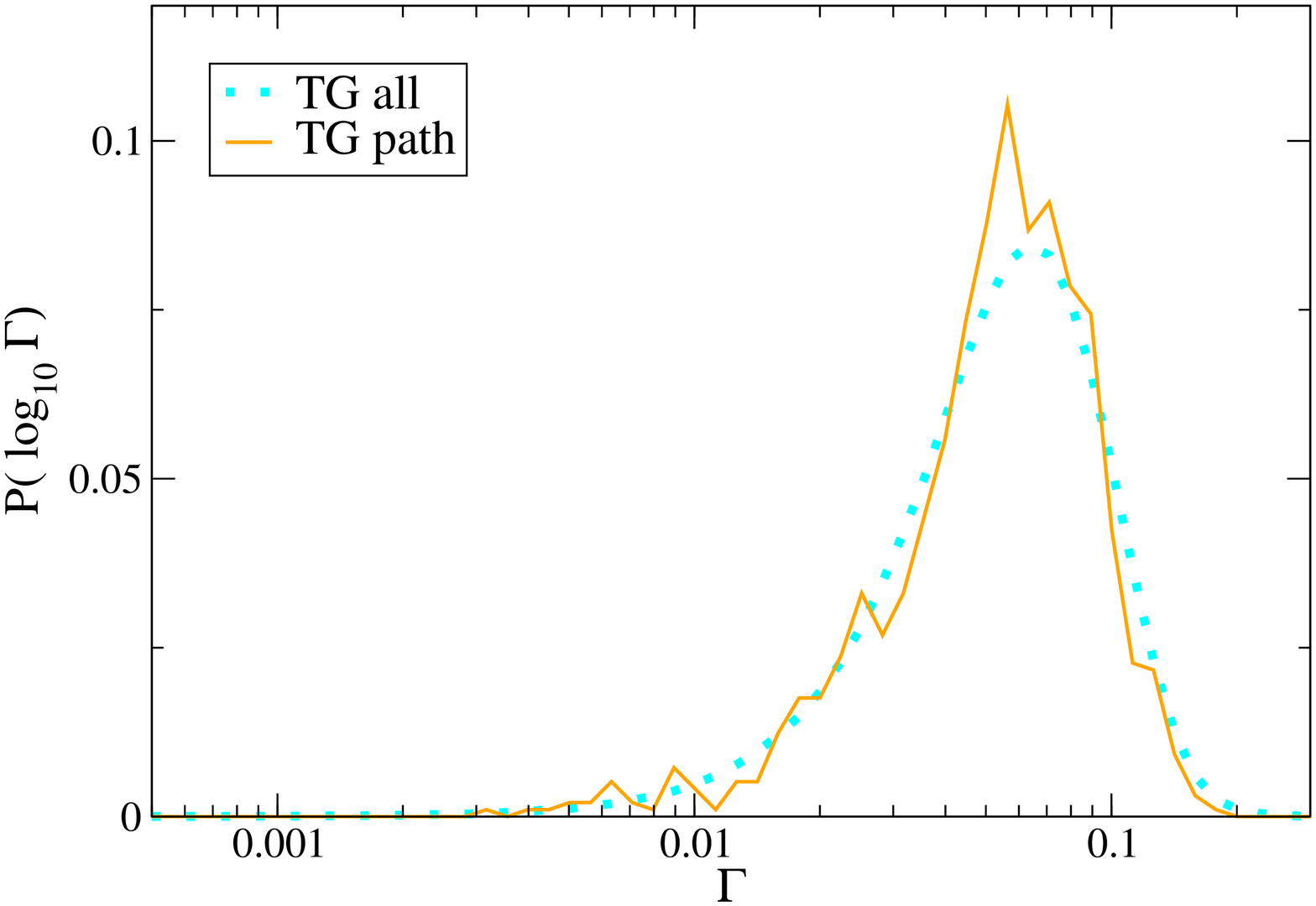}
(s) \includegraphics[width=0.2\columnwidth]{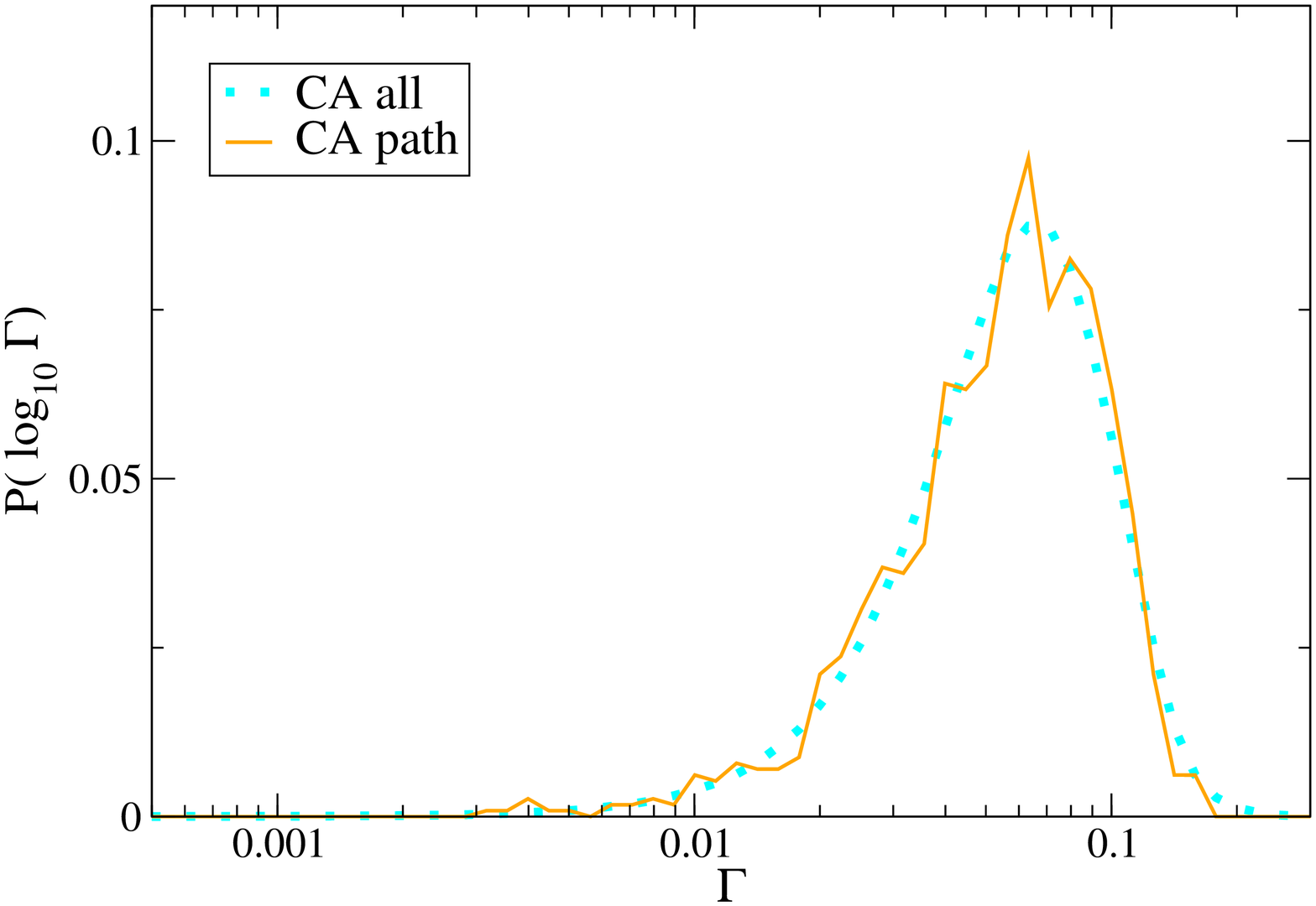}
(t) \includegraphics[width=0.2\columnwidth]{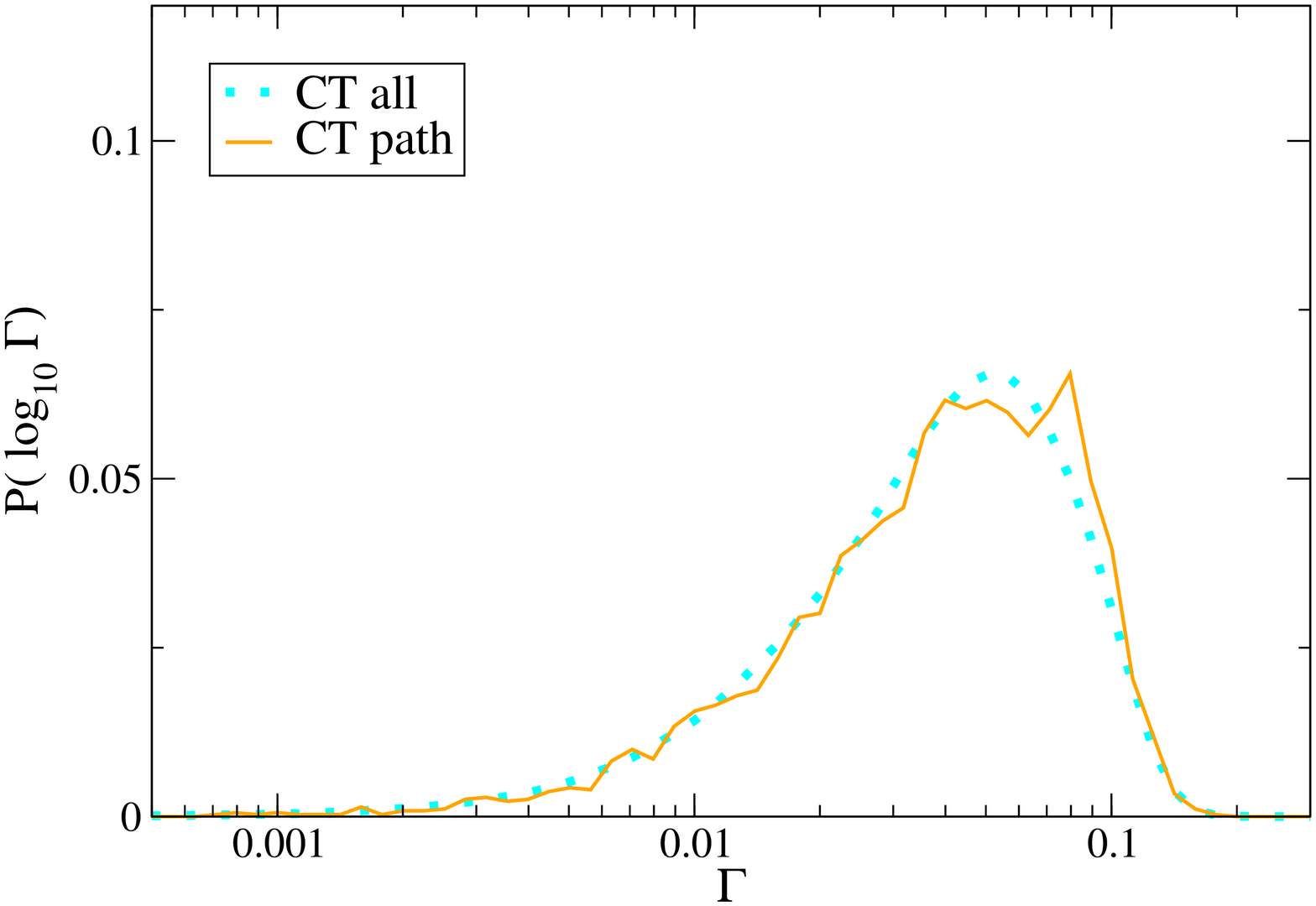}
(u) \includegraphics[width=0.2\columnwidth]{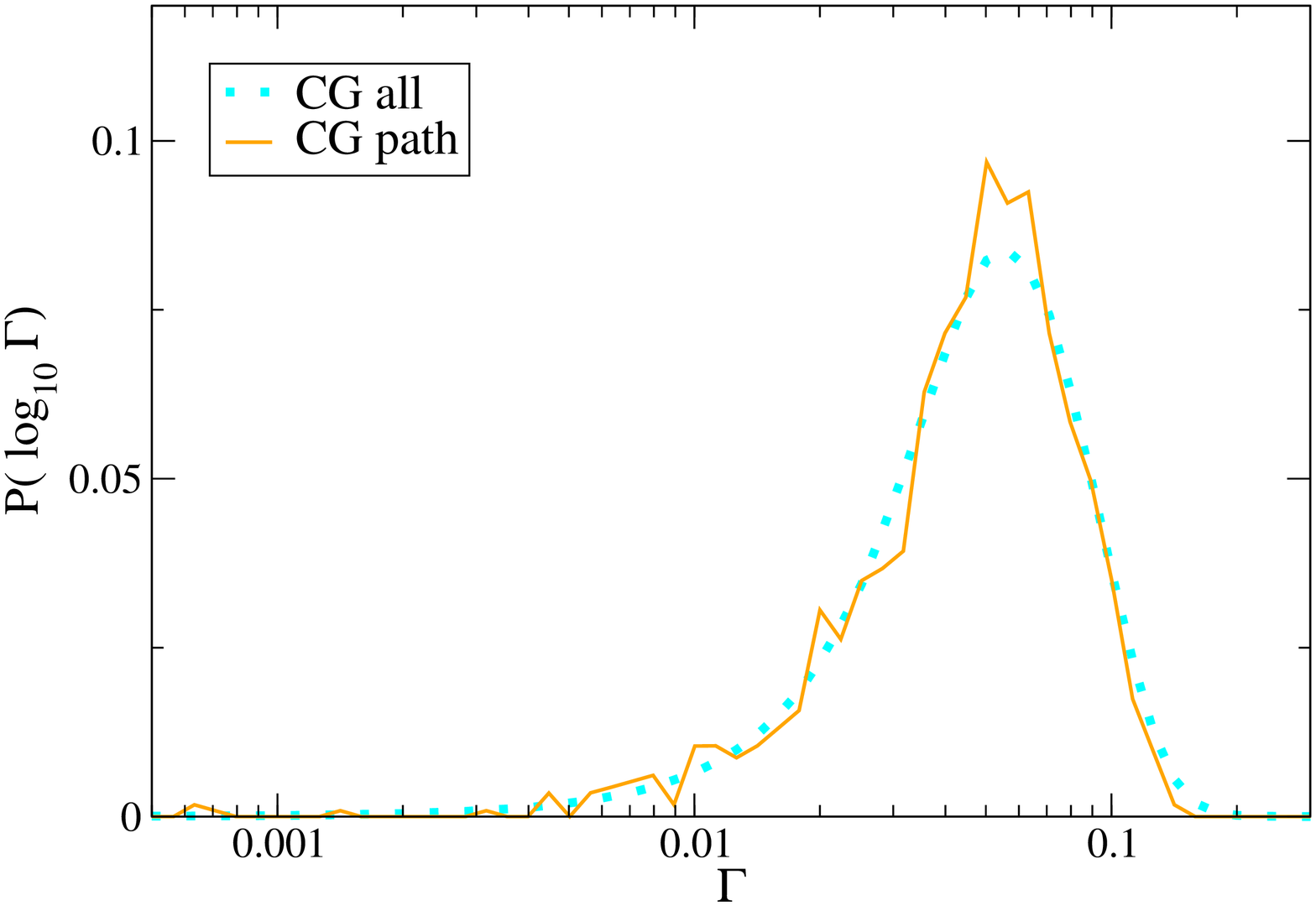}
(v) \includegraphics[width=0.2\columnwidth]{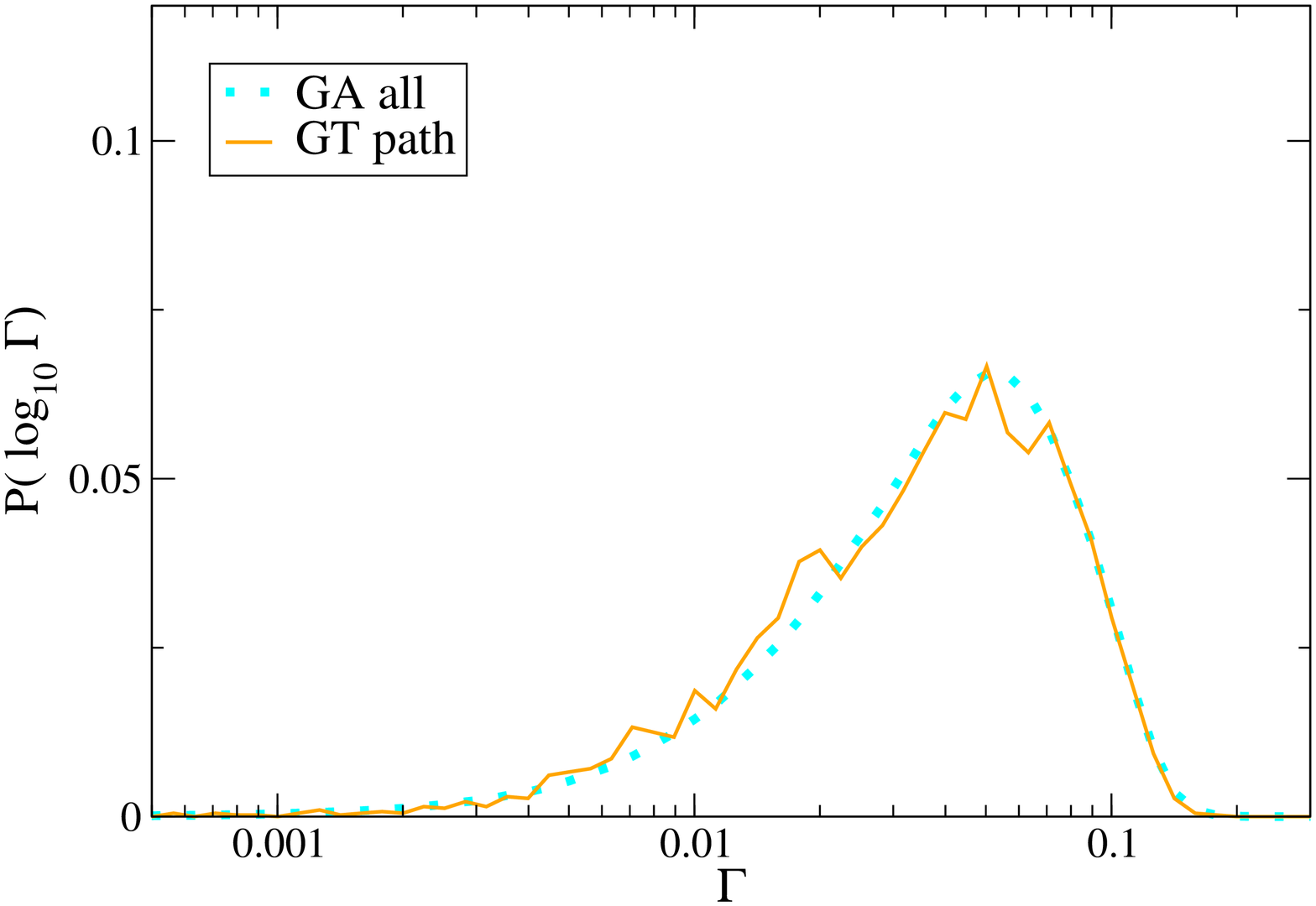}
(w) \includegraphics[width=0.2\columnwidth]{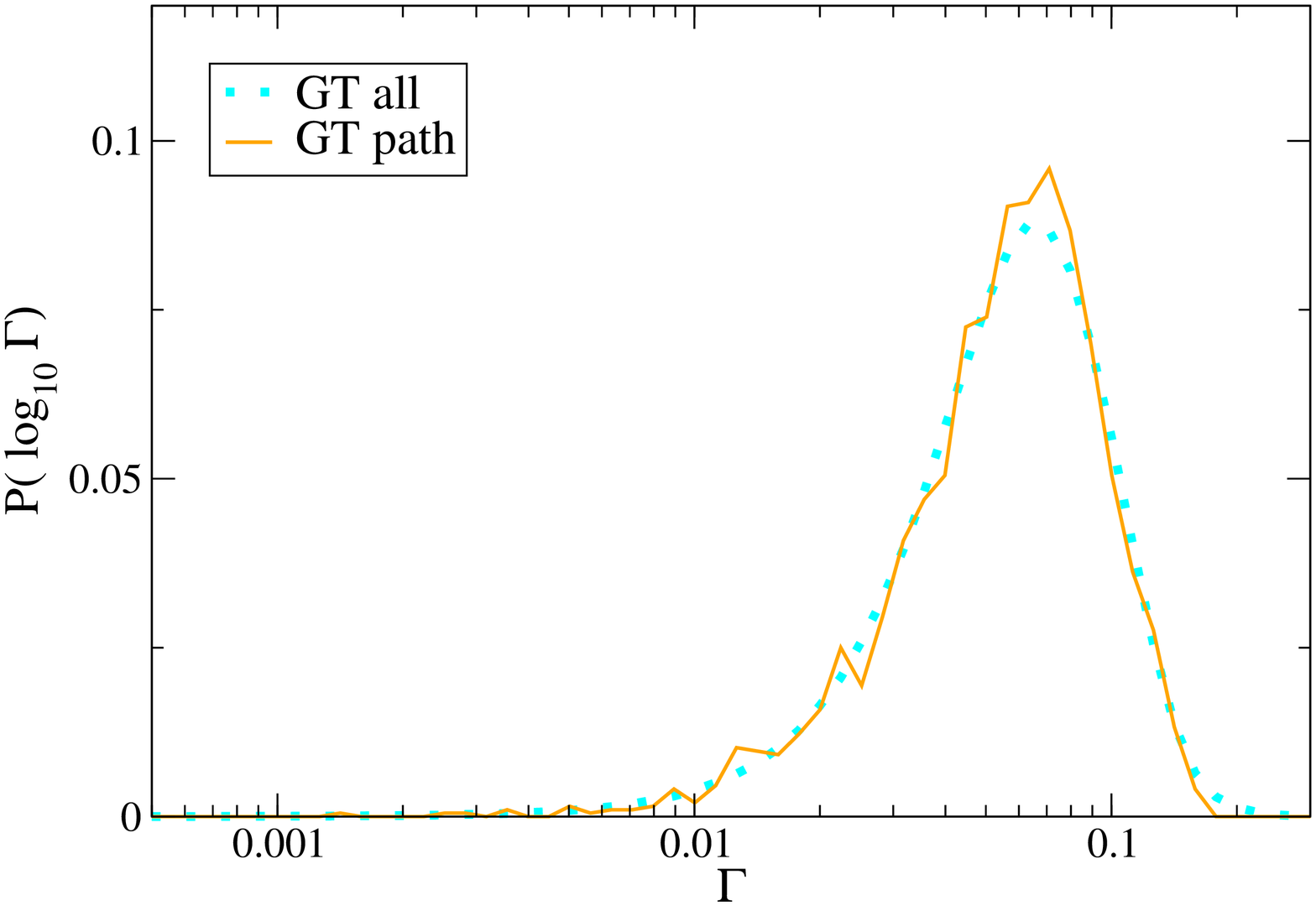}
(x) \includegraphics[width=0.2\columnwidth]{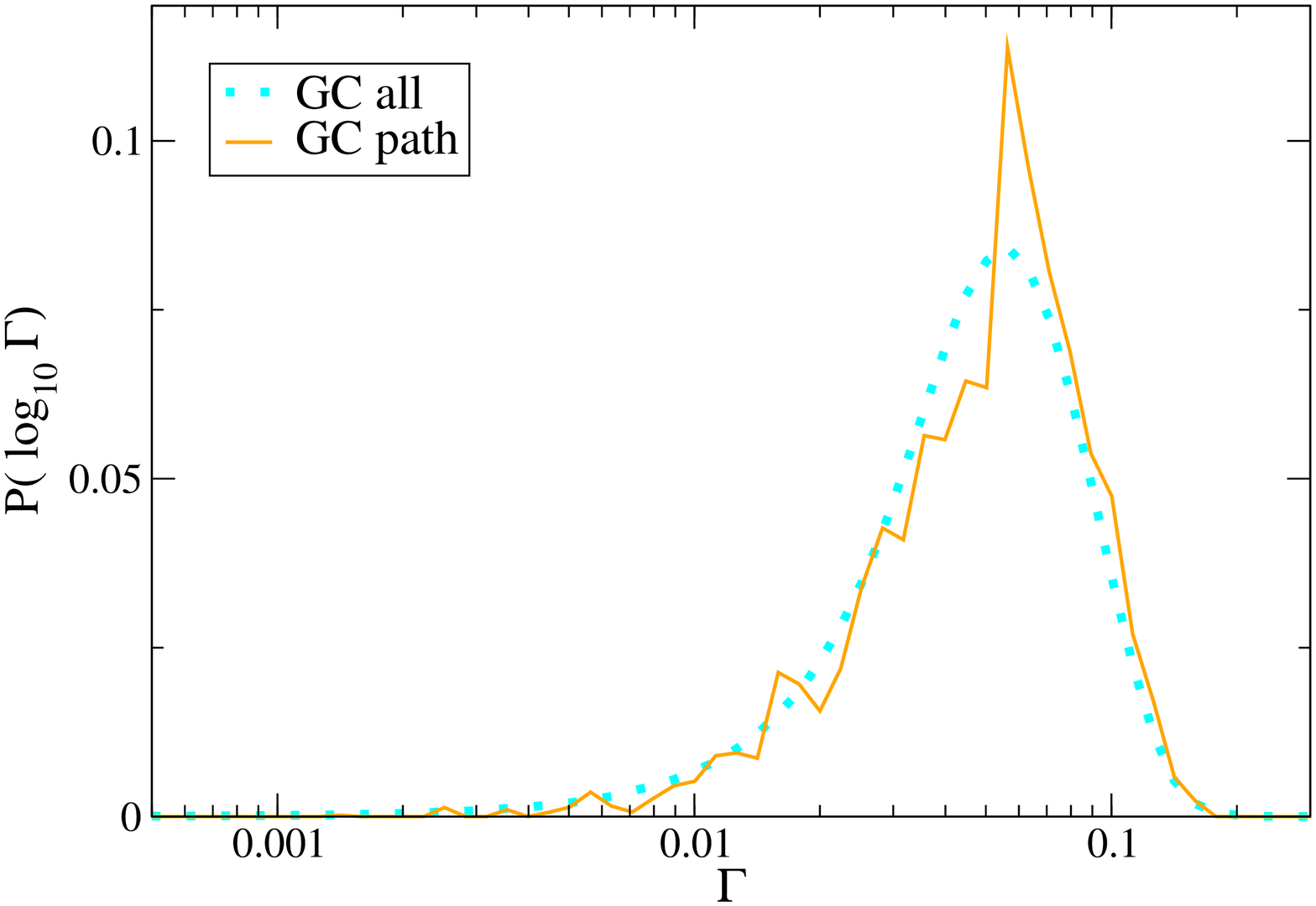}

\caption{(Supplementary) Panels a-l: 1D model, results divided into the
  twelve subtypes of mutation. The shift for pathogenic mutations is clearly
  present in every case.  Panels m-x: 2L model, results divided into the
  twelve subtypes of mutation. There is no consistent shift for pathogenic
  mutations.  }
\label{fig-sub-gamma-12type}\label{fig-S2}
\end{figure}
\clearpage
\begin{figure}
\centering
(a)\includegraphics[width=0.45\columnwidth]{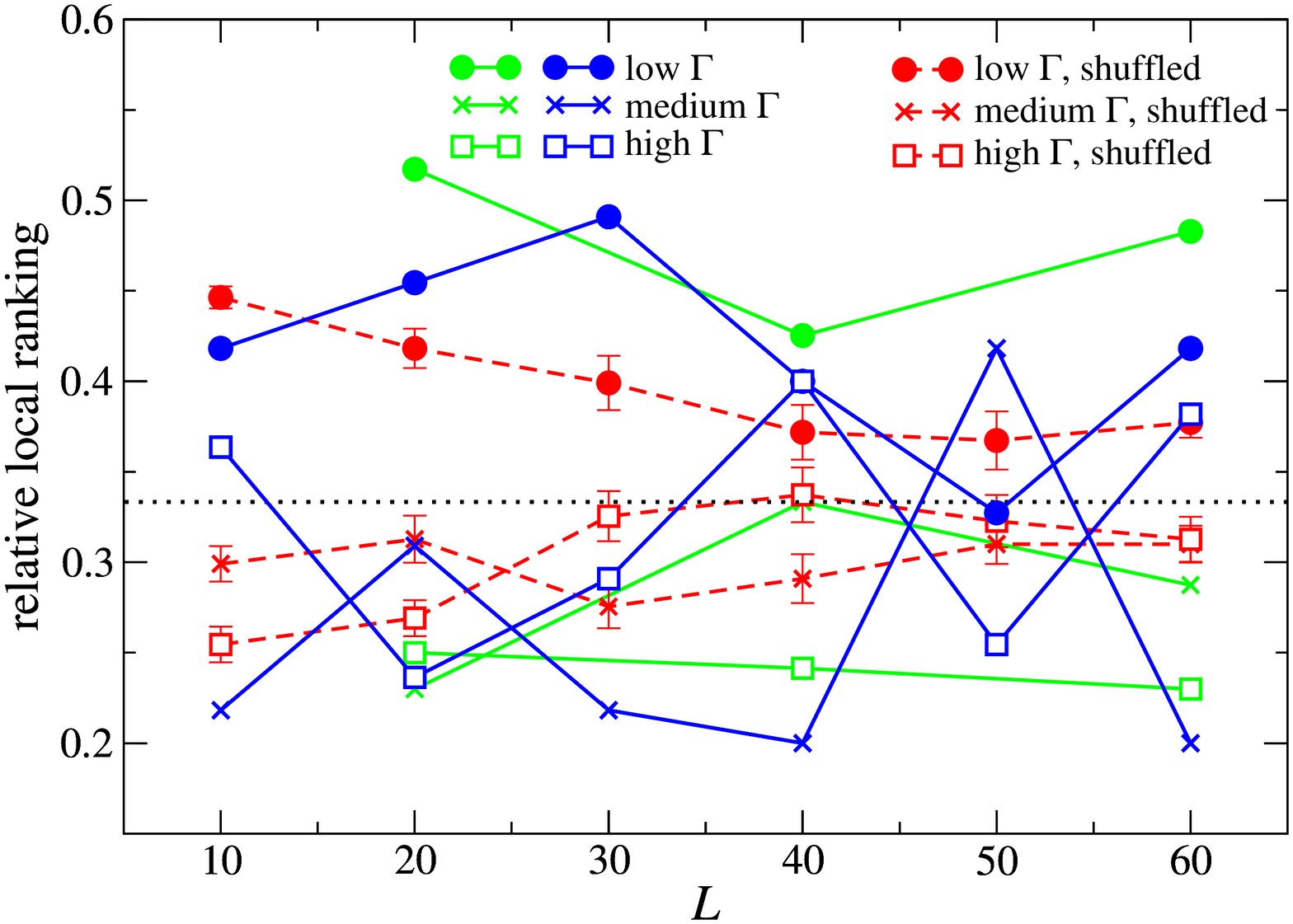}
(b)\includegraphics[width=0.45\columnwidth]{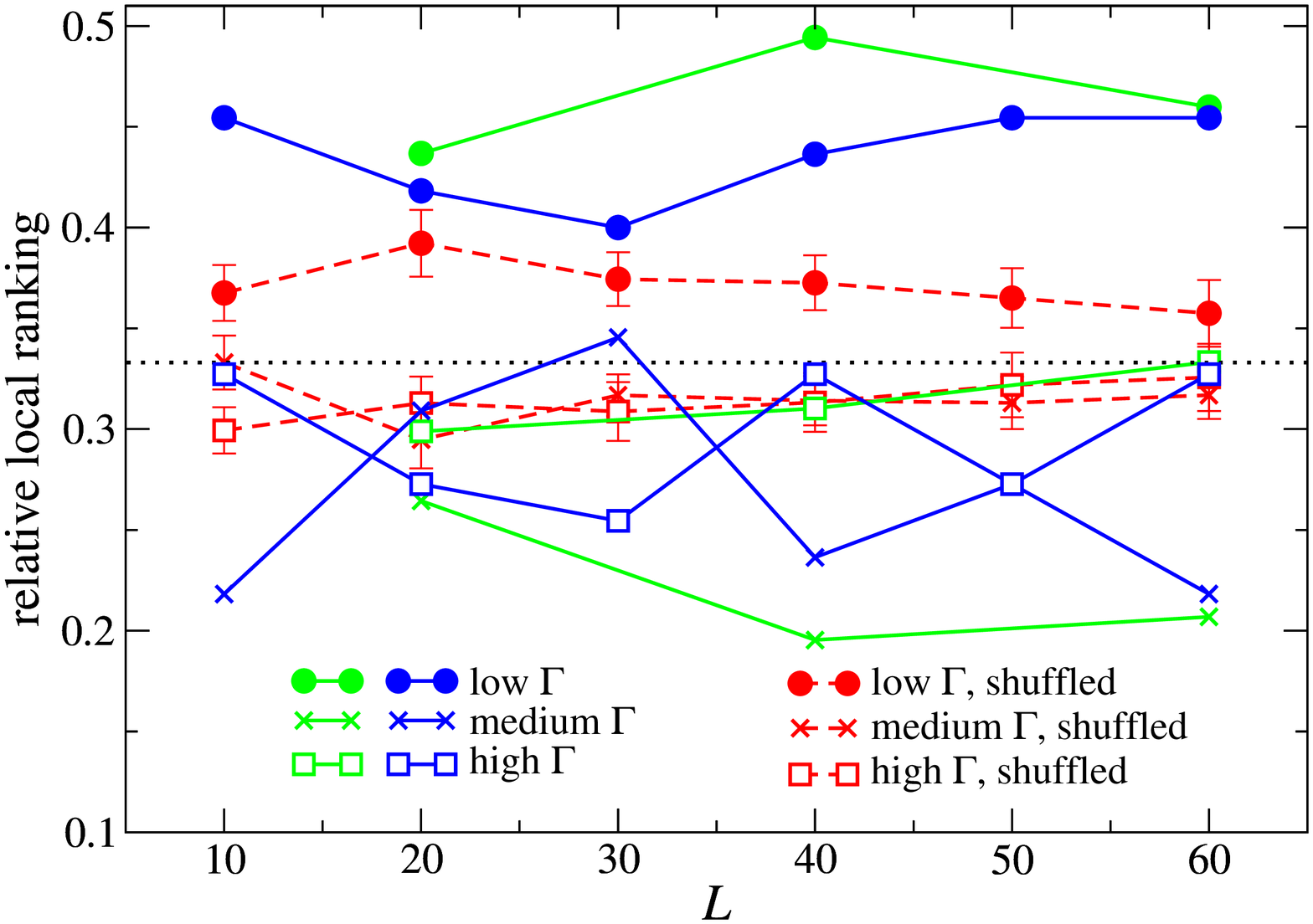}
(c)\includegraphics[width=0.45\columnwidth]{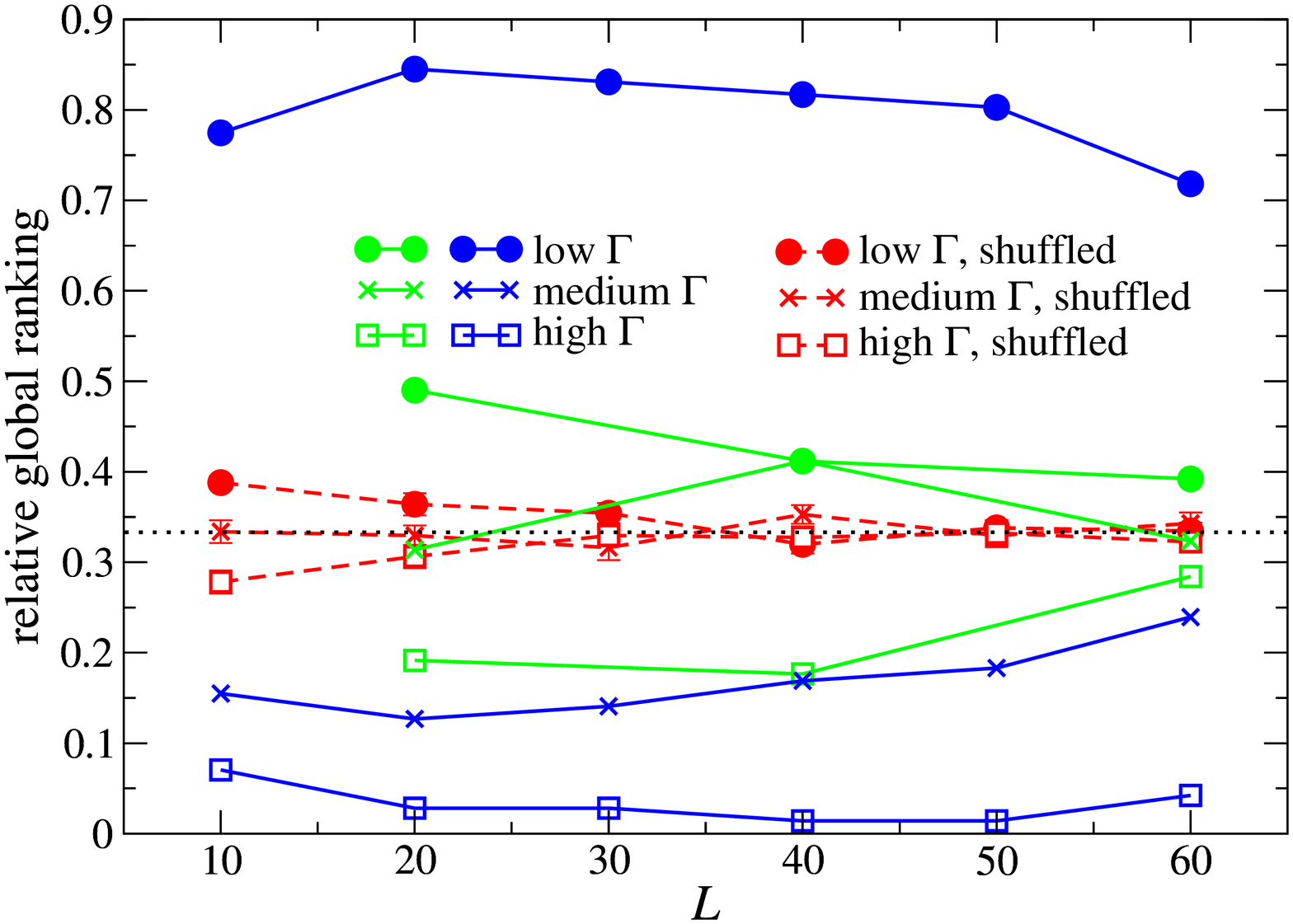}
(d)\includegraphics[width=0.45\columnwidth]{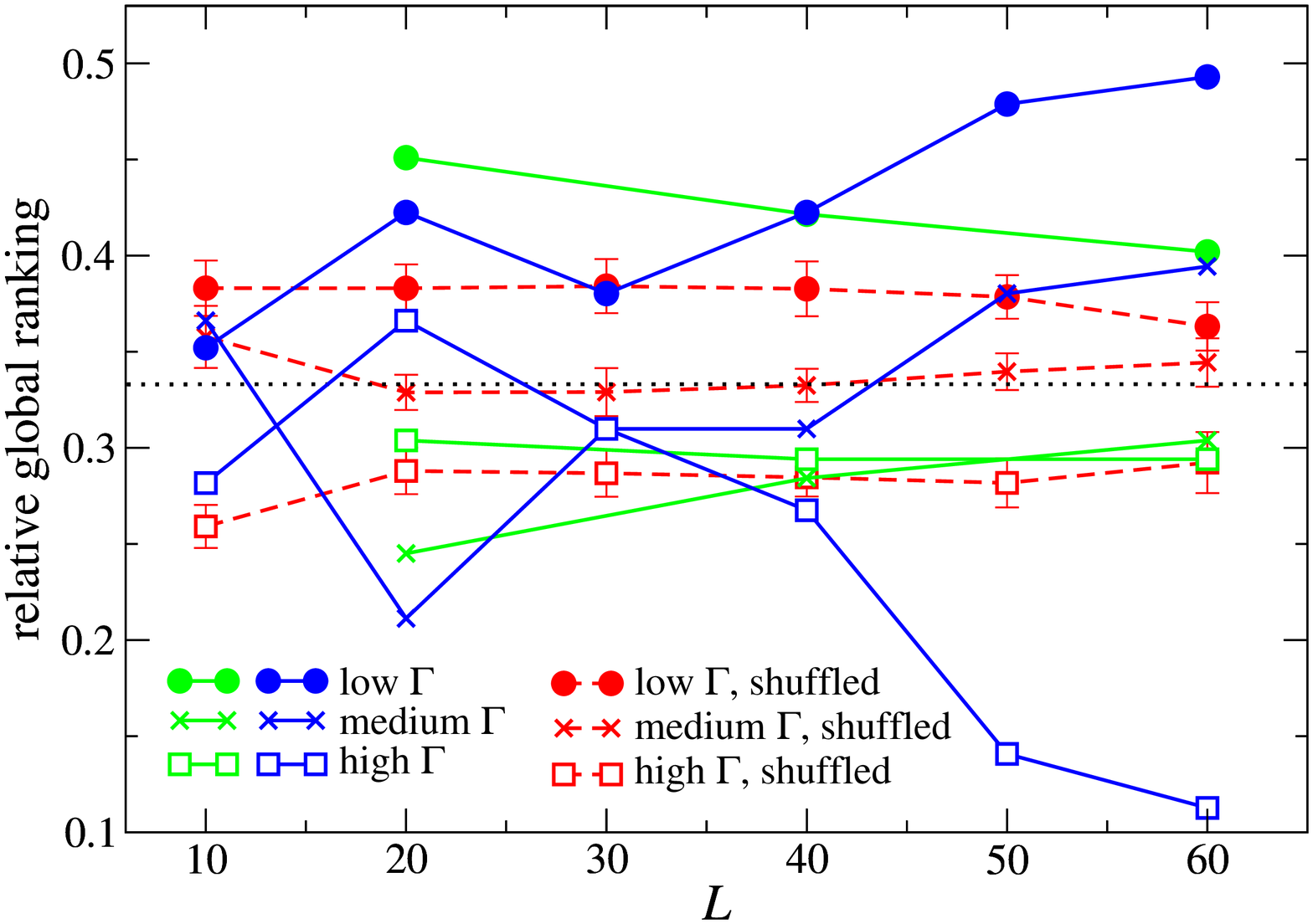}
\caption{(Supplementary) Distribution of the {\em local} (a+b) and {\em
    global} (c+d) ranking results of pathogenic mutations of $p16$ (CDKN2A)
  (blue solid lines) and CYP21A2 (green) as a function of window lengths
  $L$. The dashed lines indicate averaged results for $20$ randomly shuffled
  $p16$ sequences. The left/right columns distinguish results for the 1D/2-leg
  models. The dashed horizontal line shows the $33\%$ mark expected for a
  completely random sequence. All lines are guides to the eyes only. Error
  bars are within symbol size.}
\label{fig-sub-L_ranking-p16}\label{fig-S3}
\end{figure}
\clearpage
\begin{figure}
\centering
(a)\includegraphics[width=0.45\columnwidth]{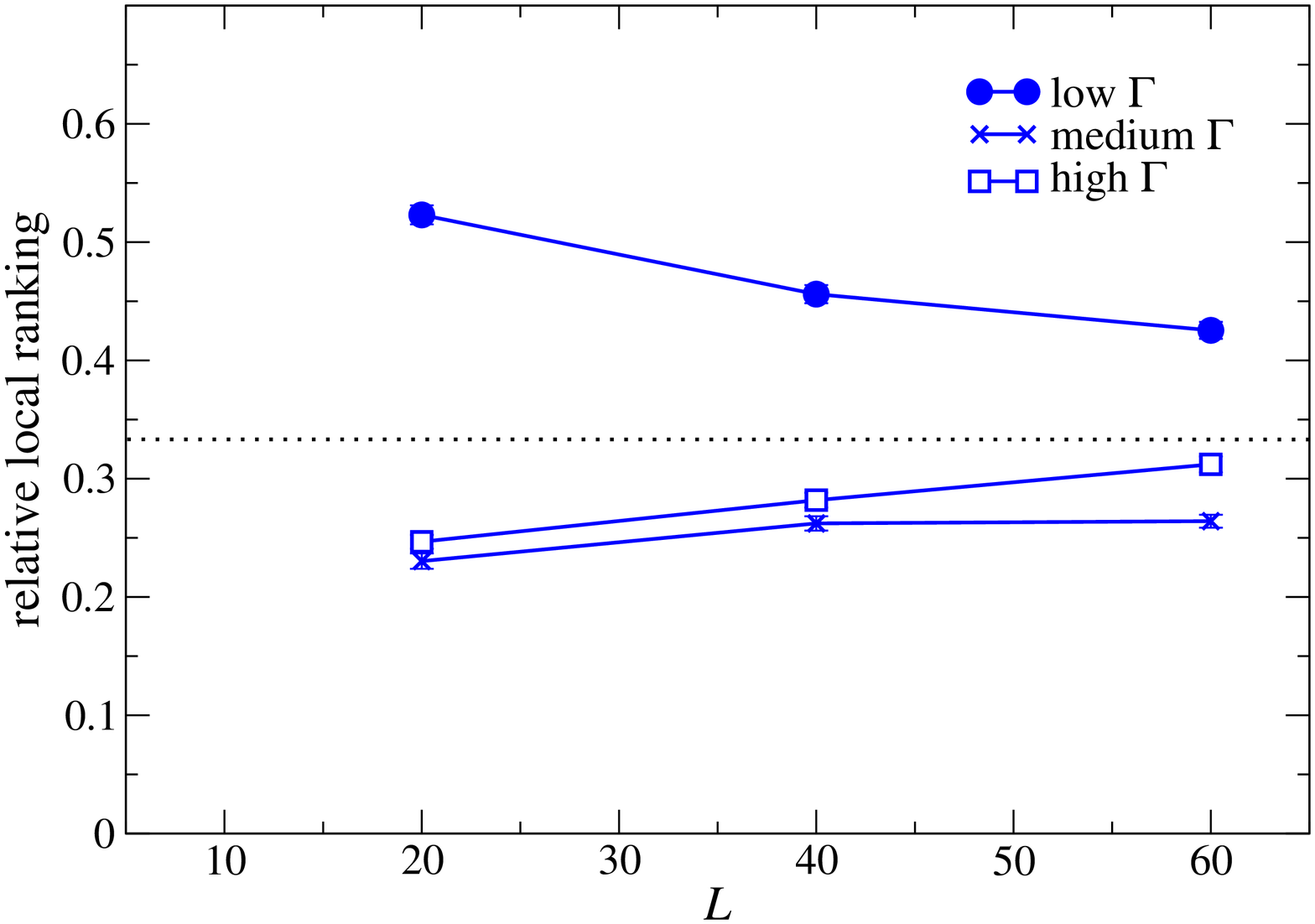}
(b)\includegraphics[width=0.45\columnwidth]{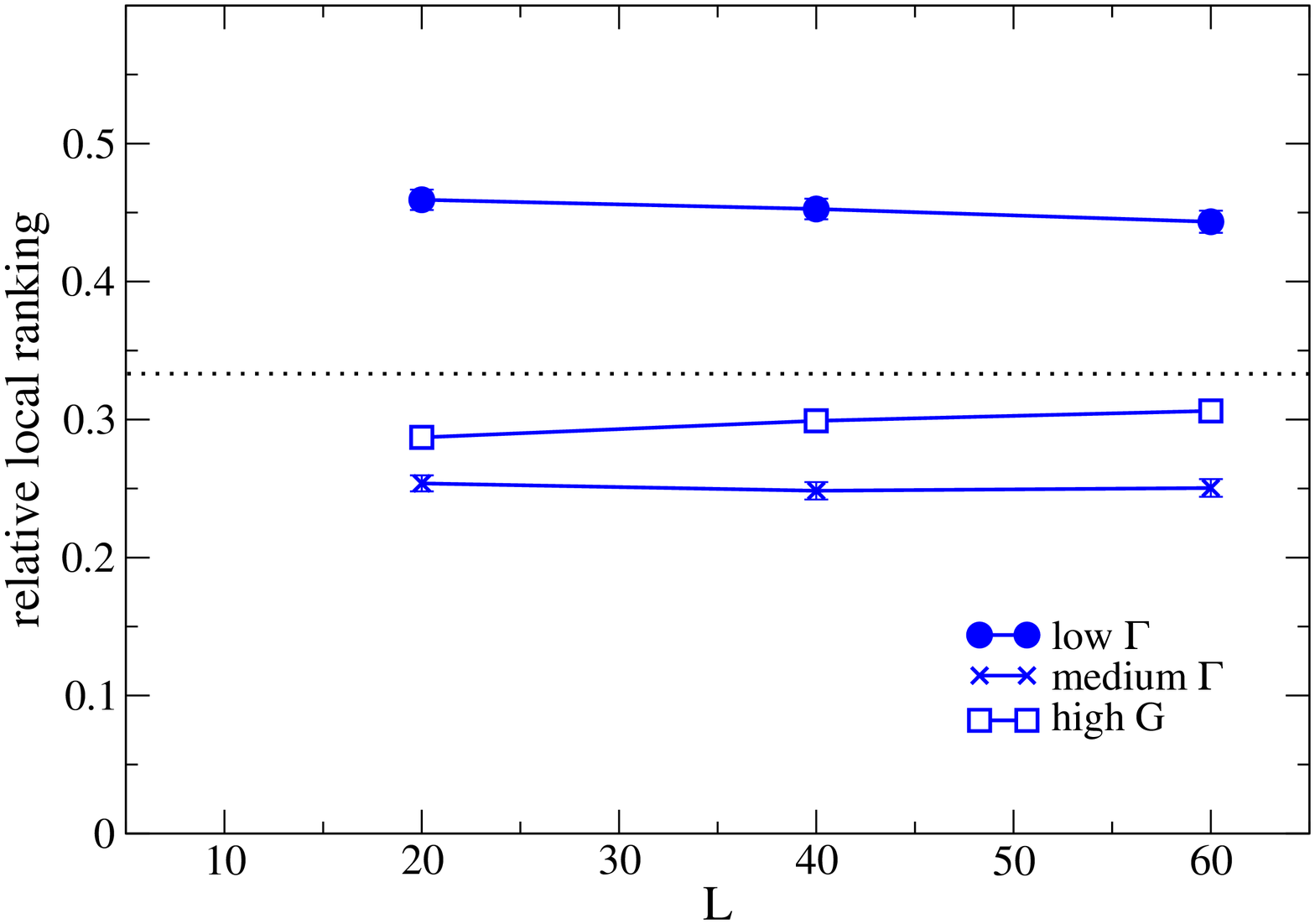}
(c)\includegraphics[width=0.45\columnwidth]{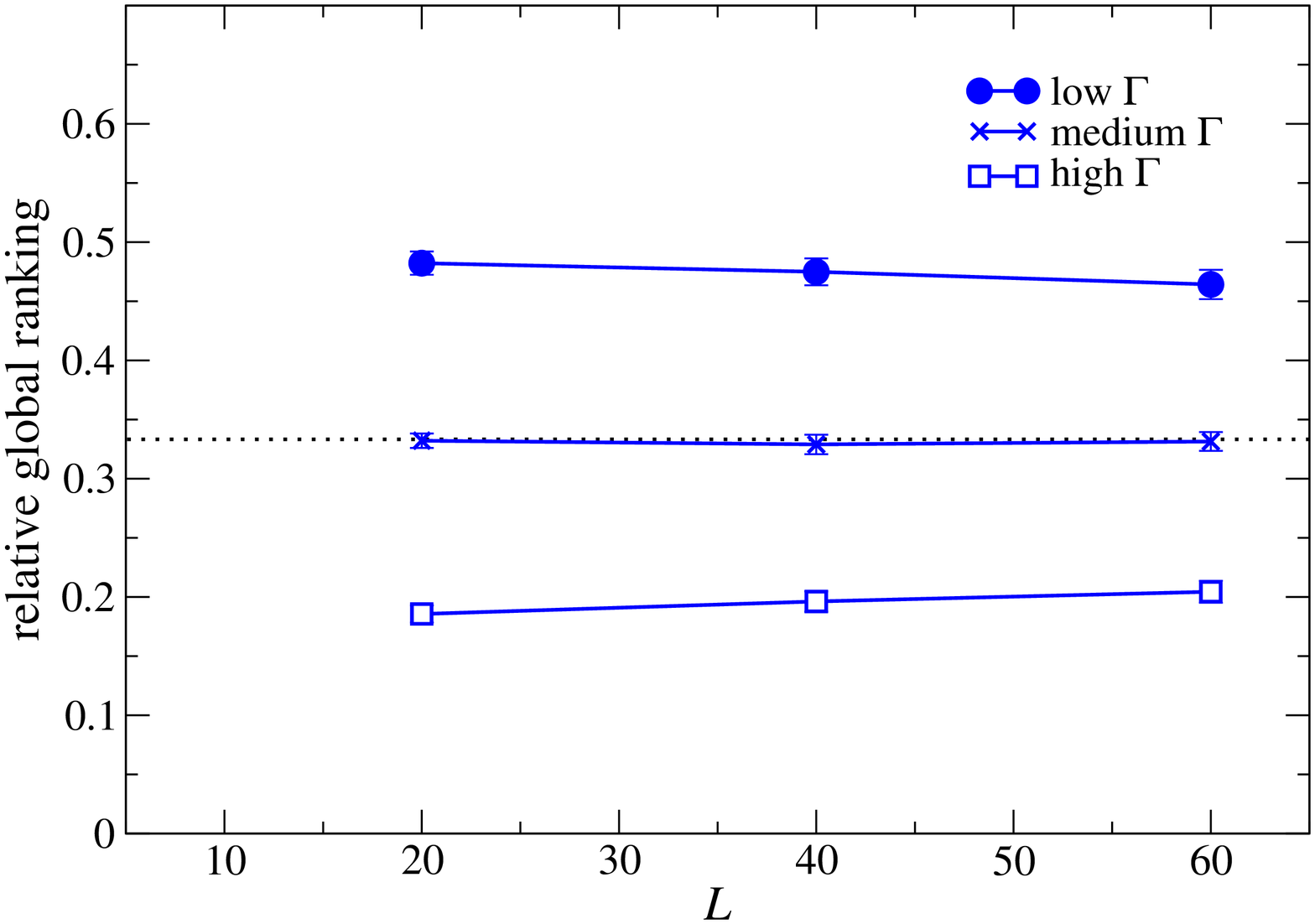}
(d)\includegraphics[width=0.45\columnwidth]{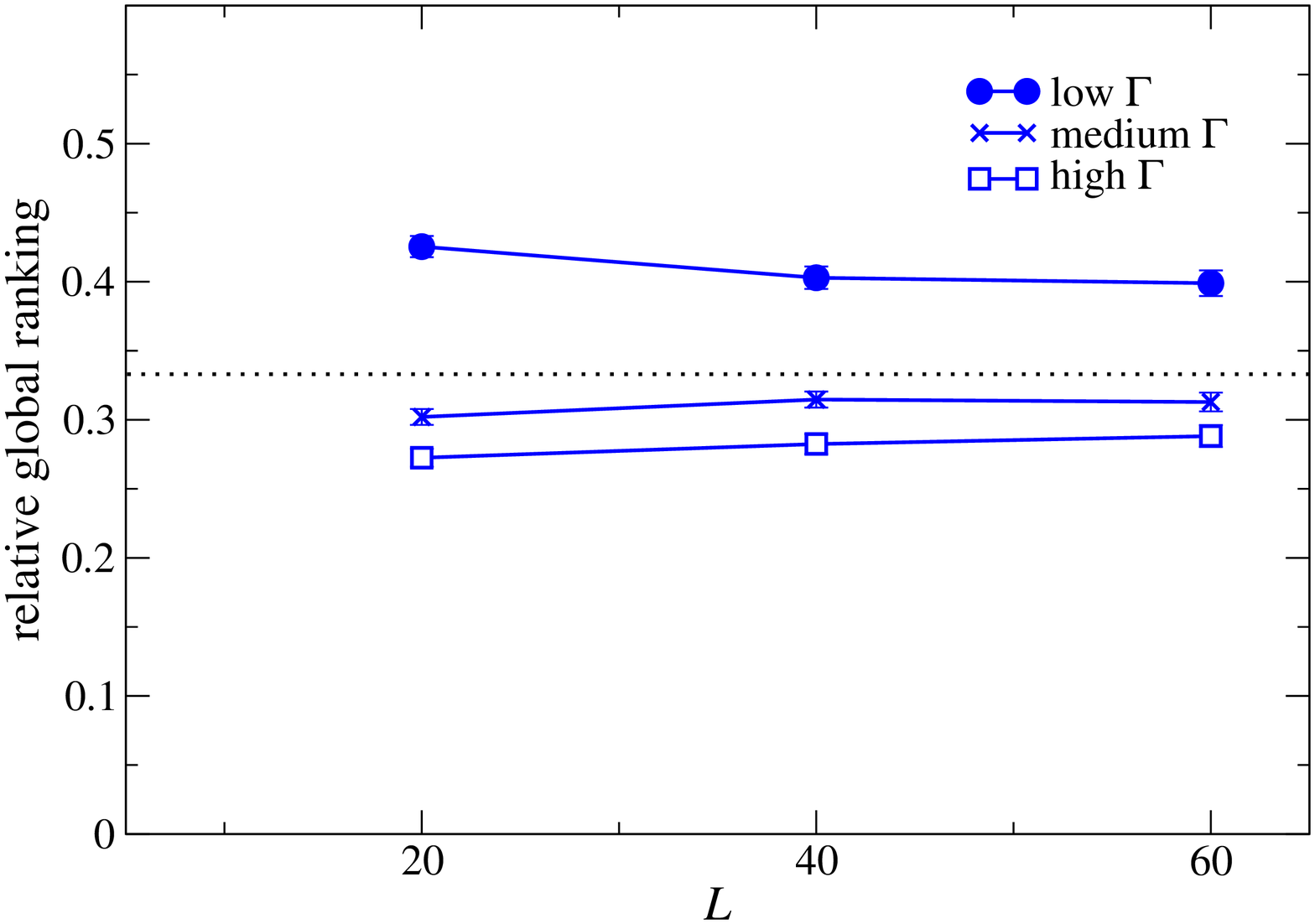}
\caption{(Supplementary) Distribution of the {\em local} (a+b) and {\em
    global} (c+d) ranking results of {\em all} $19882$ pathogenic mutations of
  the $162$ genes as a function of window lengths $L$. The left/right columns
  distinguish results for the 1D/2-leg models. The dashed horizontal lines
  show the $33\%$ mark of a completely random sequence. All lines are guides
  to the eyes only.}
\label{fig-sub-L_ranking-all}\label{fig-S4}
\end{figure}
\clearpage
\begin{figure}
\centering
\includegraphics[width=0.99\columnwidth,angle=-270,scale=0.41]{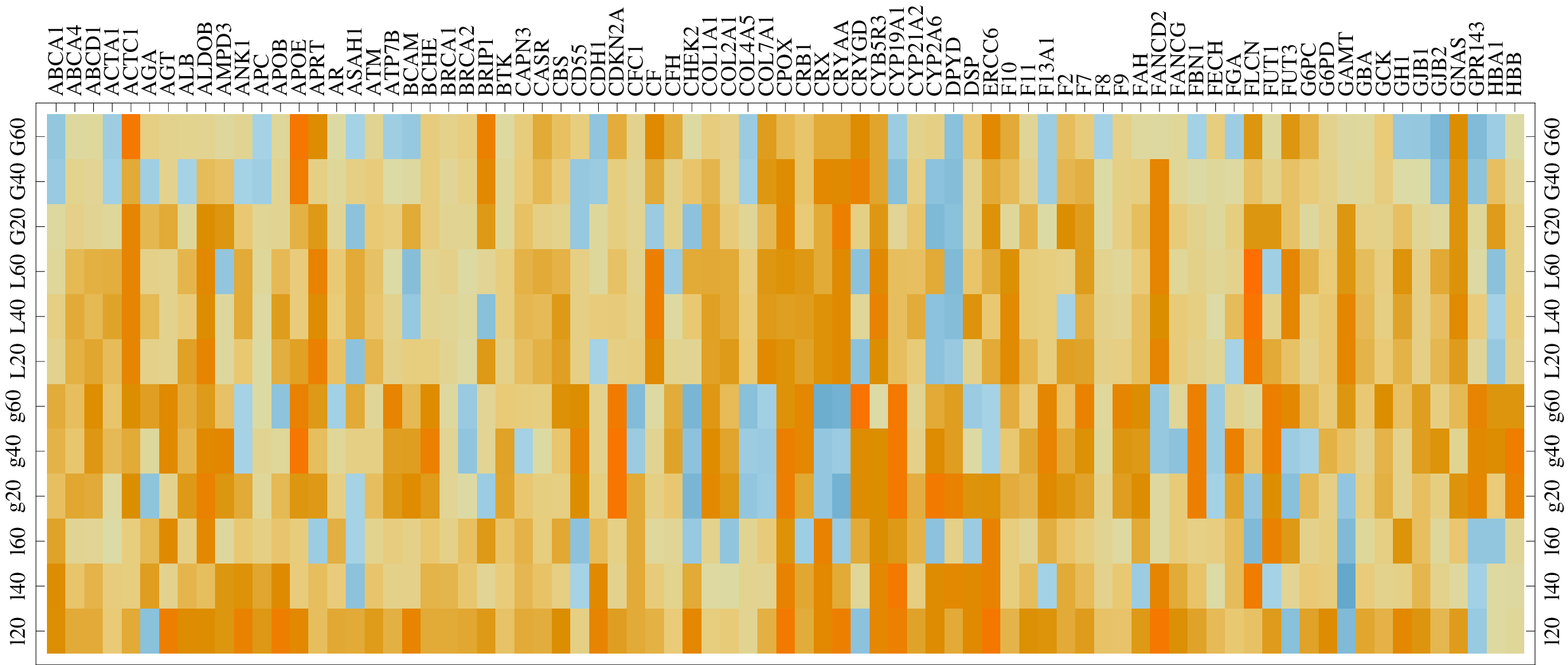}
\includegraphics[width=0.99\columnwidth,angle=-270,scale=0.406]{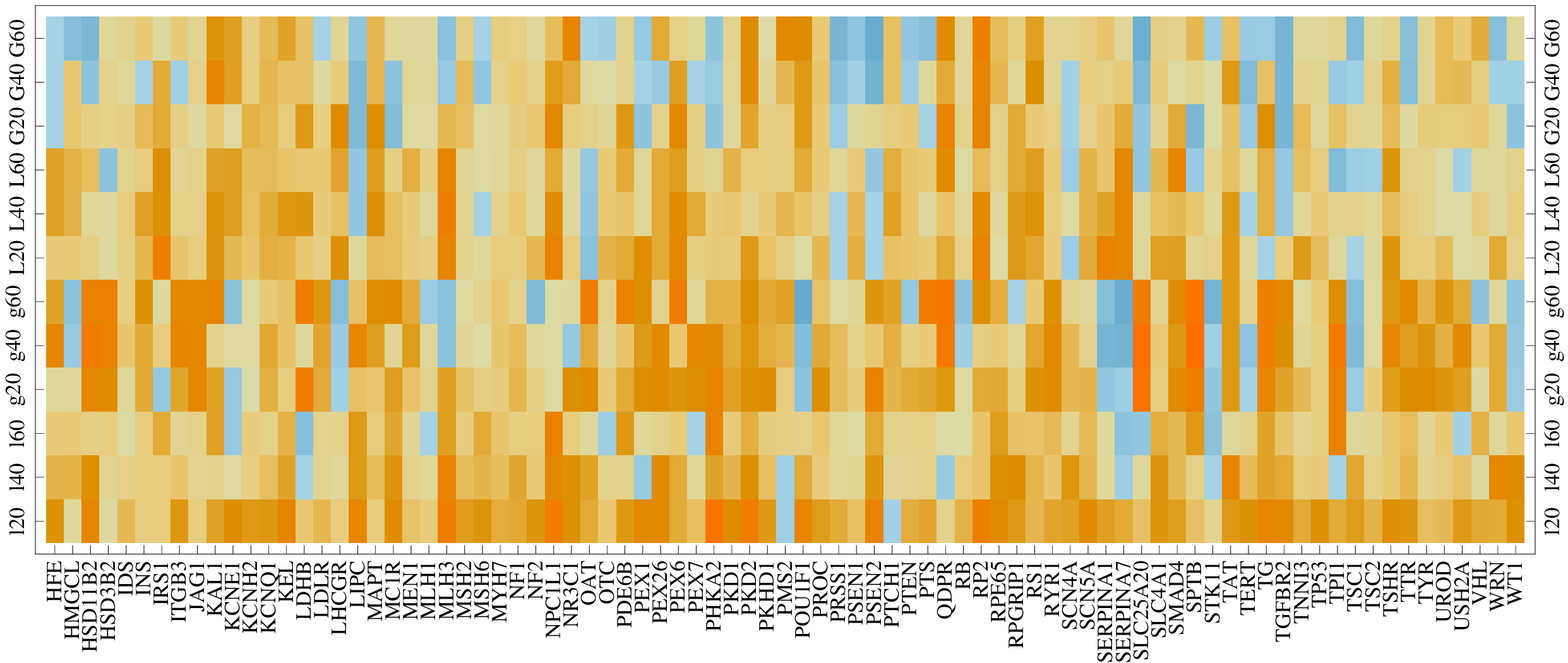}
\caption{(Supplementary) Numerical representation of the $12$ criteria
  for all $162$ genes, i.e.\ deviation from the $0.33$ line for the {\em
    local} rankings (l$i$, L$i$) and the {\em global} rankings (g$i$, G$i$)
  corresponding to the sorted prevalence for $L=20, 40$ and $60$,
  respectively. The lower case (l,g) indicates results for the 1D model,
  uppercase (L,G) refers to the 2-leg model. The genes are named according to
  the usage in the DNA databases.\cite{OMIM,SteBM03,PetMKI07,LohG04} The
  orange shading corresponds to an agreement with the CT hypothesis while the
  blue shading denotes disagreement. The first (last) column in the top
  (bottom) row gives the scale from $0$ to $1$ with $0.33$ corresponding to
  the white square.
}
\label{fig-sub-all}\label{fig-S5}
\end{figure}
\clearpage
\begin{figure}
\centering
\includegraphics[width=0.99\columnwidth,angle=-270,scale=0.35]{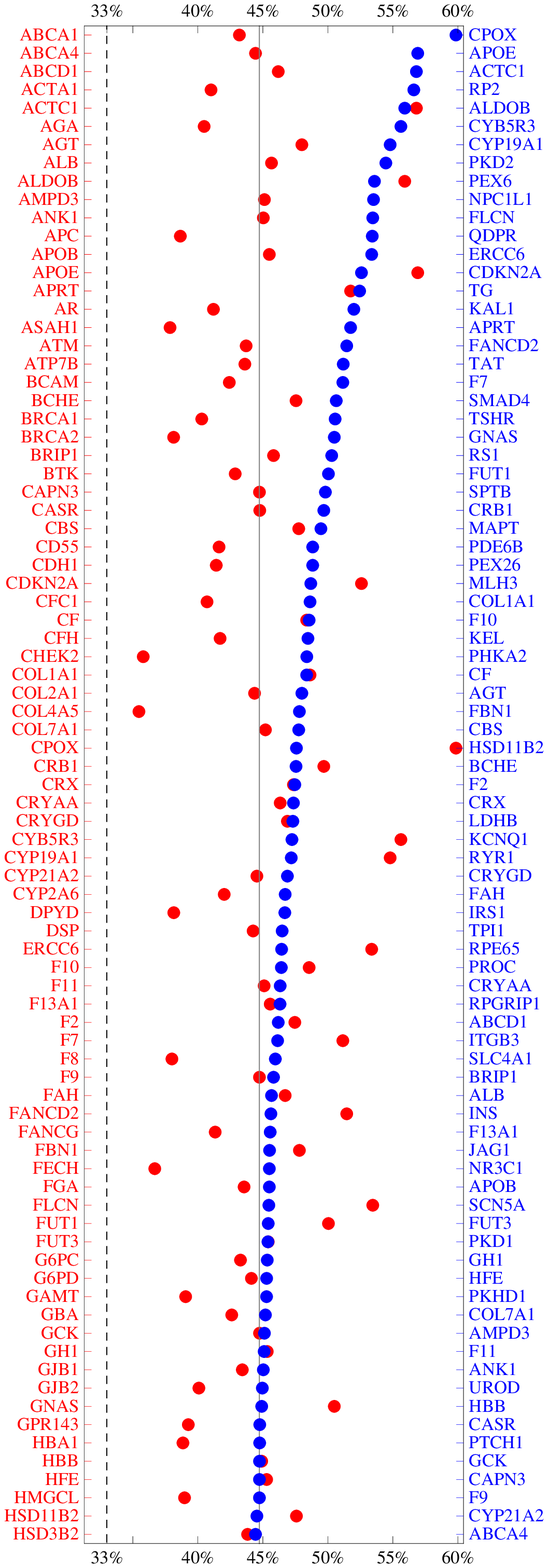}
\includegraphics[width=0.99\columnwidth,angle=-270,scale=0.355]{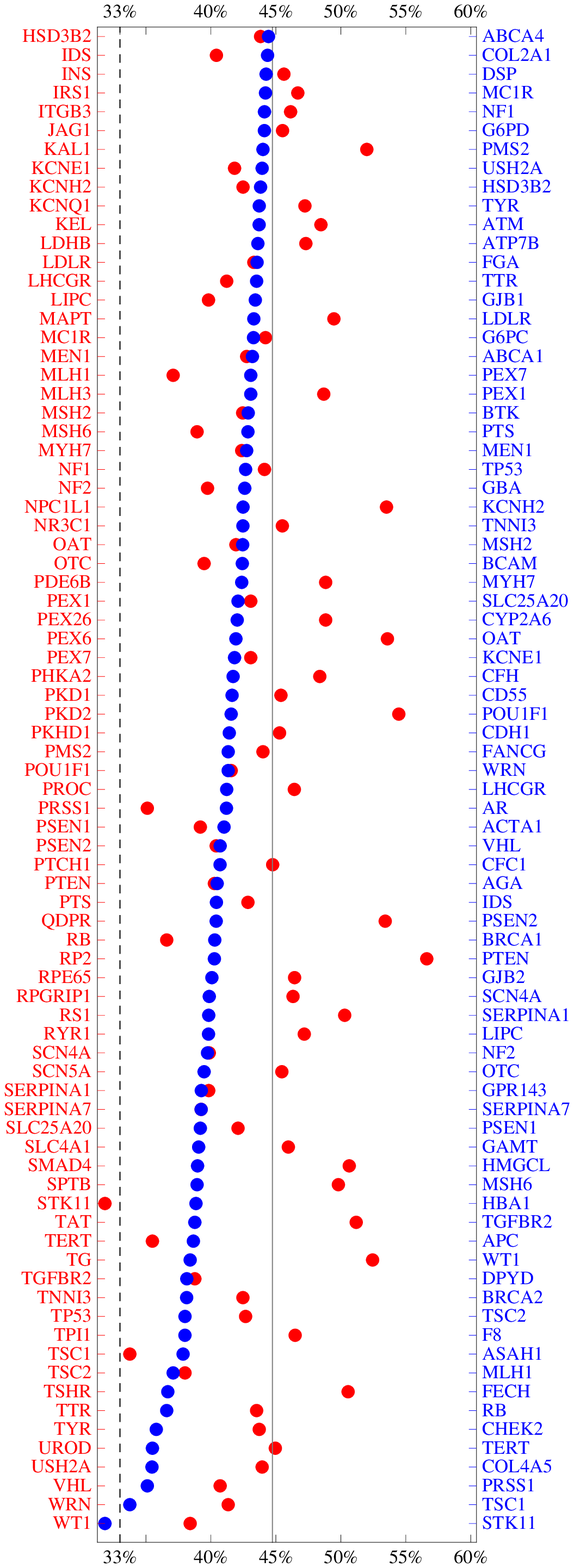}
\caption{(Supplementary) Graphs of the {\em average} over all $12$
  criteria as displayed in Fig.~\ref{fig-sub-all}. The red data points and
  gene names correspond to an alphabetic ordering of genes, whereas the blue
  points and labels are ordered according to the magnitude of the average. A
  larger average denotes a better agreement with our hypothesis. Points which
  lie below the dashed $33\%$ line show genes which on average fail. The
  average over all genes is denoted by the solid line. Results for HSD3B2
  (unsorted) and ABCA4 (sorted) have been duplicated in both rows.}
\label{fig-sub-all-avg}\label{fig-S6}
\end{figure}
\clearpage
\begin{figure}
\centering
(a)\includegraphics[width=0.45\columnwidth]{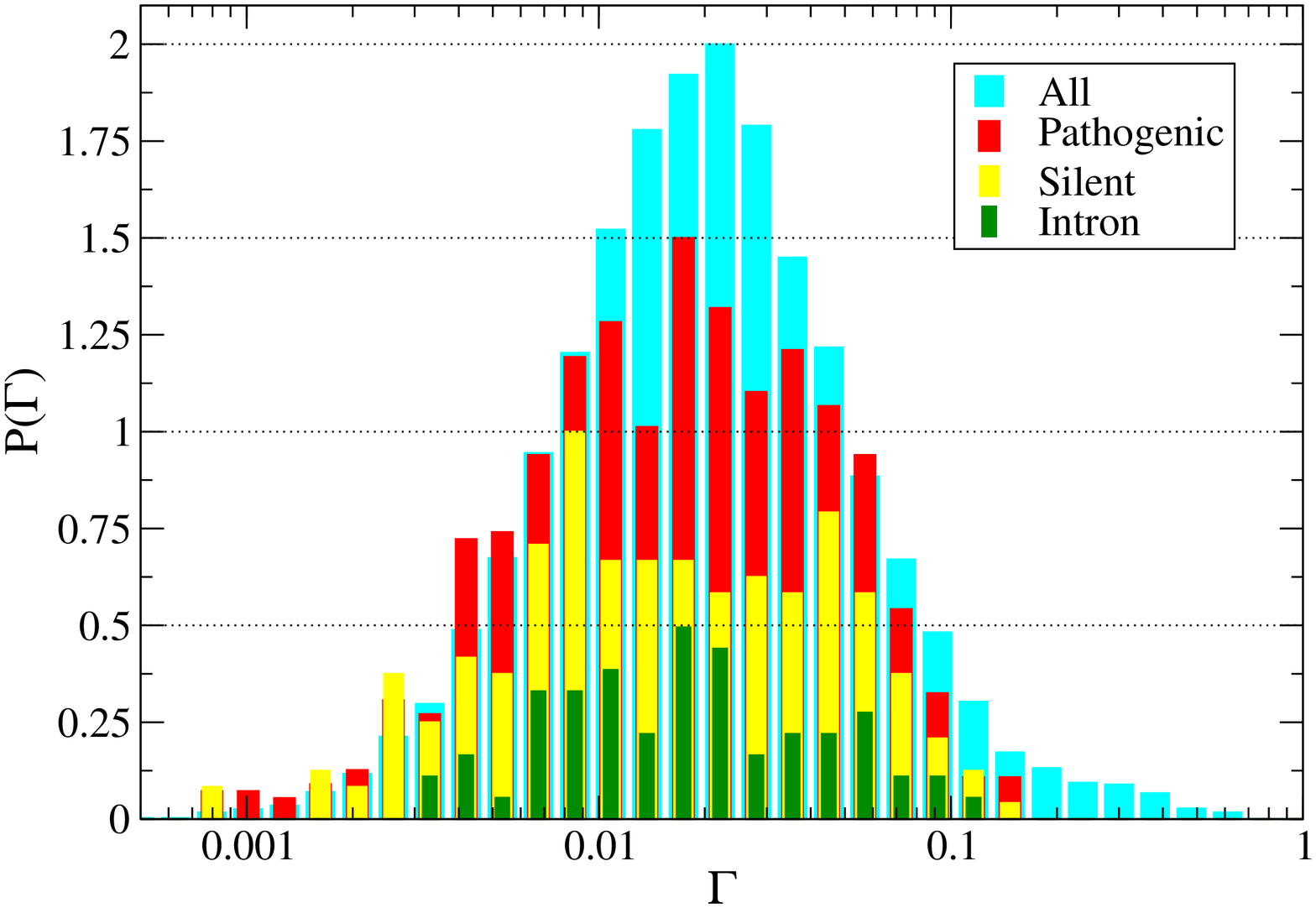}
(b)\includegraphics[width=0.45\columnwidth]{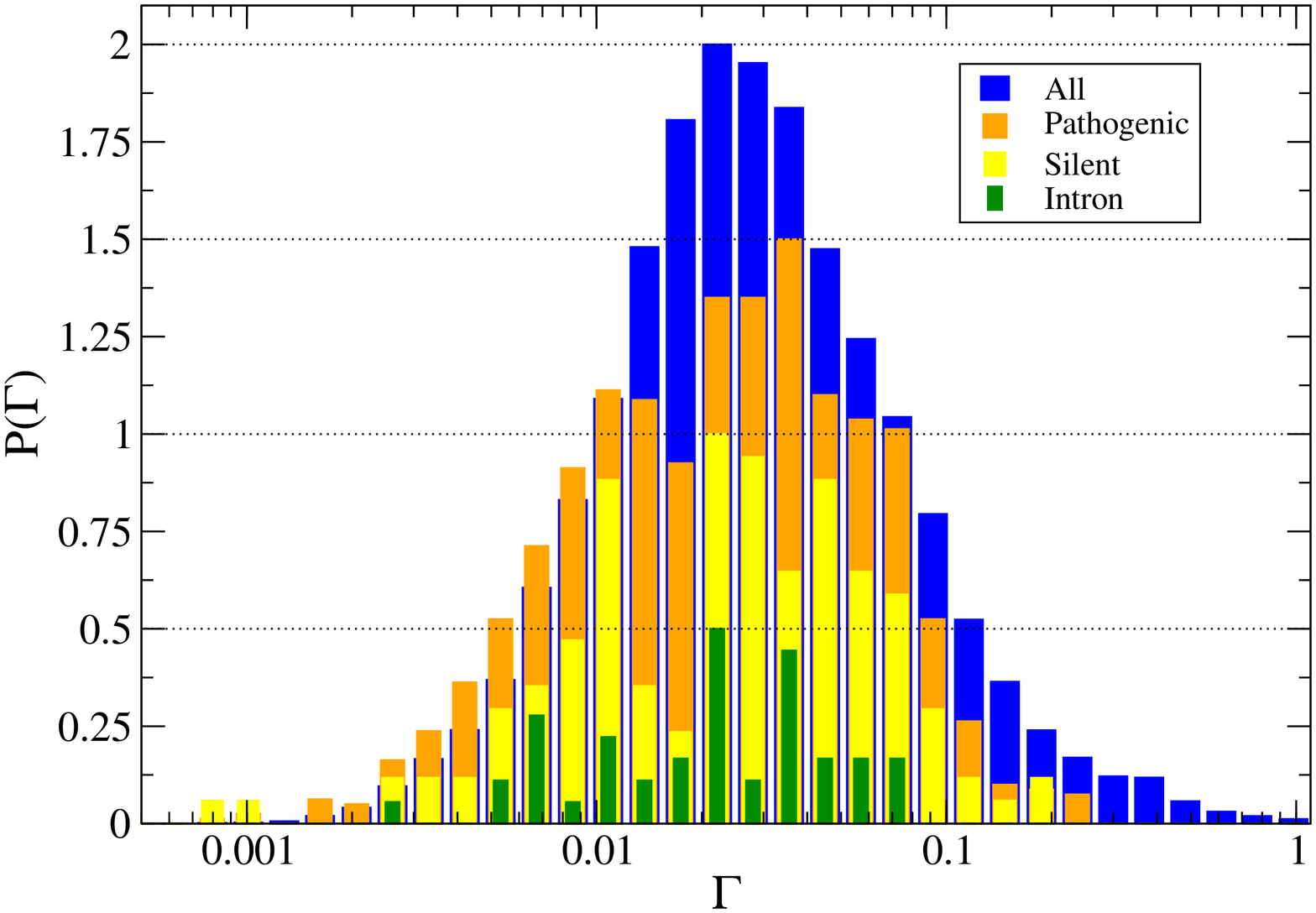}
\caption{Histograms of $\Gamma$ distributions for (a) transitions and (b) transversions in TP53, simulated using the 1D model and $L=20$. Histograms are shown for all possible mutations and for pathogenic, silent and intronic subsets. The maximum heights of the populations are scaled to be $2$, $1.5$, $1$ and $0.5$ to ease comparison. The scales factors are indicated by the dotted horizontal lines.}
  \label{fig-neutral}
\end{figure}

\clearpage
\setlength\LTleft{-10pt}
\setlength\LTright{-10pt}
{\small
\begin{longtable}{|l|rr|rrr|rrr|rrr|rrr|}

\caption{(Supplementary) List of the $162$ genes with their lengths
  (bps), number of all point mutations ($N_{pa}$), and their numbers of the
  $12$ types of point mutations. For example, $N_{At}$ means the number of
  $A\to T$ substitution.}
\label{tab-sub-genes}\label{tab-S1} \\
\toprule

Name & Length & N & N & N & N & N & N & N & N & N & N & N & N & N \\\kill
Name & Length & $N_{pa}$ & $N_{At}$ & $N_{Ac}$ & $N_{Ag}$ & $N_{Ta}$ & $N_{Tc}$ & $N_{Tg}$ & $N_{Ca}$ & $N_{Ct}$ & $N_{Cg}$ & $N_{Ga}$ & $N_{Gt}$ & $N_{Gc}$ \\
\midrule
\endfirsthead
\midrule
Name & Length & $N_{pa}$ & $N_{At}$ & $N_{Ac}$ & $N_{Ag}$ & $N_{Ta}$ & $N_{Tc}$ & $N_{Tg}$ & $N_{Ca}$ & $N_{Ct}$ & $N_{Cg}$ & $N_{Ga}$ & $N_{Gt}$ & $N_{Gc}$ \\
\midrule
\endhead 
\midrule
\endfoot 
\endlastfoot

ABCA1	&	147154	&	87	&	0	&	4	&	9	&	2	&	7	&	2	&	4	 &	 24	&	3	&	 18	&	7	&	7	\\
ABCA4	&	128313	&	382	&	11	&	9	&	21	&	13	&	51	&	21	&	27	 &	 73	&	19	&	 99	&	23	&	15	\\
ABCD1	&	19894	&	223	&	8	&	7	&	14	&	6	&	31	&	3	&	15	 &	 46	&	17	&	 47	&	13	&	16	\\
ACTA1	&	2852	&	164	&	10	&	7	&	22	&	5	&	13	&	6	&	13	 &	 12	&	11	&	 29	&	17	&	19	\\
ACTC1	&	7631	&	14	&	0	&	1	&	3	&	0	&	0	&	0	&	0	 &	 4	&	1	&	 2	&	1	&	2	\\
AGA	&	11668	&	19	&	0	&	0	&	0	&	1	&	3	&	1	&	0	&	 2	 &	0	&	8	 &	2	&	2	\\
AGT	&	11673	&	10	&	0	&	0	&	1	&	0	&	1	&	1	&	0	&	 5	 &	0	&	1	 &	0	&	1	\\
ALB	&	17127	&	63	&	3	&	2	&	13	&	2	&	1	&	0	&	1	&	 6	 &	1	&	24	 &	4	&	6	\\
ALDOB	&	14448	&	28	&	0	&	0	&	0	&	1	&	9	&	1	&	3	 &	 5	&	3	&	 3	&	1	&	2	\\
AMPD3	&	56903	&	11	&	0	&	1	&	0	&	1	&	1	&	0	&	0	 &	 6	&	1	&	 0	&	0	&	1	\\
ANK1	&	144397	&	18	&	0	&	0	&	1	&	0	&	2	&	0	&	1	 &	 7	&	1	&	 4	&	2	&	0	\\
APC	&	108353	&	222	&	10	&	0	&	4	&	18	&	1	&	8	&	21	&	 83	 &	28	&	18	 &	28	&	3	\\
APOB	&	42645	&	51	&	0	&	0	&	2	&	4	&	1	&	1	&	3	 &	 26	&	2	&	 8	&	3	&	1	\\
APOE	&	3612	&	33	&	0	&	1	&	1	&	0	&	2	&	0	&	2	 &	 9	&	2	&	 9	&	2	&	5	\\
APRT	&	2466	&	13	&	2	&	0	&	1	&	0	&	3	&	0	&	0	 &	 1	&	0	&	 4	&	1	&	1	\\
AR	&	180246	&	299	&	11	&	6	&	24	&	11	&	31	&	12	&	22	&	 53	 &	25	&	56	 &	31	&	17	\\
ASAH1	&	28574	&	12	&	1	&	0	&	3	&	1	&	0	&	0	&	1	 &	 0	&	3	&	 1	&	0	&	2	\\
ATM	&	146268	&	169	&	8	&	3	&	20	&	9	&	11	&	15	&	5	&	 55	 &	10	&	19	 &	8	&	6	\\
ATP7B	&	78826	&	315	&	10	&	14	&	25	&	14	&	27	&	10	&	17	 &	 62	&	16	&	 68	&	30	&	22	\\
BCAM	&	12341	&	14	&	1	&	0	&	1	&	1	&	0	&	0	&	1	 &	 4	&	1	&	 5	&	0	&	0	\\
BCHE	&	64562	&	58	&	6	&	2	&	6	&	3	&	6	&	3	&	2	 &	 12	&	0	&	 8	&	5	&	5	\\
BRCA1	&	81155	&	301	&	12	&	6	&	30	&	14	&	29	&	23	&	12	 &	 63	&	15	&	 38	&	50	&	9	\\
BRCA2	&	84193	&	162	&	12	&	9	&	20	&	8	&	11	&	8	&	12	 &	 33	&	13	&	 15	&	19	&	2	\\
BRIP1	&	180771	&	13	&	1	&	0	&	0	&	0	&	1	&	1	&	0	 &	 3	&	2	&	 2	&	1	&	2	\\
BTK	&	36741	&	329	&	15	&	14	&	29	&	19	&	47	&	23	&	26	&	 44	 &	14	&	48	 &	32	&	18	\\
CAPN3	&	64215	&	213	&	2	&	9	&	18	&	5	&	23	&	6	&	10	 &	 45	&	19	&	 48	&	14	&	14	\\
CASR	&	102813	&	144	&	2	&	5	&	12	&	4	&	21	&	7	&	8	 &	 20	&	10	&	 38	&	12	&	5	\\
CBS	&	23121	&	107	&	2	&	1	&	6	&	4	&	10	&	0	&	4	&	 24	 &	7	&	39	 &	2	&	8	\\
CD55	&	38983	&	14	&	0	&	0	&	1	&	2	&	0	&	1	&	0	 &	 4	&	0	&	 3	&	1	&	2	\\
CDH1	&	98250	&	30	&	0	&	1	&	2	&	0	&	2	&	2	&	0	 &	 9	&	1	&	 8	&	4	&	1	\\
CDKN2A	&	26740	&	71	&	1	&	3	&	4	&	2	&	6	&	6	&	5	 &	 12	&	3	&	 11	&	8	&	10	\\
CFC1	&	6748	&	10	&	0	&	0	&	0	&	0	&	1	&	0	&	0	 &	 4	&	1	&	 4	&	0	&	0	\\
CF	&	188699	&	828	&	35	&	31	&	103	&	50	&	85	&	54	&	47	&	 117	 &	41	&	 136	&	84	&	45	\\
CFH	&	95494	&	83	&	3	&	3	&	8	&	6	&	10	&	5	&	2	&	 10	 &	6	&	14	 &	13	&	3	\\
CHEK2	&	54092	&	20	&	1	&	1	&	2	&	0	&	1	&	0	&	2	 &	 4	&	0	&	 7	&	1	&	1	\\
COL1A1	&	17544	&	292	&	0	&	2	&	2	&	0	&	1	&	2	&	1	 &	 21	&	4	&	 134	&	79	&	46	\\
COL2A1	&	31538	&	124	&	0	&	1	&	2	&	1	&	1	&	1	&	5	 &	 26	&	0	&	 53	&	19	&	15	\\
COL4A5	&	257622	&	244	&	2	&	0	&	4	&	2	&	2	&	5	&	4	 &	 20	&	1	&	 117	&	55	&	32	\\
COL7A1	&	31088	&	265	&	0	&	3	&	6	&	2	&	1	&	0	&	1	 &	 56	&	7	&	 122	&	34	&	33	\\
CPOX	&	14152	&	36	&	0	&	2	&	1	&	0	&	3	&	1	&	0	 &	 14	&	2	&	 9	&	3	&	1	\\
CRB1	&	210178	&	91	&	3	&	1	&	2	&	8	&	16	&	7	&	3	 &	 11	&	2	&	 22	&	11	&	5	\\
CRX	&	21483	&	18	&	0	&	1	&	1	&	0	&	0	&	1	&	2	&	 4	 &	0	&	8	 &	0	&	1	\\
CRYAA	&	3773	&	10	&	0	&	0	&	0	&	0	&	0	&	0	&	1	 &	 5	&	0	&	 3	&	1	&	0	\\
CRYGD	&	2882	&	12	&	0	&	1	&	0	&	0	&	0	&	0	&	4	 &	 3	&	1	&	 2	&	1	&	0	\\
CYB5R3	&	30587	&	35	&	0	&	0	&	3	&	0	&	6	&	2	&	2	 &	 12	&	0	&	 10	&	0	&	0	\\
CYP19A1	&	129126	&	13	&	0	&	0	&	0	&	0	&	2	&	1	&	0	 &	 5	&	0	&	 5	&	0	&	0	\\
CYP21A2	&	3338	&	102	&	4	&	4	&	5	&	7	&	8	&	4	&	6	 &	 23	&	2	&	 25	&	4	&	10	\\
CYP2A6	&	6897	&	12	&	1	&	0	&	1	&	1	&	2	&	0	&	0	 &	 2	&	0	&	 2	&	2	&	1	\\
DPYD	&	843317	&	34	&	2	&	3	&	7	&	2	&	0	&	1	&	2	 &	 7	&	0	&	 5	&	4	&	1	\\
DSP	&	45077	&	20	&	0	&	0	&	2	&	1	&	1	&	1	&	0	&	 6	 &	1	&	6	 &	1	&	1	\\
ERCC6	&	80364	&	18	&	1	&	0	&	2	&	1	&	1	&	1	&	0	 &	 10	&	1	&	 1	&	0	&	0	\\
F10	&	26731	&	81	&	1	&	4	&	5	&	1	&	6	&	2	&	4	&	 11	 &	3	&	33	 &	5	&	6	\\
F11	&	23718	&	131	&	2	&	5	&	6	&	3	&	17	&	3	&	9	&	 28	 &	2	&	29	 &	13	&	14	\\
F13A1	&	176614	&	55	&	1	&	0	&	2	&	0	&	6	&	4	&	4	 &	 12	&	3	&	 14	&	8	&	1	\\
F2	&	20301	&	42	&	0	&	3	&	3	&	0	&	1	&	1	&	0	&	 11	 &	1	&	17	 &	3	&	2	\\
F7	&	14891	&	164	&	4	&	1	&	13	&	1	&	17	&	4	&	9	&	 30	 &	6	&	55	 &	13	&	11	\\
F8	&	186936	&	1168	&	79	&	47	&	124	&	56	&	117	&	78	&	55	 &	 153	&	72	 &	198	&	112	&	77	\\
F9	&	32723	&	707	&	31	&	26	&	55	&	58	&	69	&	52	&	42	&	 54	 &	28	&	135	 &	95	&	62	\\
FAH	&	33342	&	26	&	2	&	1	&	2	&	0	&	1	&	3	&	2	&	 6	 &	0	&	5	 &	4	&	0	\\
FANCD2	&	75502	&	14	&	0	&	0	&	0	&	0	&	3	&	3	&	0	 &	 4	&	0	&	 4	&	0	&	0	\\
FANCG	&	6179	&	16	&	0	&	0	&	0	&	0	&	2	&	1	&	0	 &	 7	&	0	&	 2	&	2	&	2	\\
FBN1	&	237414	&	640	&	18	&	12	&	52	&	32	&	88	&	37	&	21	 &	 63	&	32	&	 173	&	68	&	44	\\
FECH	&	38454	&	49	&	2	&	1	&	2	&	3	&	7	&	3	&	1	 &	 11	&	1	&	 11	&	4	&	3	\\
FGA	&	7618	&	45	&	3	&	1	&	3	&	3	&	1	&	2	&	3	&	 12	 &	2	&	7	 &	7	&	1	\\
FLCN	&	24971	&	11	&	0	&	0	&	1	&	0	&	0	&	0	&	0	 &	 4	&	2	&	 3	&	1	&	0	\\
FUT1	&	7380	&	22	&	0	&	0	&	1	&	2	&	2	&	1	&	2	 &	 5	&	1	&	 4	&	1	&	3	\\
FUT3	&	8587	&	11	&	0	&	0	&	0	&	1	&	0	&	2	&	2	 &	 0	&	0	&	 5	&	0	&	1	\\
G6PC	&	12572	&	66	&	2	&	2	&	3	&	2	&	8	&	3	&	3	 &	 13	&	2	&	 15	&	5	&	8	\\
G6PD	&	16182	&	163	&	3	&	3	&	21	&	4	&	15	&	4	&	8	 &	 27	&	15	&	 39	&	13	&	11	\\
GAMT	&	4465	&	11	&	0	&	2	&	0	&	0	&	1	&	0	&	0	 &	 1	&	1	&	 3	&	1	&	2	\\
GBA	&	10246	&	259	&	8	&	11	&	25	&	8	&	32	&	19	&	14	&	 42	 &	10	&	53	 &	19	&	18	\\
GCK	&	45153	&	255	&	5	&	13	&	15	&	7	&	32	&	8	&	19	&	 40	 &	11	&	64	 &	23	&	18	\\
GH1	&	1636	&	35	&	2	&	2	&	7	&	0	&	3	&	1	&	1	&	 5	 &	2	&	7	 &	3	&	2	\\
GJB1	&	10004	&	240	&	4	&	5	&	25	&	18	&	31	&	12	&	10	 &	 39	&	24	&	 39	&	17	&	16	\\
GJB2	&	5513	&	208	&	8	&	9	&	19	&	5	&	28	&	8	&	12	 &	 23	&	15	&	 49	&	19	&	13	\\
GNAS	&	71456	&	51	&	2	&	2	&	2	&	1	&	6	&	2	&	1	 &	 17	&	4	&	 9	&	3	&	2	\\
GPR143	&	40464	&	43	&	2	&	0	&	3	&	2	&	4	&	3	&	4	 &	 6	&	1	&	 10	&	4	&	4	\\
HBA1	&	842	&	73	&	2	&	5	&	9	&	2	&	5	&	2	&	7	&	 6	 &	9	&	8	 &	7	&	11	\\
HBB	&	1606	&	263	&	15	&	20	&	20	&	21	&	23	&	16	&	22	&	 26	 &	18	&	38	 &	20	&	24	\\
HFE	&	9612	&	27	&	1	&	2	&	0	&	0	&	4	&	1	&	0	&	 3	 &	2	&	7	 &	3	&	4	\\
HMGCL	&	23583	&	27	&	2	&	0	&	2	&	0	&	3	&	1	&	1	 &	 4	&	1	&	 8	&	3	&	2	\\
HSD11B2	&	6421	&	24	&	1	&	0	&	1	&	0	&	3	&	2	&	1	 &	 12	&	1	&	 3	&	0	&	0	\\
HSD3B2	&	7879	&	32	&	0	&	1	&	1	&	1	&	2	&	3	&	3	 &	 8	&	3	&	 6	&	2	&	2	\\
IDS	&	26493	&	203	&	15	&	8	&	15	&	2	&	16	&	13	&	17	&	 31	 &	19	&	32	 &	20	&	15	\\
INS	&	1431	&	30	&	0	&	0	&	2	&	0	&	3	&	2	&	1	&	 3	 &	6	&	6	 &	4	&	3	\\
IRS1	&	64538	&	14	&	0	&	1	&	3	&	0	&	1	&	0	&	1	 &	 2	&	1	&	 3	&	0	&	2	\\
ITGB3	&	58870	&	53	&	2	&	2	&	3	&	1	&	10	&	4	&	1	 &	 12	&	1	&	 11	&	5	&	1	\\
JAG1	&	36257	&	131	&	2	&	0	&	3	&	6	&	11	&	6	&	11	 &	 30	&	12	&	 28	&	16	&	6	\\
KAL1	&	203313	&	25	&	0	&	0	&	1	&	2	&	1	&	1	&	1	 &	 9	&	2	&	 6	&	1	&	1	\\
KCNE1	&	65586	&	17	&	0	&	0	&	1	&	0	&	2	&	0	&	1	 &	 5	&	0	&	 6	&	1	&	1	\\
KCNH2	&	32966	&	266	&	8	&	11	&	27	&	5	&	19	&	12	&	15	 &	 61	&	9	&	 43	&	35	&	21	\\
KCNQ1	&	404120	&	226	&	3	&	2	&	19	&	8	&	24	&	5	&	12	 &	 44	&	13	&	 61	&	11	&	24	\\
KEL	&	21303	&	33	&	2	&	0	&	3	&	1	&	3	&	0	&	0	&	 9	 &	1	&	13	 &	0	&	1	\\
LDHB	&	22501	&	11	&	1	&	1	&	1	&	0	&	1	&	2	&	1	 &	 1	&	0	&	 2	&	0	&	1	\\
LDLR	&	44450	&	741	&	23	&	31	&	48	&	31	&	84	&	35	&	51	 &	 88	&	48	&	 168	&	92	&	42	\\
LHCGR	&	68951	&	37	&	2	&	3	&	3	&	3	&	7	&	3	&	2	 &	 7	&	1	&	 3	&	2	&	1	\\
LIPC	&	136898	&	11	&	0	&	1	&	2	&	0	&	0	&	1	&	0	 &	 2	&	0	&	 4	&	0	&	1	\\
MAPT	&	133924	&	36	&	3	&	2	&	2	&	0	&	3	&	2	&	2	 &	 6	&	1	&	 9	&	5	&	1	\\
MC1R	&	2360	&	24	&	0	&	1	&	1	&	0	&	4	&	0	&	3	 &	 8	&	0	&	 5	&	1	&	1	\\
MEN1	&	7779	&	239	&	10	&	7	&	8	&	9	&	26	&	11	&	19	 &	 44	&	14	&	 38	&	33	&	20	\\
MLH1	&	57359	&	275	&	16	&	15	&	26	&	18	&	19	&	17	&	18	 &	 42	&	20	&	 36	&	28	&	20	\\
MLH3	&	37769	&	17	&	0	&	1	&	5	&	0	&	1	&	0	&	0	 &	 2	&	1	&	 4	&	2	&	1	\\
MSH2	&	80098	&	238	&	16	&	11	&	25	&	8	&	9	&	14	&	11	 &	 62	&	14	&	 30	&	25	&	13	\\
MSH6	&	23872	&	54	&	3	&	1	&	5	&	2	&	3	&	0	&	3	 &	 17	&	6	&	 7	&	4	&	3	\\
MYH7	&	22924	&	268	&	8	&	10	&	20	&	4	&	19	&	8	&	16	 &	 47	&	17	&	 80	&	16	&	23	\\
NF1	&	282701	&	338	&	22	&	4	&	24	&	20	&	35	&	26	&	14	&	 82	 &	24	&	44	 &	29	&	14	\\
NF2	&	95023	&	72	&	5	&	2	&	5	&	2	&	6	&	1	&	2	&	 25	 &	4	&	7	 &	11	&	2	\\
NPC1L1	&	28781	&	26	&	0	&	0	&	3	&	2	&	0	&	0	&	0	 &	 11	&	1	&	 8	&	1	&	0	\\
NR3C1	&	157582	&	14	&	1	&	0	&	1	&	1	&	4	&	1	&	0	 &	 1	&	0	&	 4	&	0	&	1	\\
OAT	&	21580	&	42	&	0	&	0	&	2	&	2	&	4	&	0	&	3	&	 9	 &	2	&	11	 &	5	&	4	\\
OTC	&	68968	&	276	&	16	&	11	&	28	&	9	&	31	&	18	&	17	&	 36	 &	15	&	44	 &	27	&	24	\\
PDE6B	&	45199	&	20	&	1	&	0	&	0	&	3	&	3	&	1	&	2	 &	 5	&	1	&	 4	&	0	&	0	\\
PEX1	&	41509	&	24	&	0	&	0	&	0	&	0	&	4	&	1	&	2	 &	 7	&	3	&	 6	&	0	&	1	\\
PEX26	&	11503	&	10	&	0	&	0	&	0	&	0	&	3	&	0	&	0	 &	 3	&	2	&	 2	&	0	&	0	\\
PEX6	&	15143	&	18	&	0	&	1	&	0	&	0	&	3	&	1	&	1	 &	 7	&	0	&	 5	&	0	&	0	\\
PEX7	&	91337	&	24	&	1	&	2	&	2	&	1	&	1	&	3	&	2	 &	 6	&	1	&	 4	&	1	&	0	\\
PHKA2	&	91305	&	23	&	0	&	1	&	2	&	0	&	0	&	1	&	1	 &	 11	&	0	&	 5	&	2	&	0	\\
PKD1	&	47189	&	149	&	2	&	3	&	6	&	5	&	12	&	4	&	8	 &	 59	&	10	&	 27	&	8	&	5	\\
PKD2	&	70110	&	35	&	1	&	0	&	1	&	1	&	1	&	1	&	2	 &	 17	&	0	&	 7	&	3	&	1	\\
PKHD1	&	472279	&	213	&	8	&	10	&	22	&	7	&	29	&	9	&	7	 &	 50	&	7	&	 38	&	17	&	9	\\
PMS2	&	35868	&	21	&	3	&	1	&	1	&	1	&	0	&	0	&	0	 &	 6	&	0	&	 5	&	4	&	0	\\
POU1F1	&	16954	&	22	&	1	&	0	&	2	&	1	&	3	&	1	&	1	 &	 6	&	0	&	 4	&	2	&	1	\\
PROC	&	10802	&	203	&	6	&	6	&	10	&	3	&	21	&	6	&	15	 &	 40	&	8	&	 55	&	13	&	20	\\
PRSS1	&	3592	&	26	&	1	&	2	&	2	&	2	&	2	&	0	&	2	 &	 5	&	1	&	 5	&	2	&	2	\\
PSEN1	&	83931	&	154	&	6	&	8	&	13	&	8	&	22	&	11	&	7	 &	 21	&	12	&	 19	&	16	&	11	\\
PSEN2	&	25532	&	18	&	2	&	2	&	3	&	0	&	0	&	0	&	0	 &	 5	&	1	&	 5	&	0	&	0	\\
PTCH1	&	73984	&	59	&	3	&	2	&	1	&	2	&	4	&	2	&	7	 &	 15	&	2	&	 11	&	8	&	2	\\
PTEN	&	105338	&	98	&	2	&	2	&	10	&	9	&	13	&	11	&	5	 &	 15	&	8	&	 15	&	6	&	2	\\
PTS	&	7595	&	27	&	1	&	0	&	8	&	1	&	0	&	2	&	0	&	 6	 &	2	&	4	 &	2	&	1	\\
QDPR	&	57702	&	20	&	0	&	1	&	2	&	0	&	3	&	3	&	0	 &	 3	&	0	&	 6	&	1	&	1	\\
RB	&	180388	&	226	&	9	&	8	&	18	&	12	&	16	&	11	&	10	&	 38	 &	8	&	51	 &	28	&	17	\\
RP2	&	45418	&	17	&	0	&	0	&	1	&	0	&	1	&	2	&	0	&	 5	 &	2	&	4	 &	2	&	0	\\
RPE65	&	21136	&	42	&	1	&	1	&	2	&	1	&	5	&	3	&	3	 &	 9	&	1	&	 7	&	7	&	2	\\
RPGRIP1	&	63325	&	24	&	0	&	2	&	5	&	1	&	0	&	0	&	0	 &	 7	&	0	&	 5	&	3	&	1	\\
RS1	&	32422	&	93	&	3	&	0	&	7	&	5	&	11	&	4	&	5	&	 15	 &	5	&	19	 &	7	&	12	\\
RYR1	&	153865	&	244	&	5	&	4	&	21	&	9	&	20	&	6	&	10	 &	 56	&	14	&	 63	&	17	&	19	\\
SCN4A	&	34365	&	43	&	1	&	0	&	5	&	2	&	3	&	1	&	4	 &	 7	&	3	&	 12	&	2	&	3	\\
SCN5A	&	101611	&	226	&	0	&	2	&	18	&	9	&	16	&	6	&	13	 &	 49	&	10	&	 77	&	15	&	11	\\
SERPINA1	&	12332	&	29	&	4	&	1	&	0	&	1	&	2	&	1	&	 2	 &	8	&	1	 &	9	&	0	&	0	\\
SERPINA7	&	3870	&	16	&	1	&	0	&	0	&	2	&	1	&	0	&	 1	 &	4	&	0	 &	5	&	1	&	1	\\
SLC25A20	&	41966	&	11	&	0	&	0	&	1	&	0	&	0	&	0	&	 0	 &	4	&	1	 &	3	&	1	&	1	\\
SLC4A1	&	18428	&	65	&	1	&	1	&	3	&	2	&	5	&	0	&	6	 &	 20	&	4	&	 20	&	1	&	2	\\
SMAD4	&	49535	&	20	&	0	&	1	&	1	&	0	&	0	&	2	&	1	 &	 6	&	3	&	 4	&	1	&	1	\\
SPTB	&	76865	&	18	&	0	&	0	&	2	&	2	&	2	&	2	&	0	 &	 6	&	1	&	 0	&	1	&	2	\\
STK11	&	22637	&	62	&	4	&	4	&	2	&	1	&	4	&	5	&	7	 &	 12	&	5	&	 8	&	8	&	2	\\
TAT	&	10242	&	11	&	0	&	0	&	0	&	0	&	1	&	1	&	0	&	 5	 &	1	&	2	 &	1	&	0	\\
TERT	&	41881	&	30	&	0	&	1	&	3	&	1	&	2	&	0	&	0	 &	 10	&	3	&	 8	&	0	&	2	\\
TG	&	267939	&	33	&	0	&	1	&	2	&	1	&	2	&	1	&	1	&	 7	 &	0	&	14	 &	4	&	0	\\
TGFBR2	&	87641	&	14	&	0	&	0	&	1	&	1	&	1	&	0	&	0	 &	 5	&	0	&	 3	&	1	&	2	\\
TNNI3	&	5966	&	30	&	0	&	1	&	5	&	1	&	1	&	0	&	0	 &	 8	&	2	&	 10	&	0	&	2	\\
TP53	&	20303	&	2003	&	137	&	113	&	158	&	121	&	142	&	109	&	 165	 &	284	&	 156	&	252	&	202	&	164	\\
TPI1	&	3287	&	11	&	0	&	0	&	1	&	1	&	1	&	0	&	0	 &	 1	&	0	&	 4	&	1	&	2	\\
TSC1	&	53285	&	44	&	2	&	0	&	1	&	1	&	1	&	2	&	5	 &	 19	&	5	&	 5	&	3	&	0	\\
TSC2	&	40724	&	165	&	7	&	4	&	6	&	5	&	13	&	5	&	22	 &	 48	&	18	&	 22	&	10	&	5	\\
TSHR	&	190778	&	45	&	1	&	0	&	3	&	1	&	9	&	2	&	3	 &	 8	&	1	&	 12	&	2	&	3	\\
TTR	&	6944	&	98	&	4	&	5	&	10	&	6	&	15	&	9	&	6	&	 5	 &	1	&	19	 &	11	&	7	\\
TYR	&	117888	&	205	&	10	&	10	&	22	&	6	&	16	&	6	&	16	&	 27	 &	13	&	42	 &	26	&	11	\\
UROD	&	3512	&	45	&	0	&	1	&	2	&	5	&	6	&	2	&	3	 &	 9	&	2	&	 11	&	2	&	2	\\
USH2A	&	800503	&	66	&	0	&	3	&	1	&	1	&	2	&	3	&	6	 &	 24	&	3	&	 8	&	10	&	5	\\
VHL	&	10444	&	172	&	5	&	7	&	12	&	13	&	22	&	15	&	7	&	 22	 &	21	&	17	 &	18	&	13	\\
WRN	&	140499	&	22	&	3	&	1	&	1	&	1	&	1	&	0	&	0	&	 11	 &	2	&	1	 &	1	&	0	\\
WT1	&	47763	&	56	&	1	&	2	&	5	&	1	&	6	&	3	&	3	&	 13	 &	4	&	11	 &	5	&	2	\\

\bottomrule

\end{longtable}
}
\end{document}

%
%
\setcounter{figure}{0}
\def\thefigure{S\arabic{figure}}
\def\thetable{S\arabic{table}}
\clearpage
\setcounter{page}{1}
\begin{figure}
(a)\includegraphics[width=0.45\columnwidth]{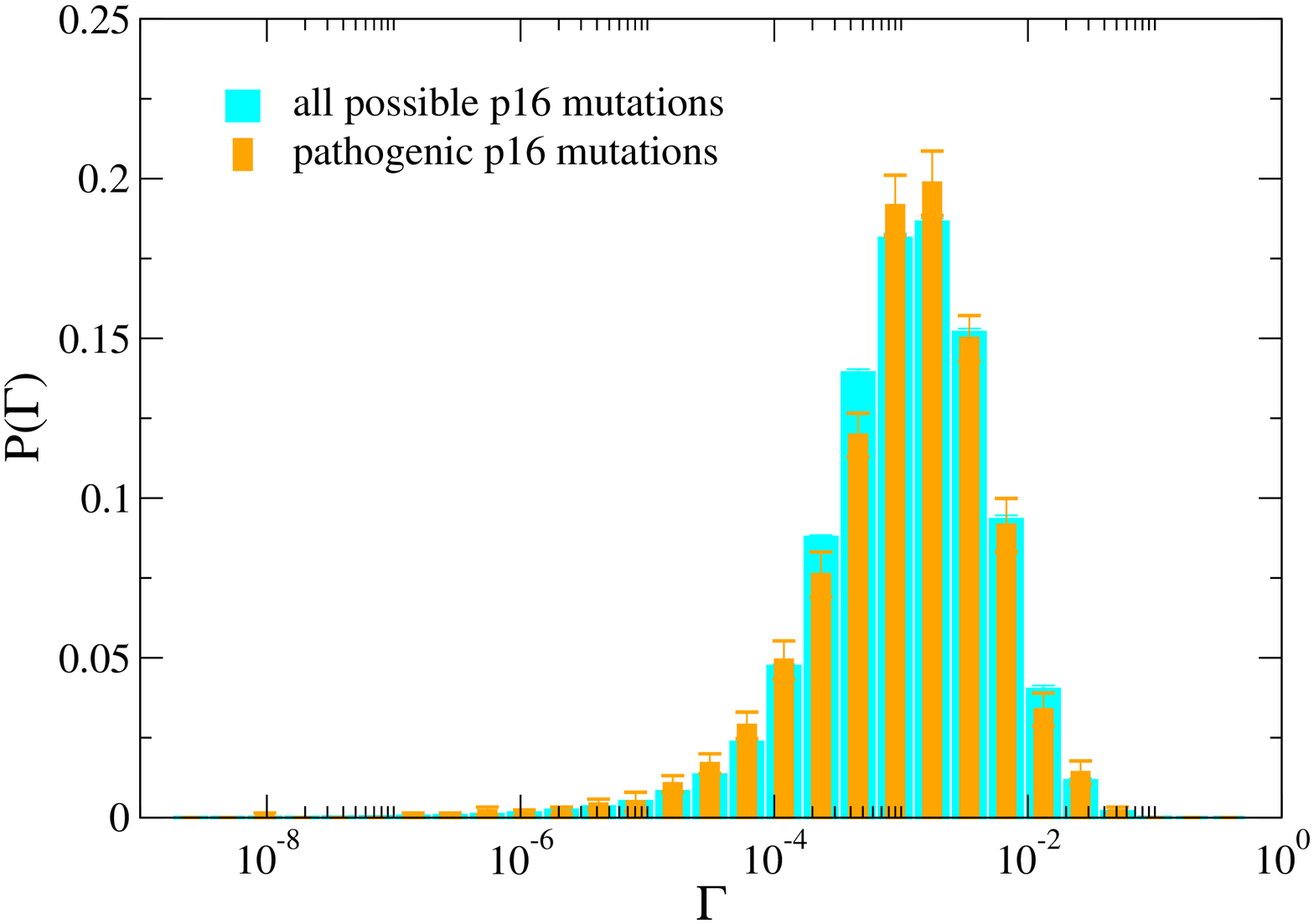}
(b)\includegraphics[width=0.45\columnwidth]{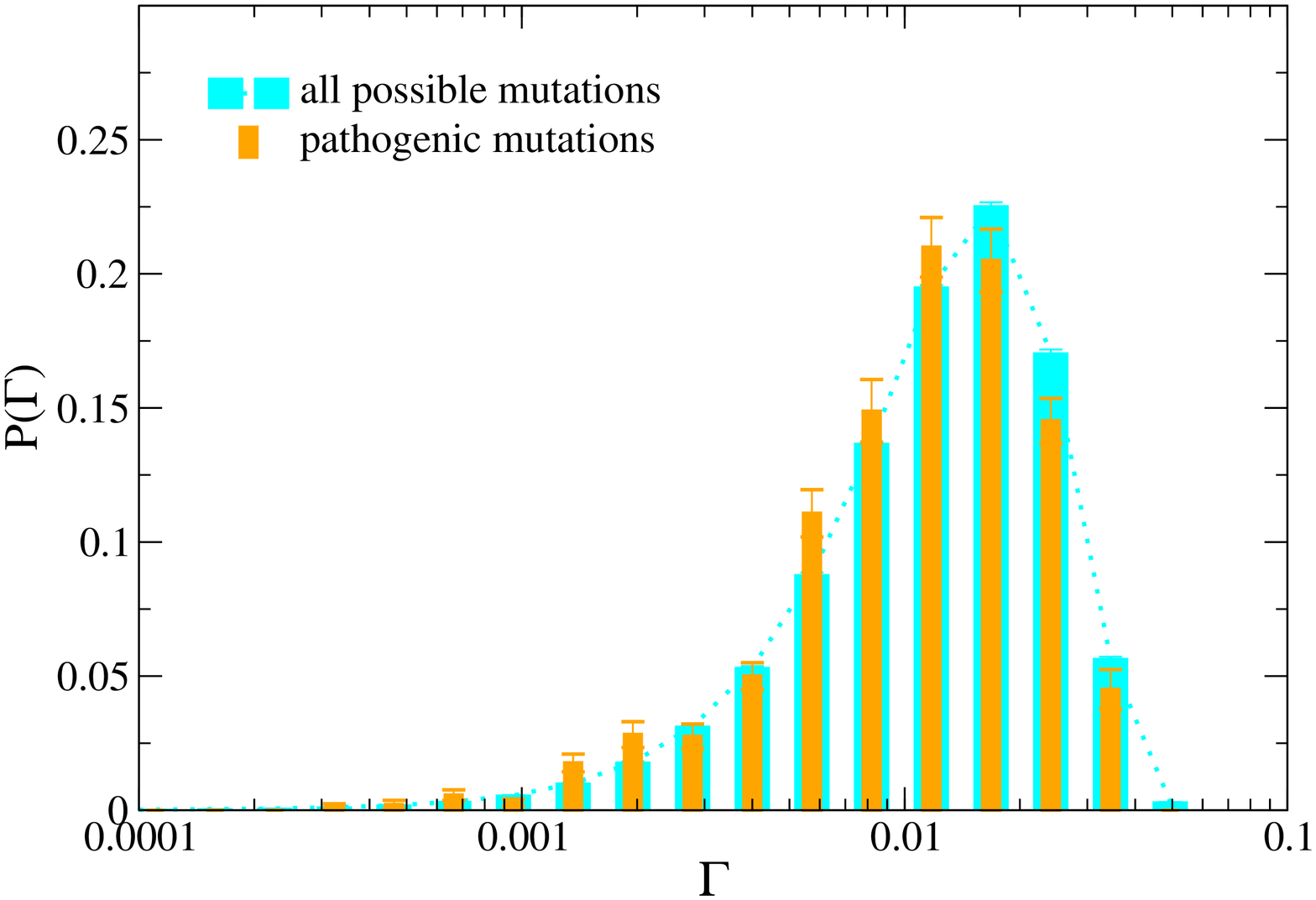}
\caption{}
\end{figure}
\clearpage
\begin{figure}
\centering

(a) \includegraphics[width=0.2\columnwidth]{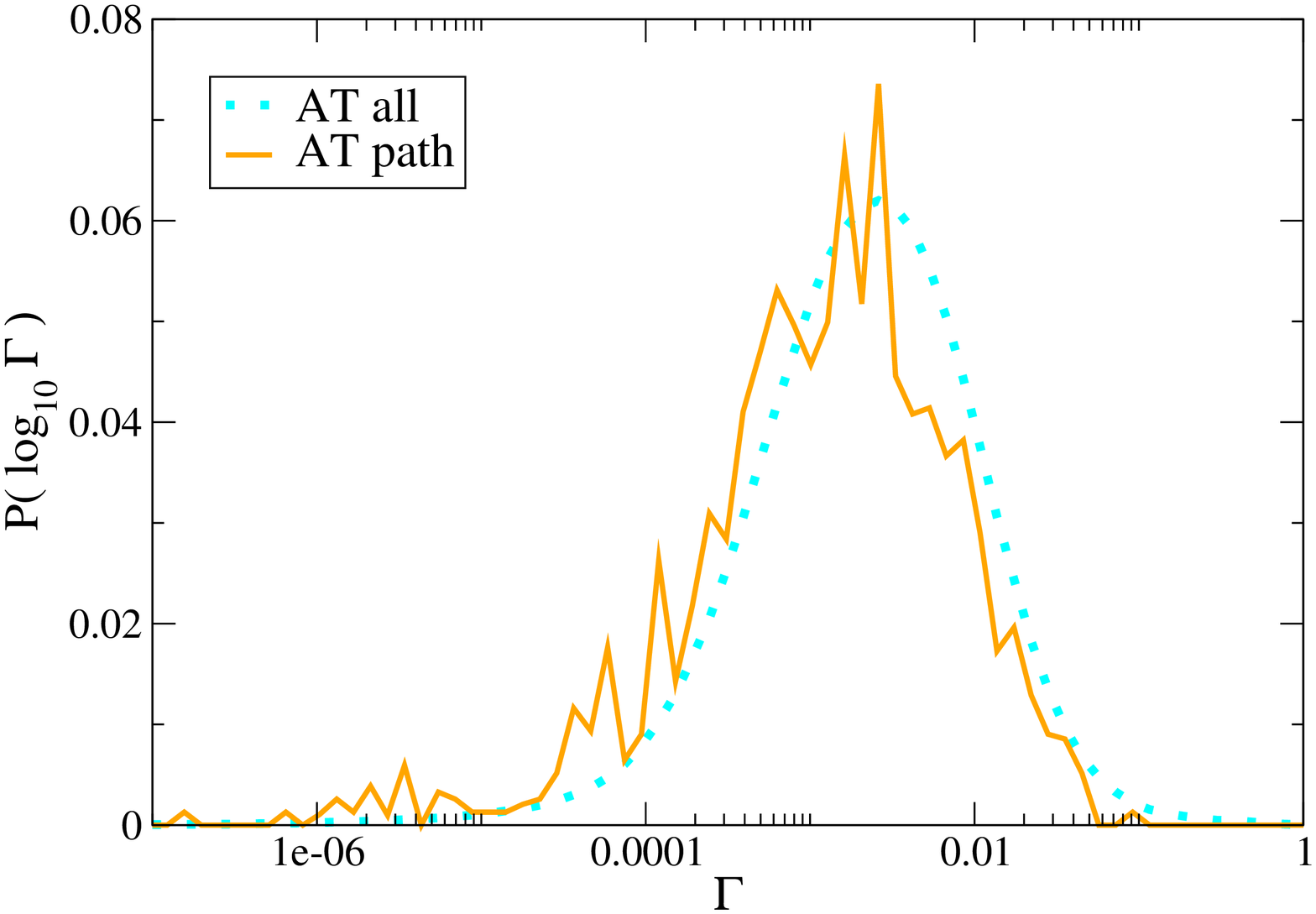}
(b) \includegraphics[width=0.2\columnwidth]{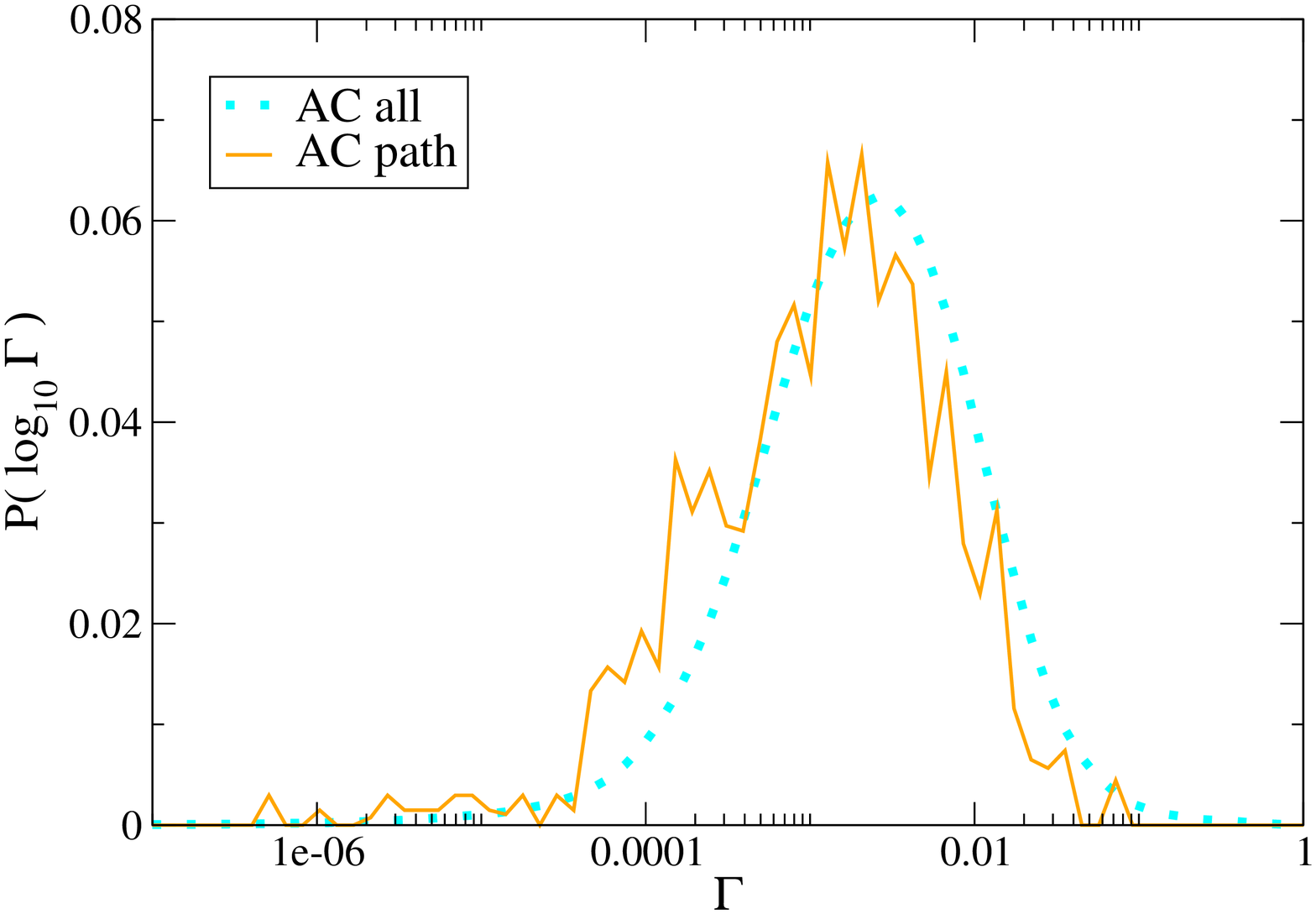}
(c) \includegraphics[width=0.2\columnwidth]{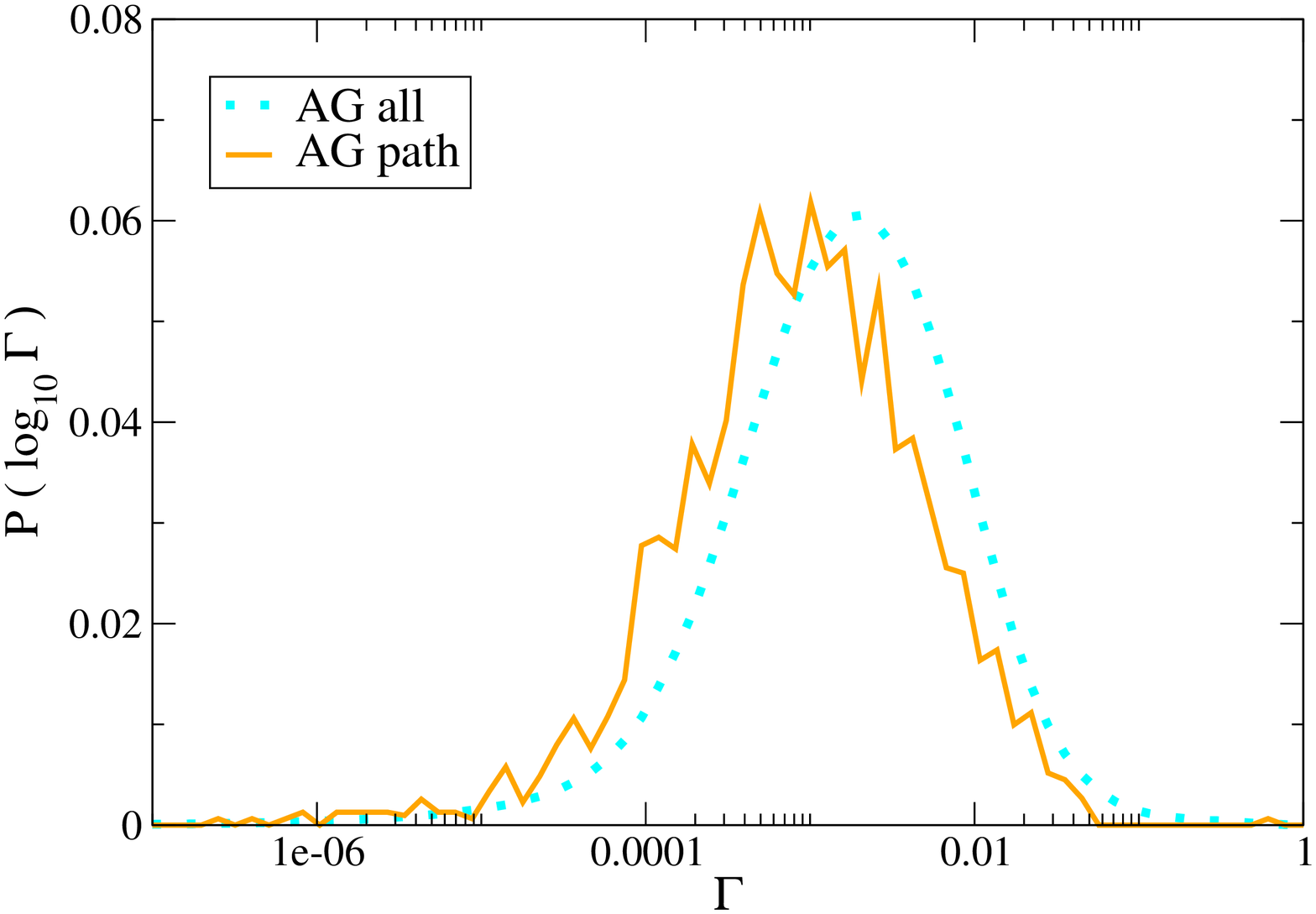}
(d) \includegraphics[width=0.2\columnwidth]{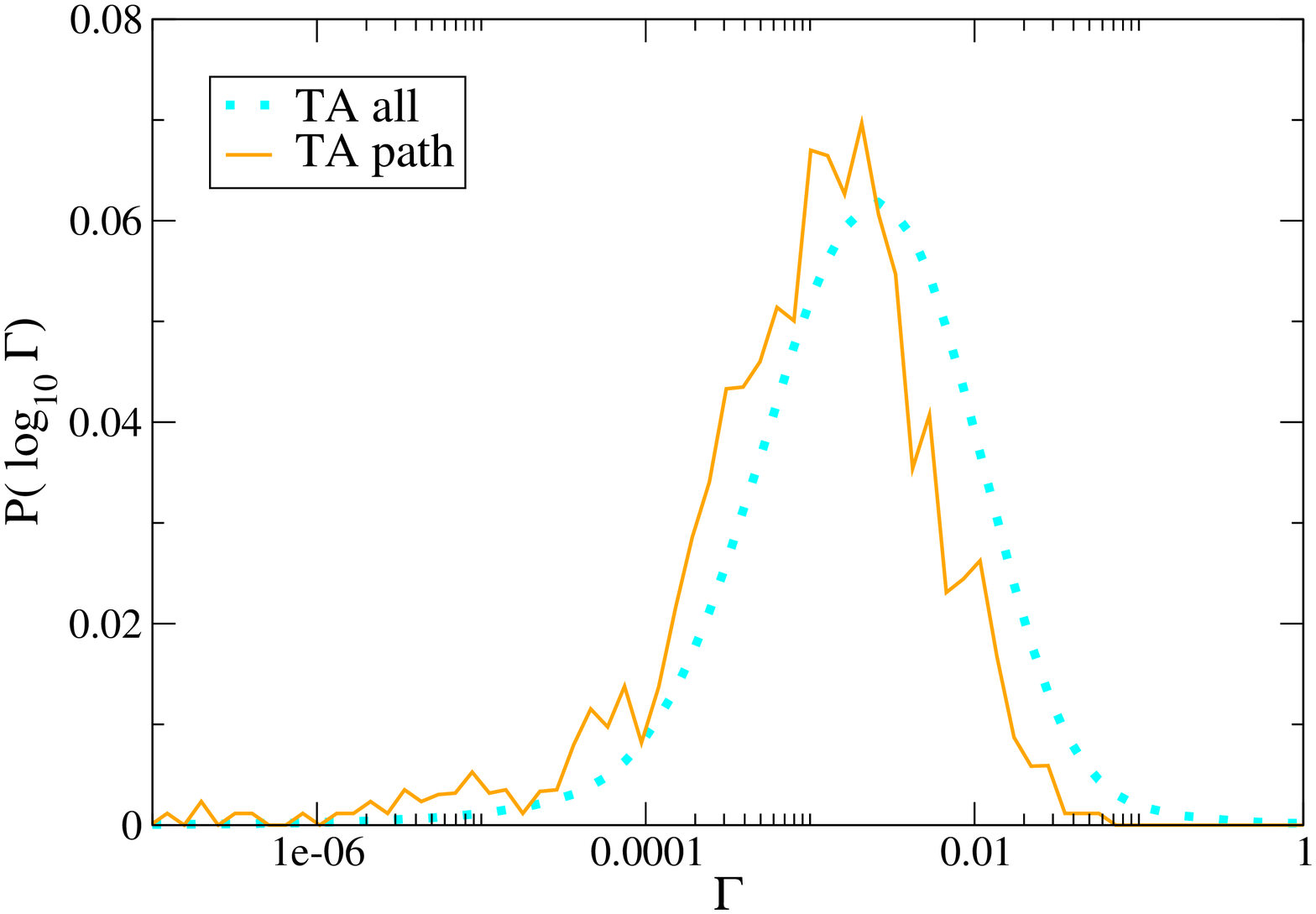}
\\
(e) \includegraphics[width=0.2\columnwidth]{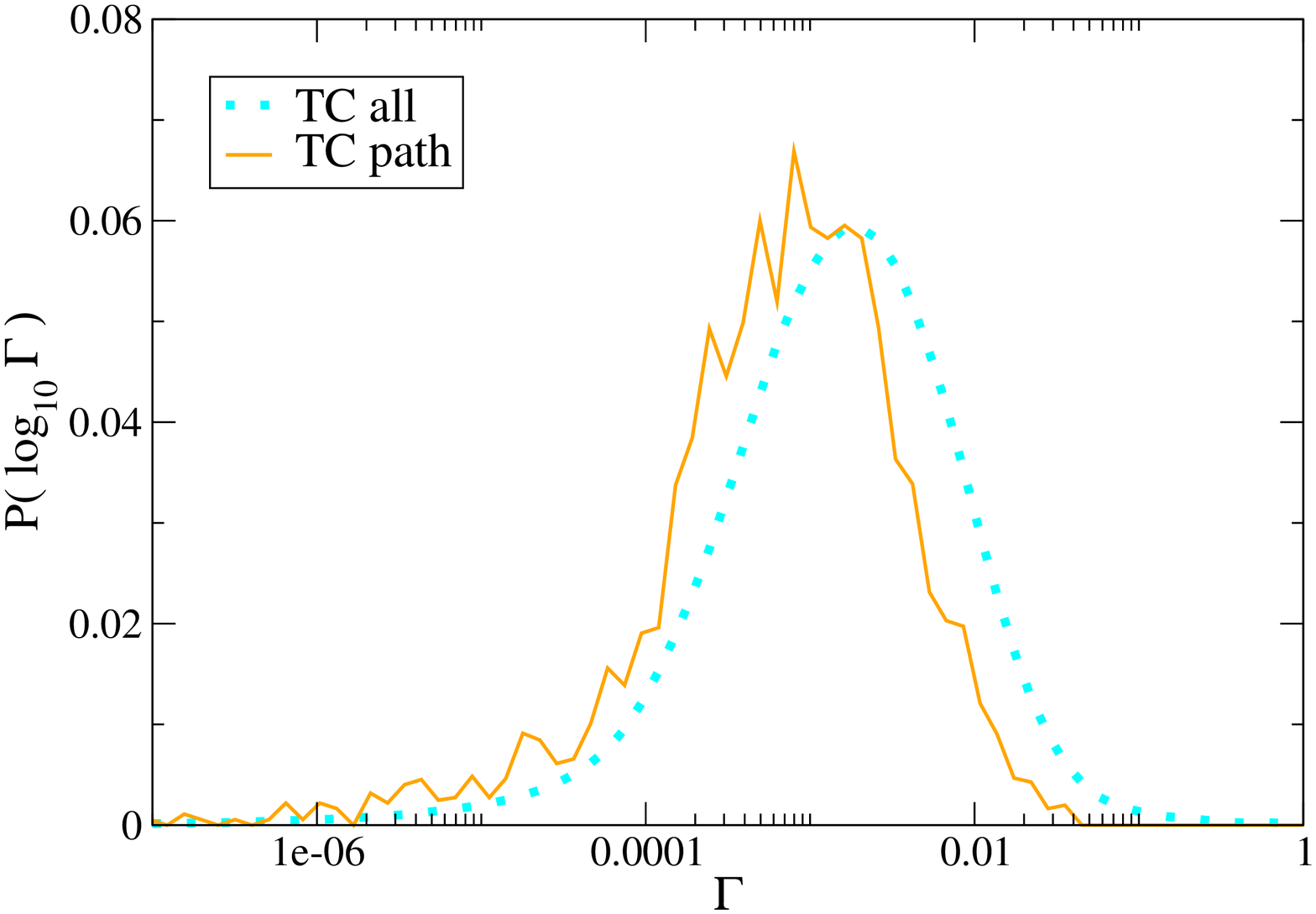}
(f) \includegraphics[width=0.2\columnwidth]{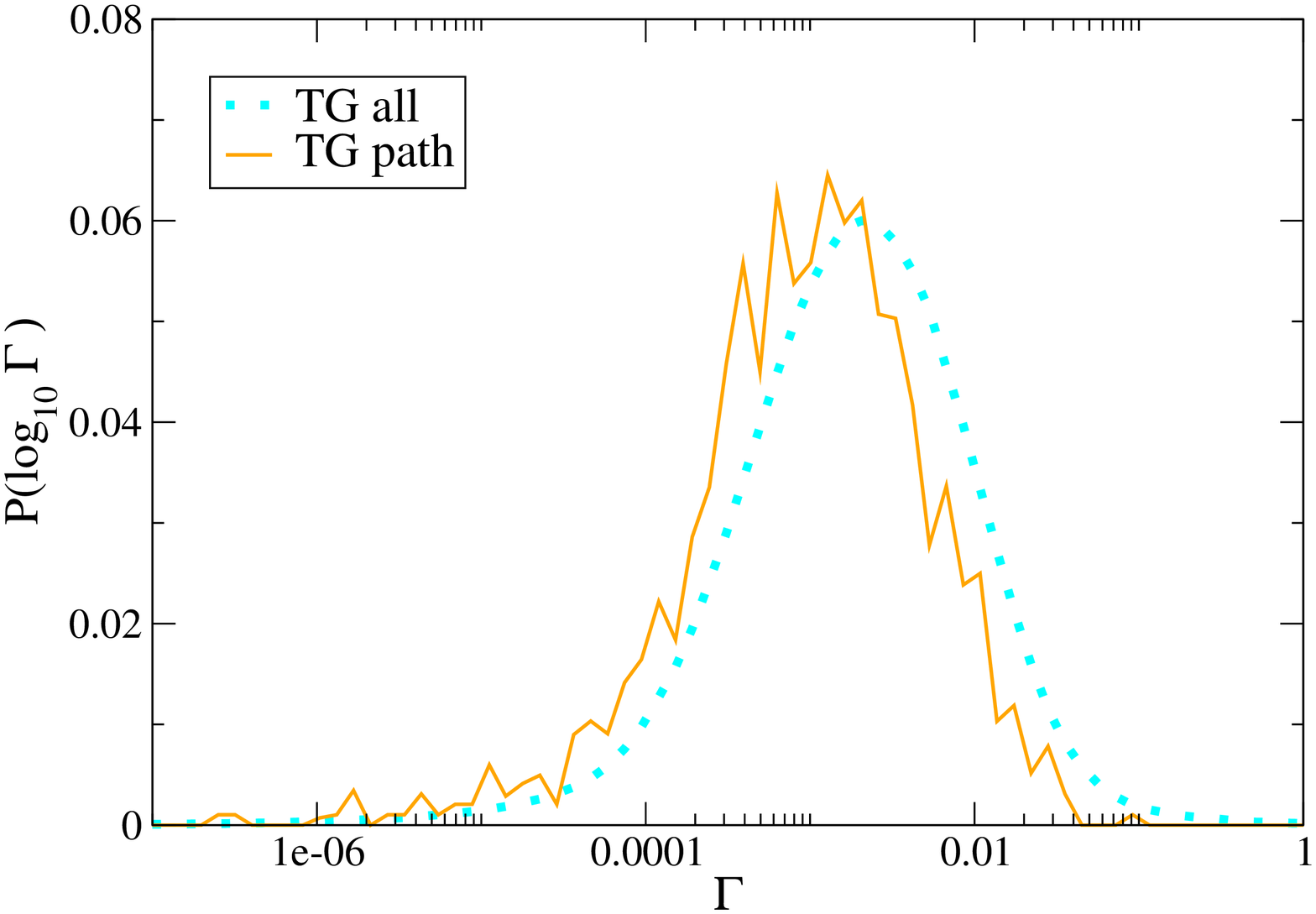}
(g) \includegraphics[width=0.2\columnwidth]{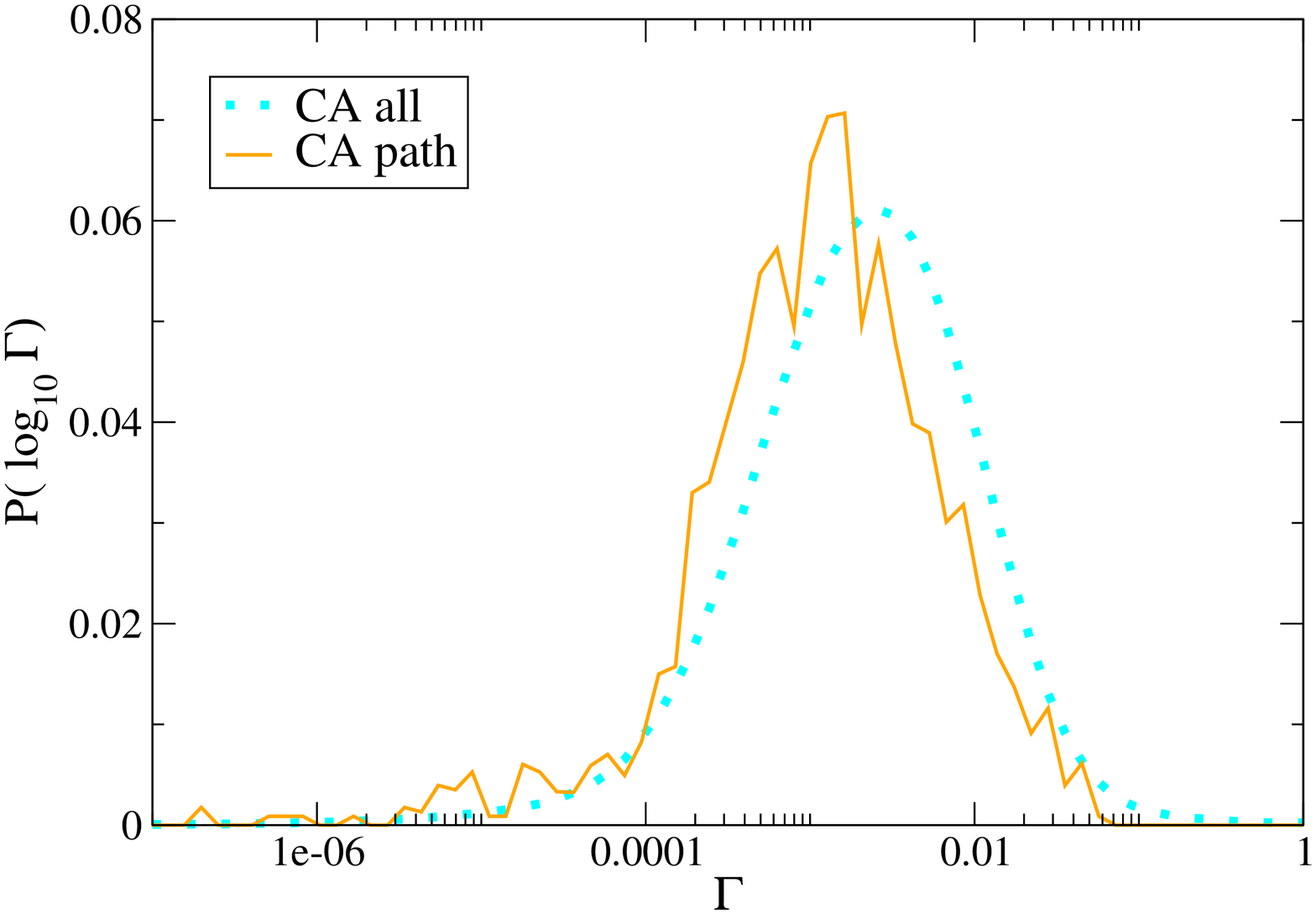}
(h) \includegraphics[width=0.2\columnwidth]{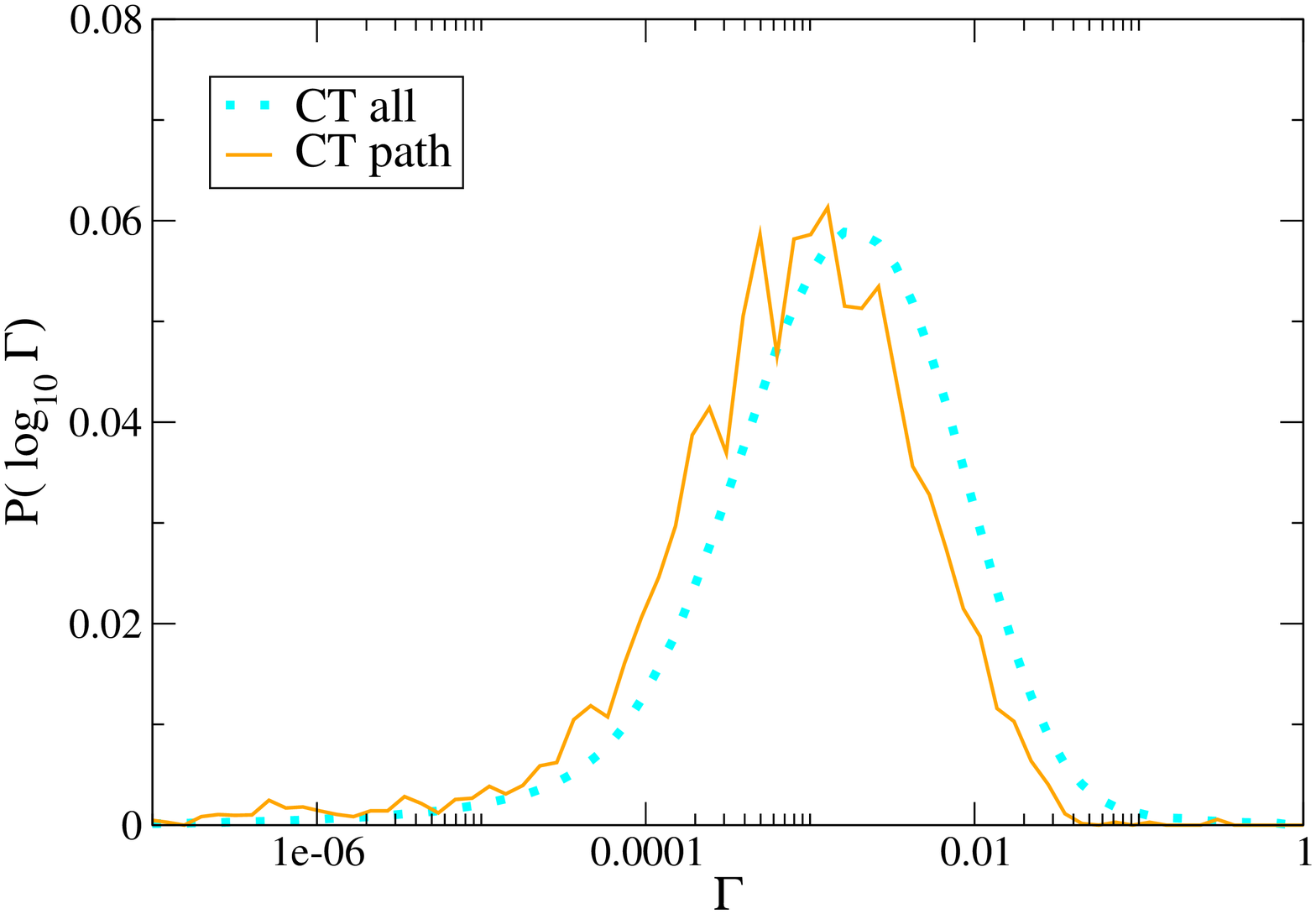}
\\
(i) \includegraphics[width=0.2\columnwidth]{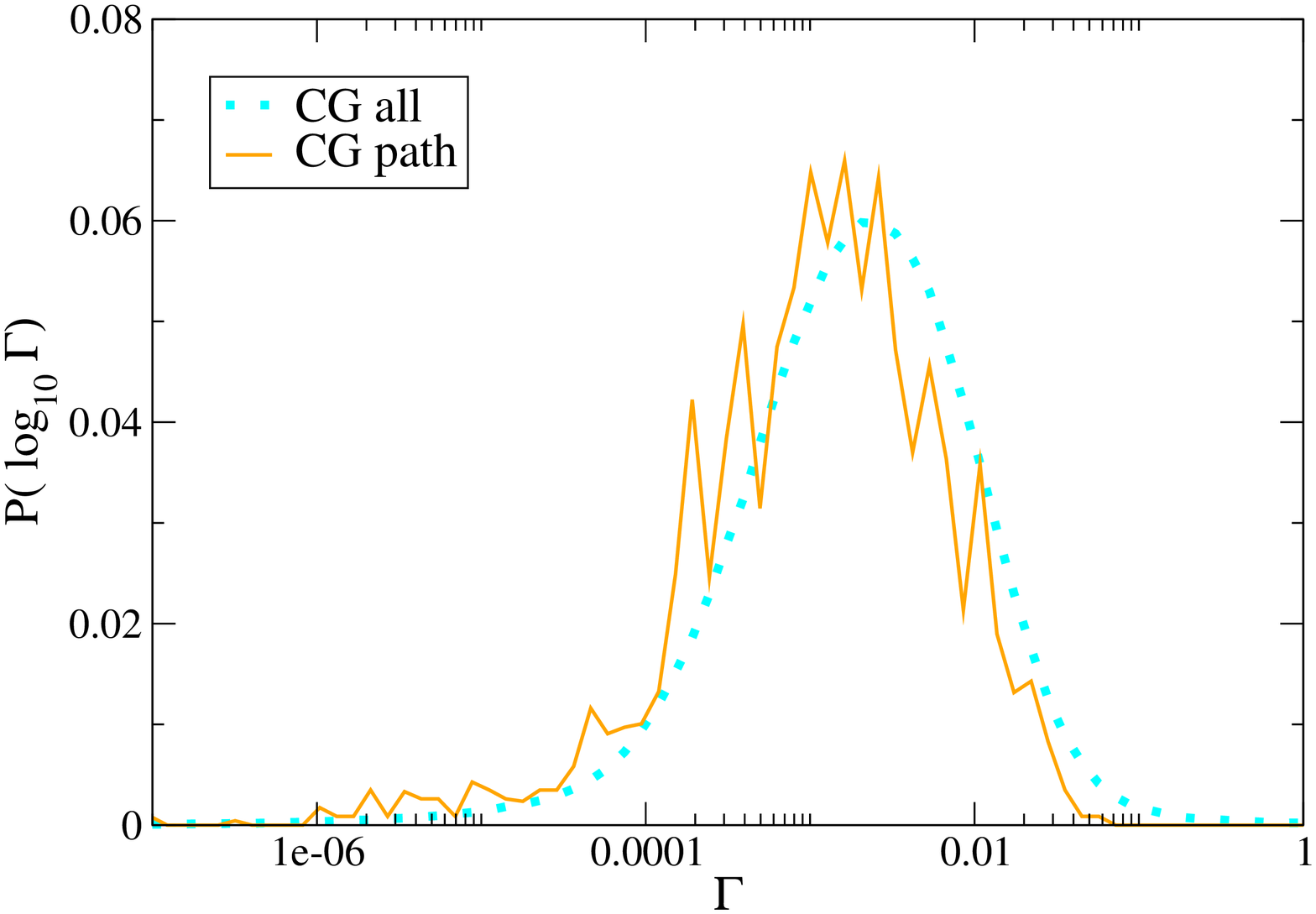}
(j) \includegraphics[width=0.2\columnwidth]{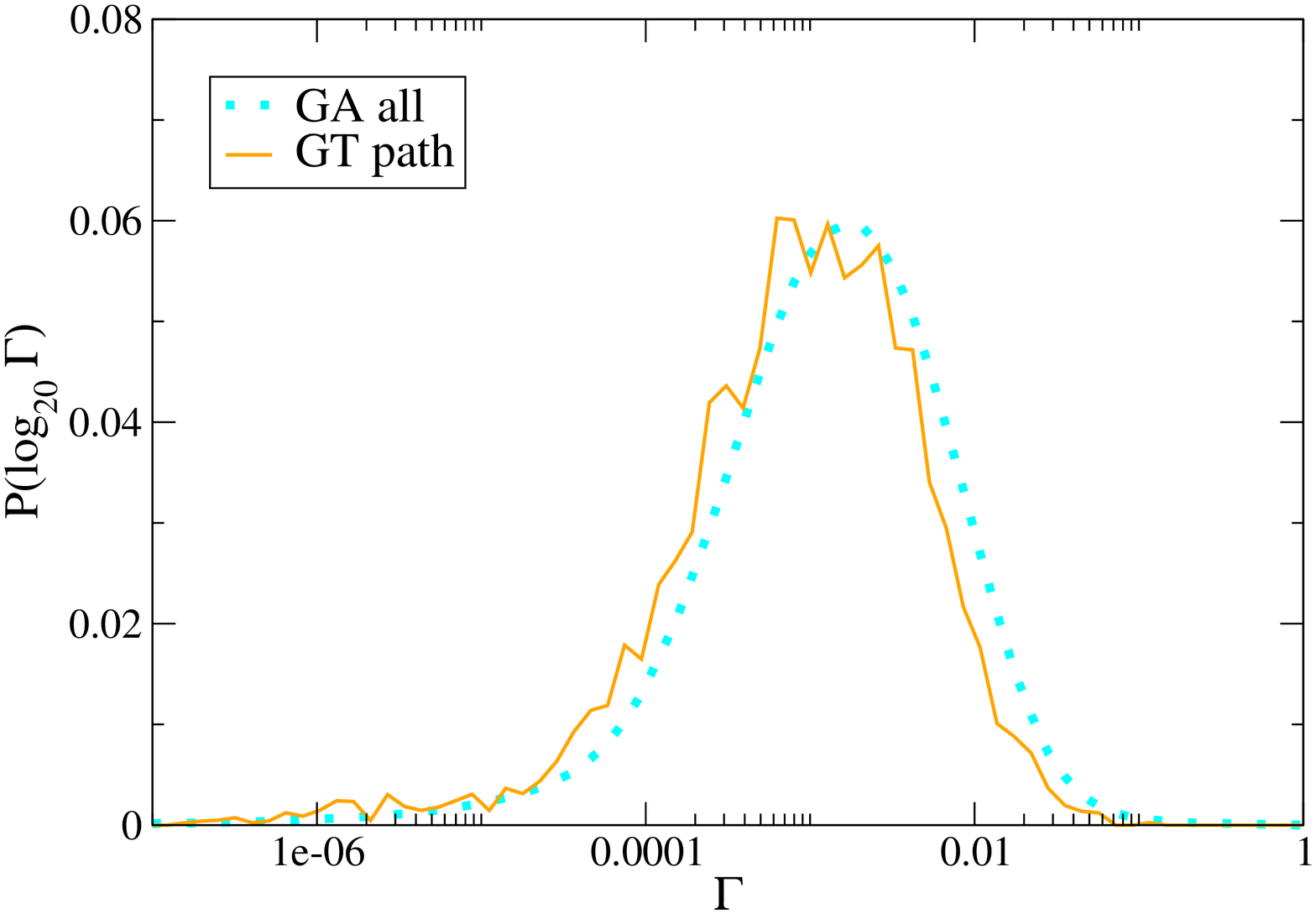}
(k) \includegraphics[width=0.2\columnwidth]{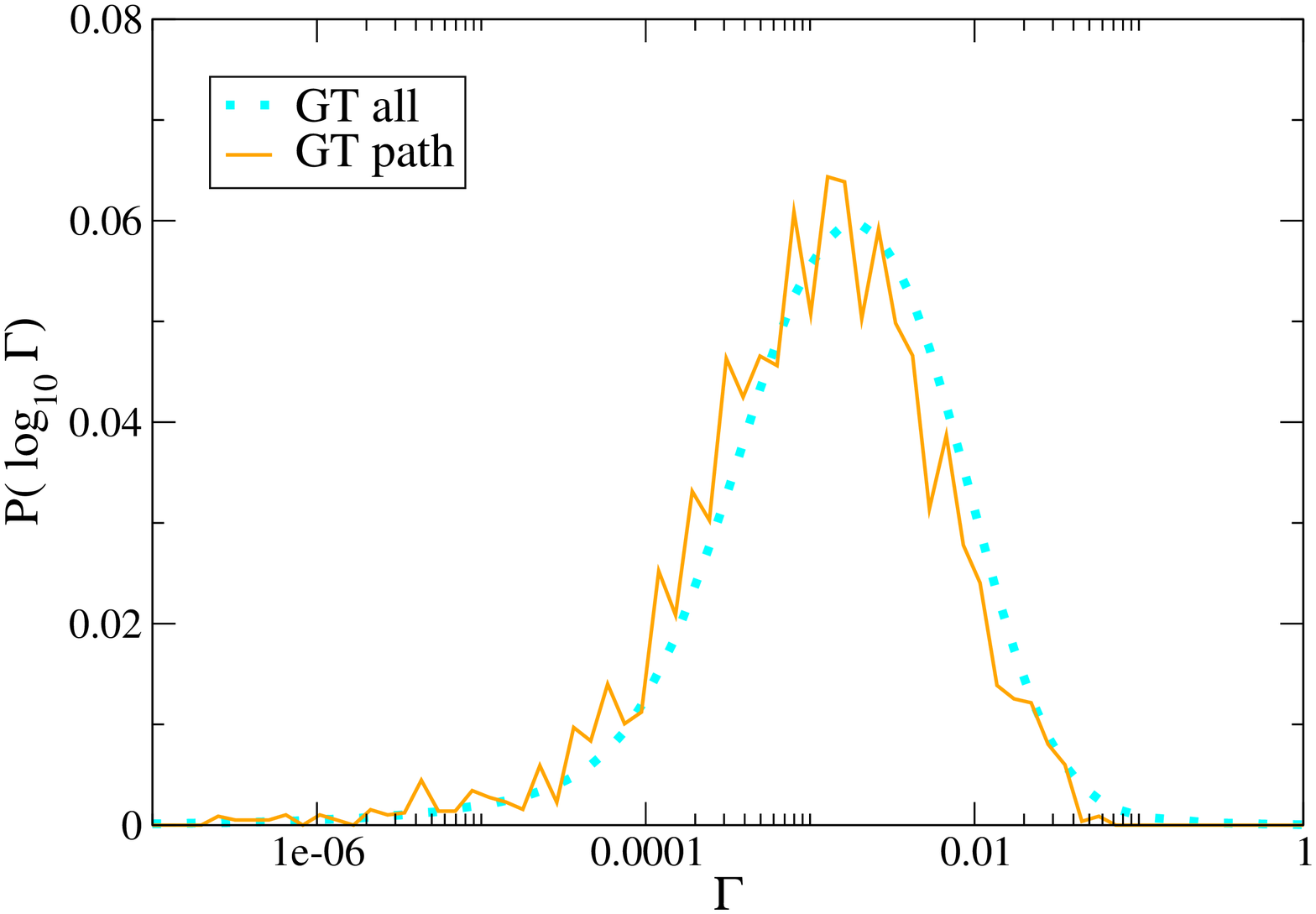}
(l) \includegraphics[width=0.2\columnwidth]{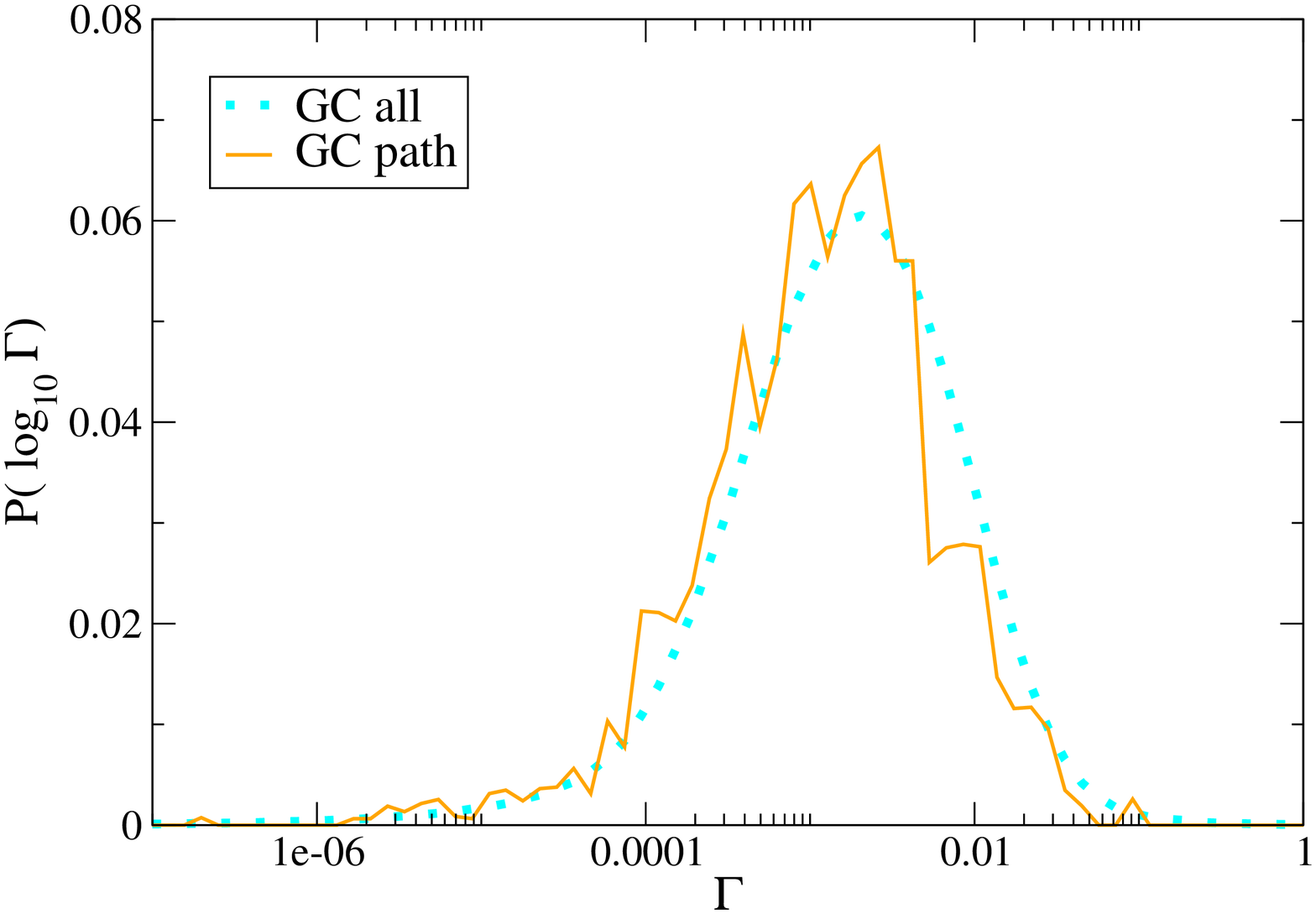}
\\
(m) \includegraphics[width=0.2\columnwidth]{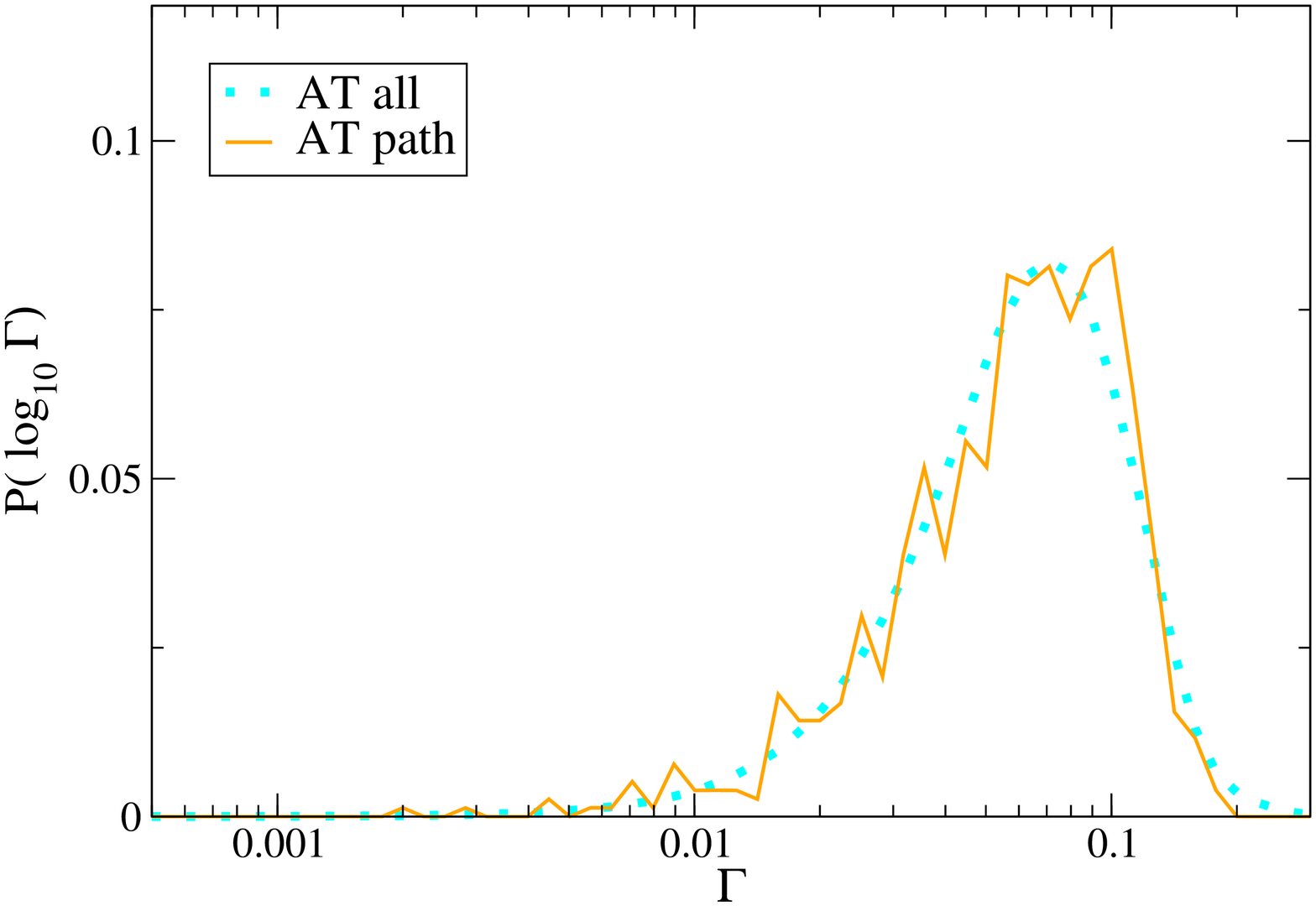}
(n) \includegraphics[width=0.2\columnwidth]{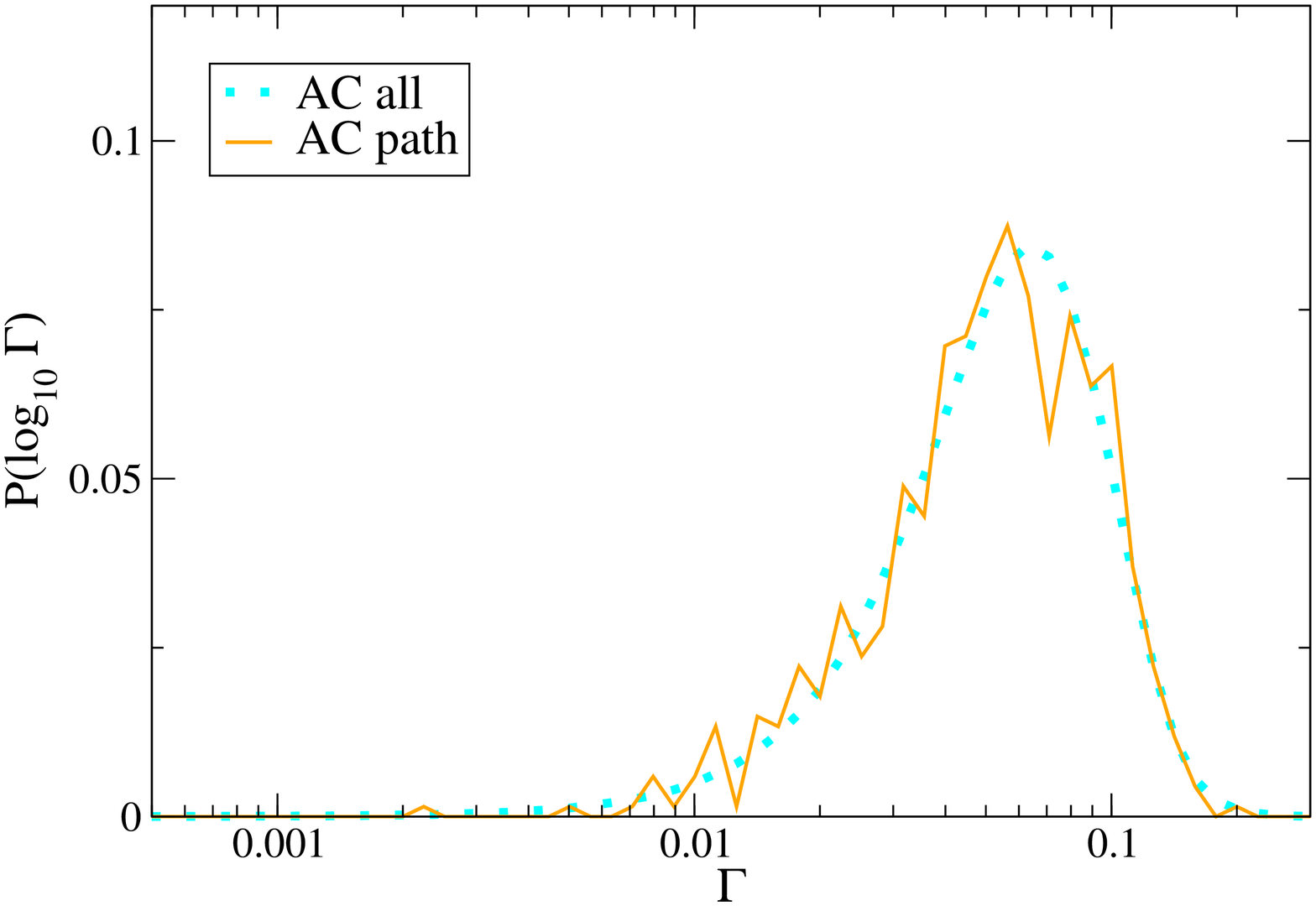}
(o) \includegraphics[width=0.2\columnwidth]{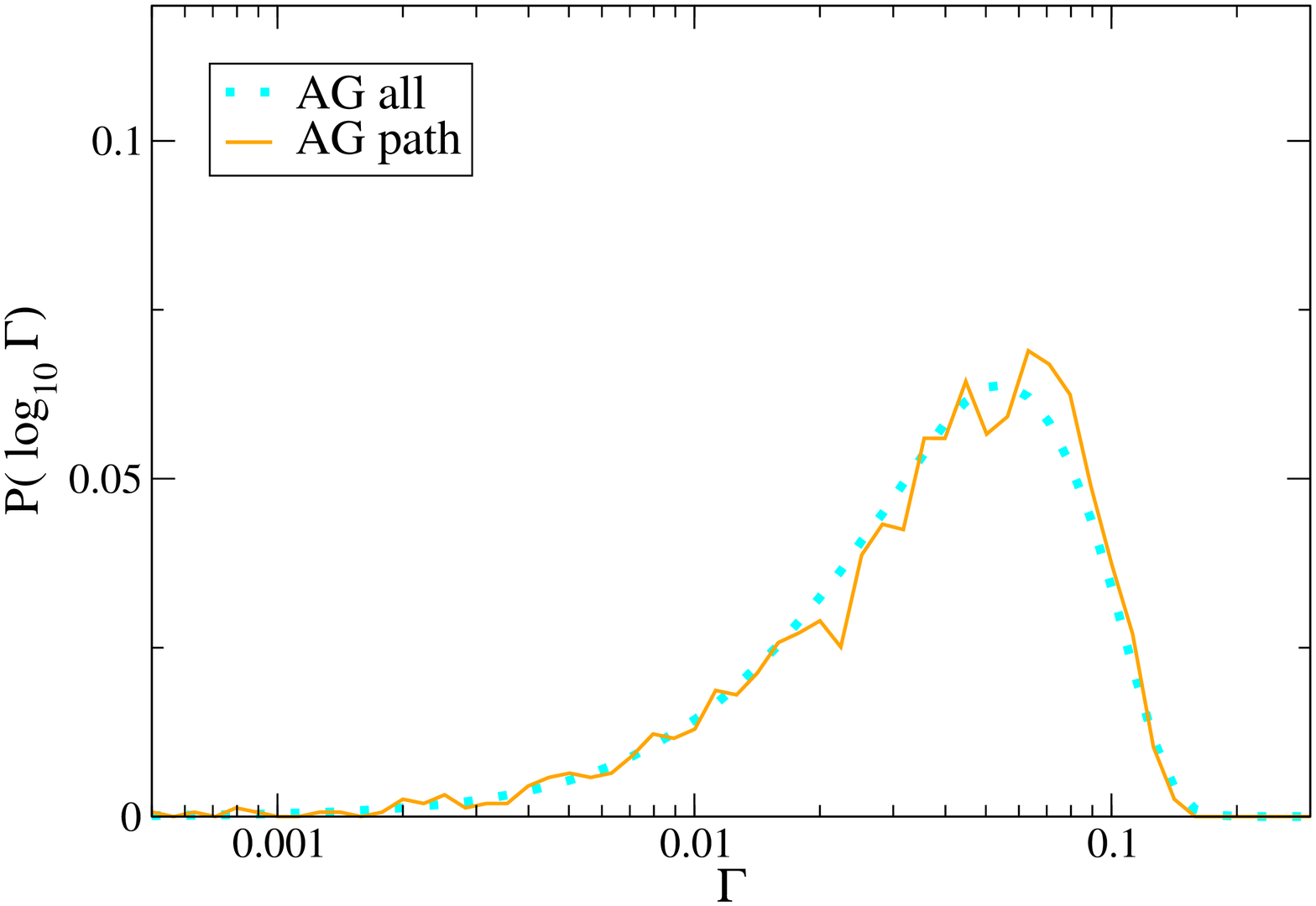}
(p) \includegraphics[width=0.2\columnwidth]{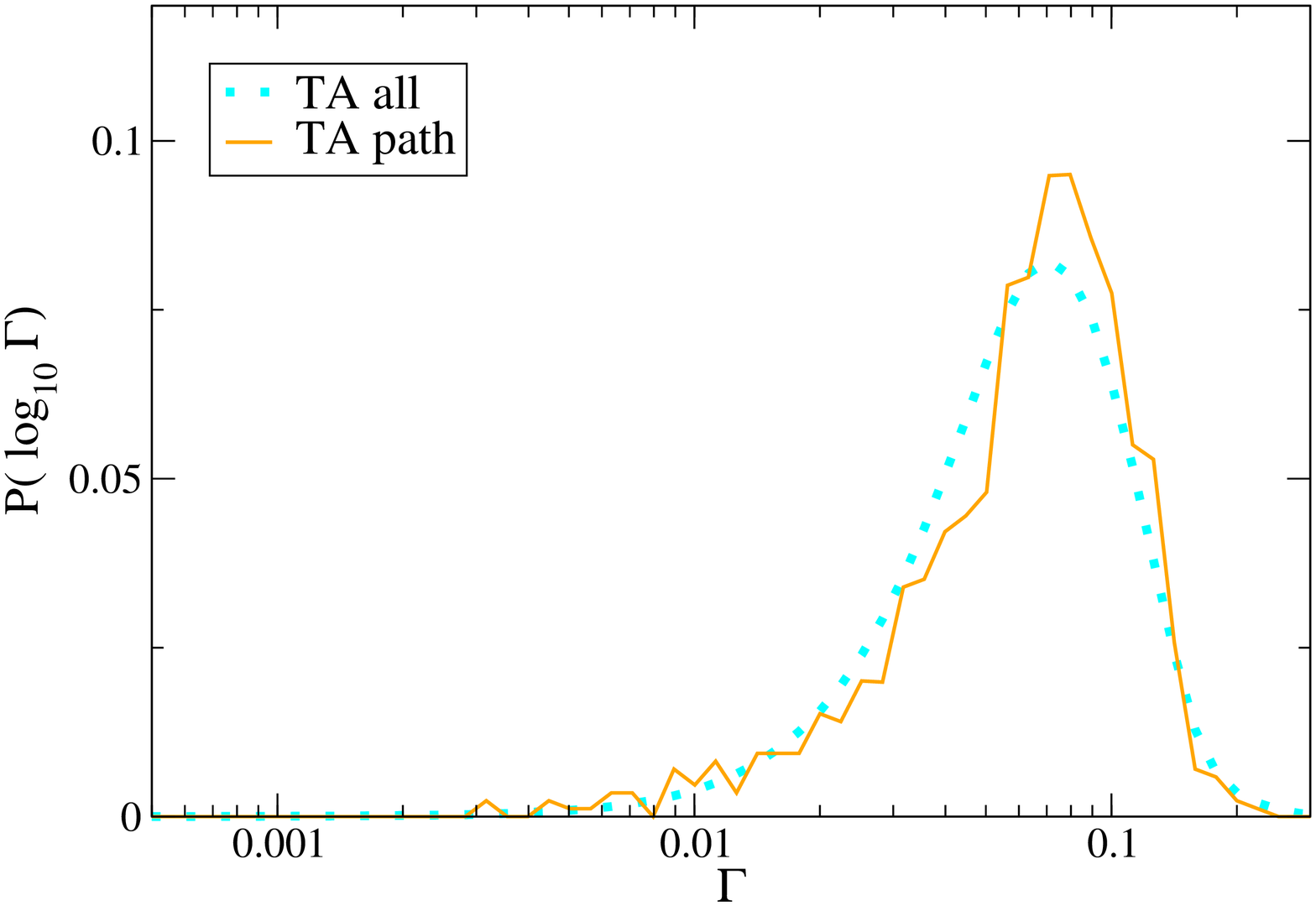}
\\
(q) \includegraphics[width=0.2\columnwidth]{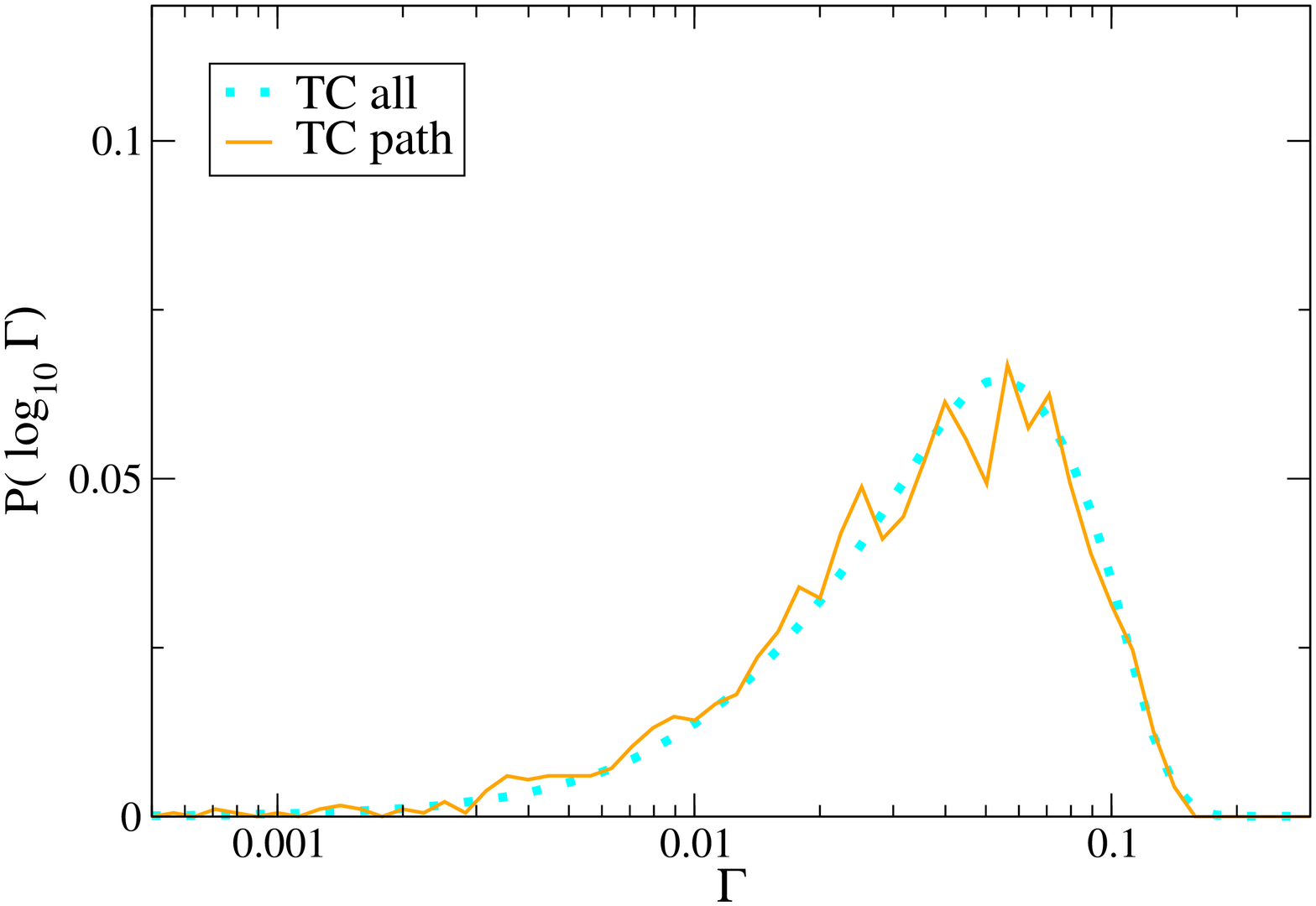}
(r) \includegraphics[width=0.2\columnwidth]{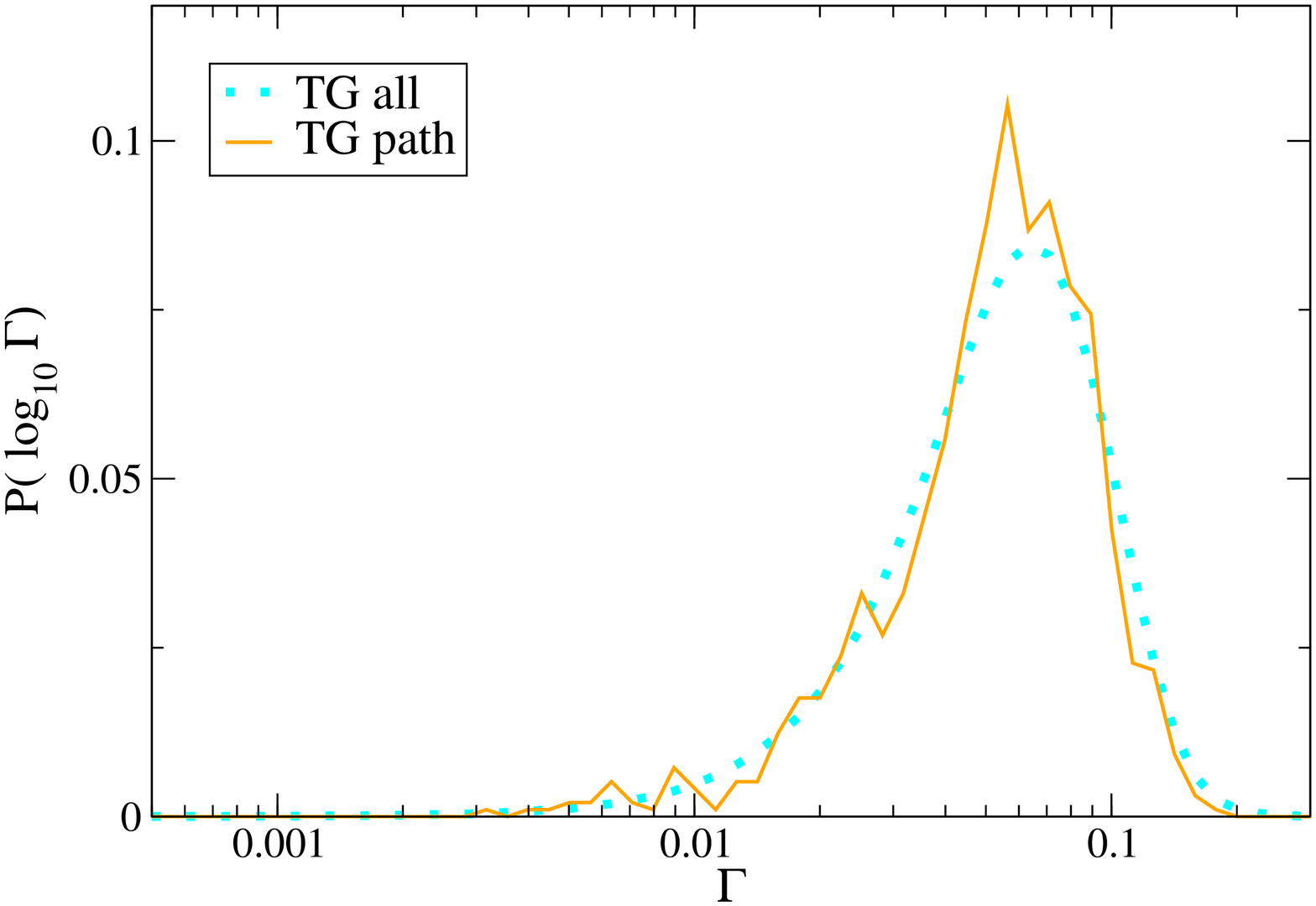}
(s) \includegraphics[width=0.2\columnwidth]{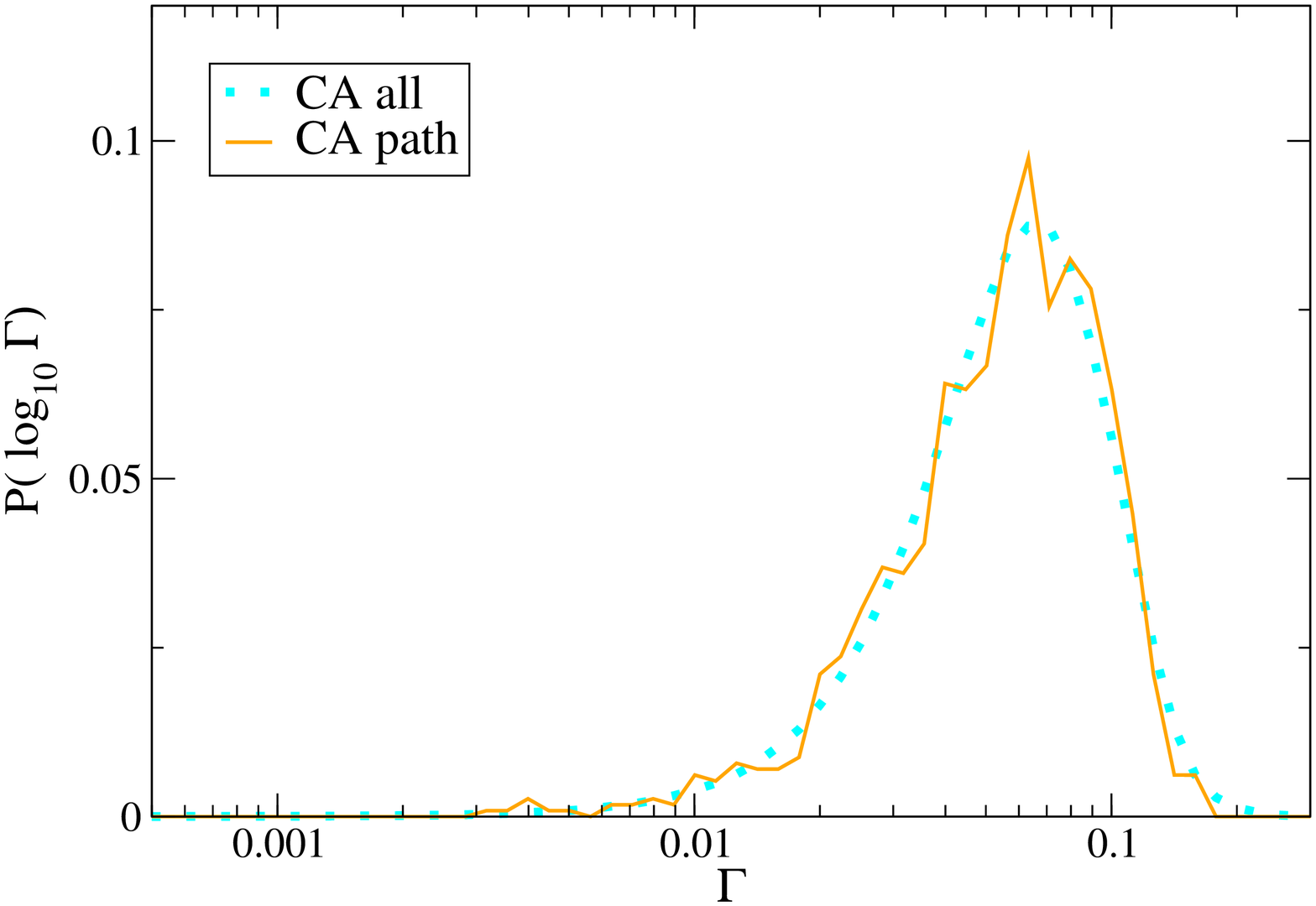}
(t) \includegraphics[width=0.2\columnwidth]{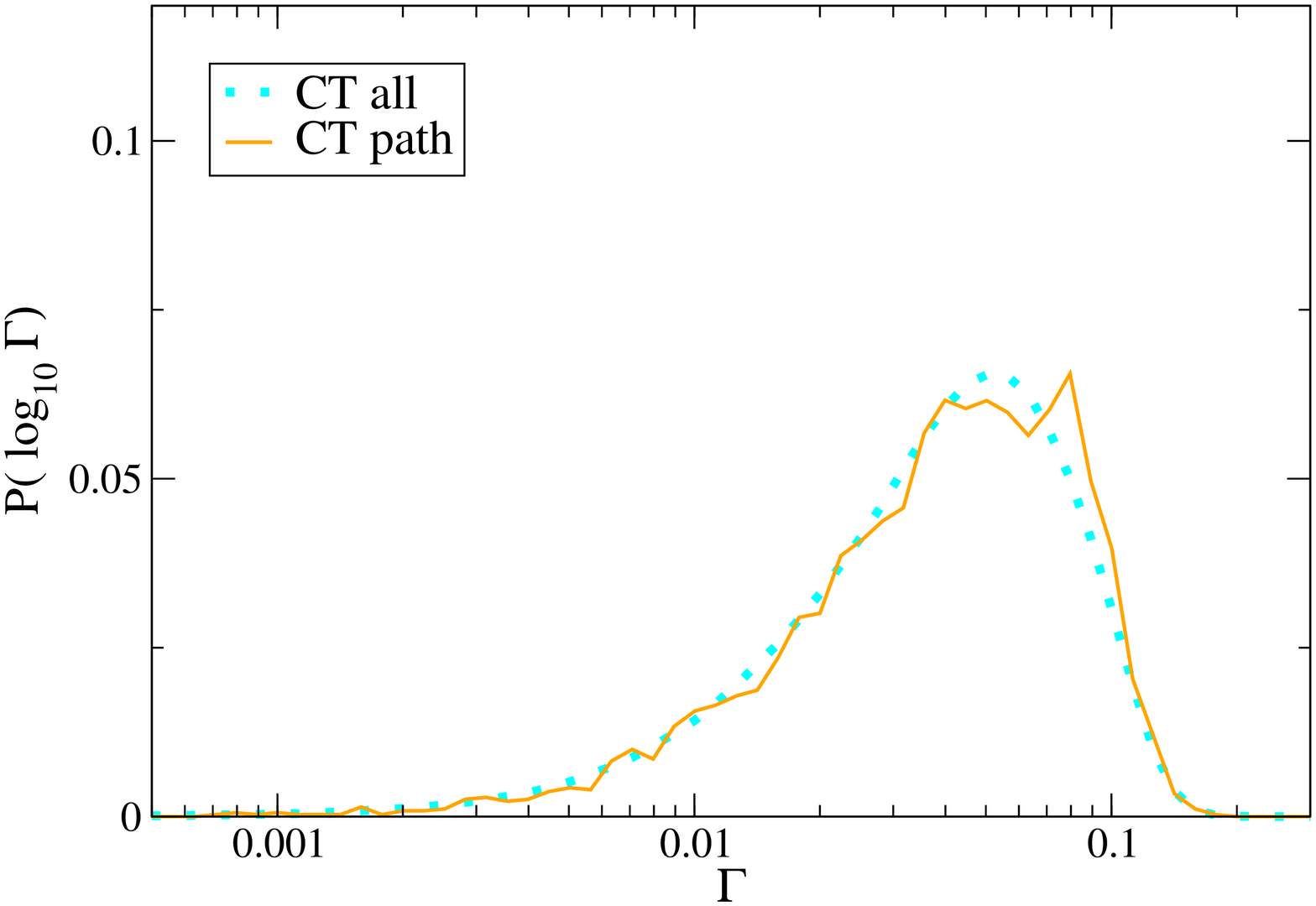}
\\
(u) \includegraphics[width=0.2\columnwidth]{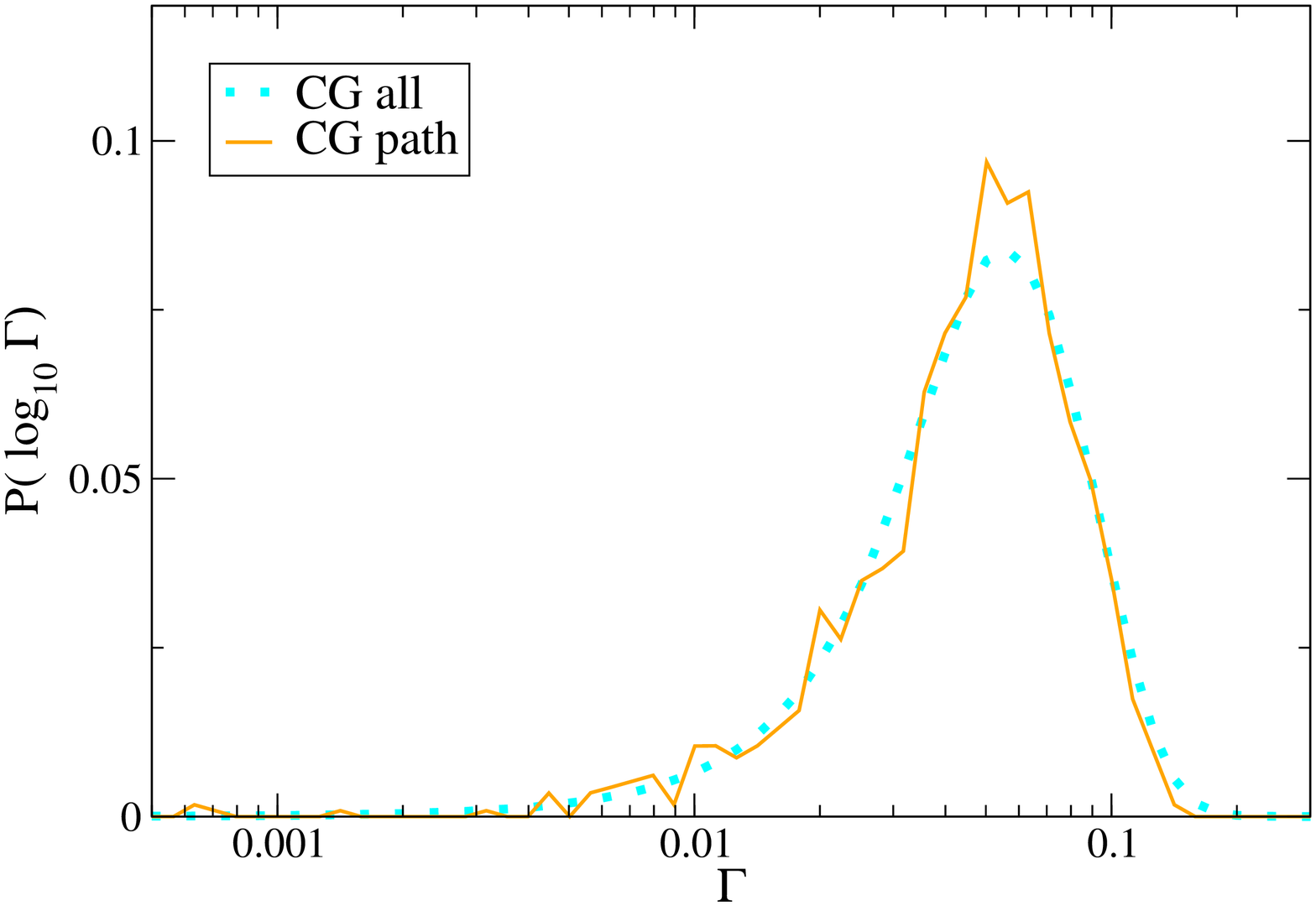}
(v) \includegraphics[width=0.2\columnwidth]{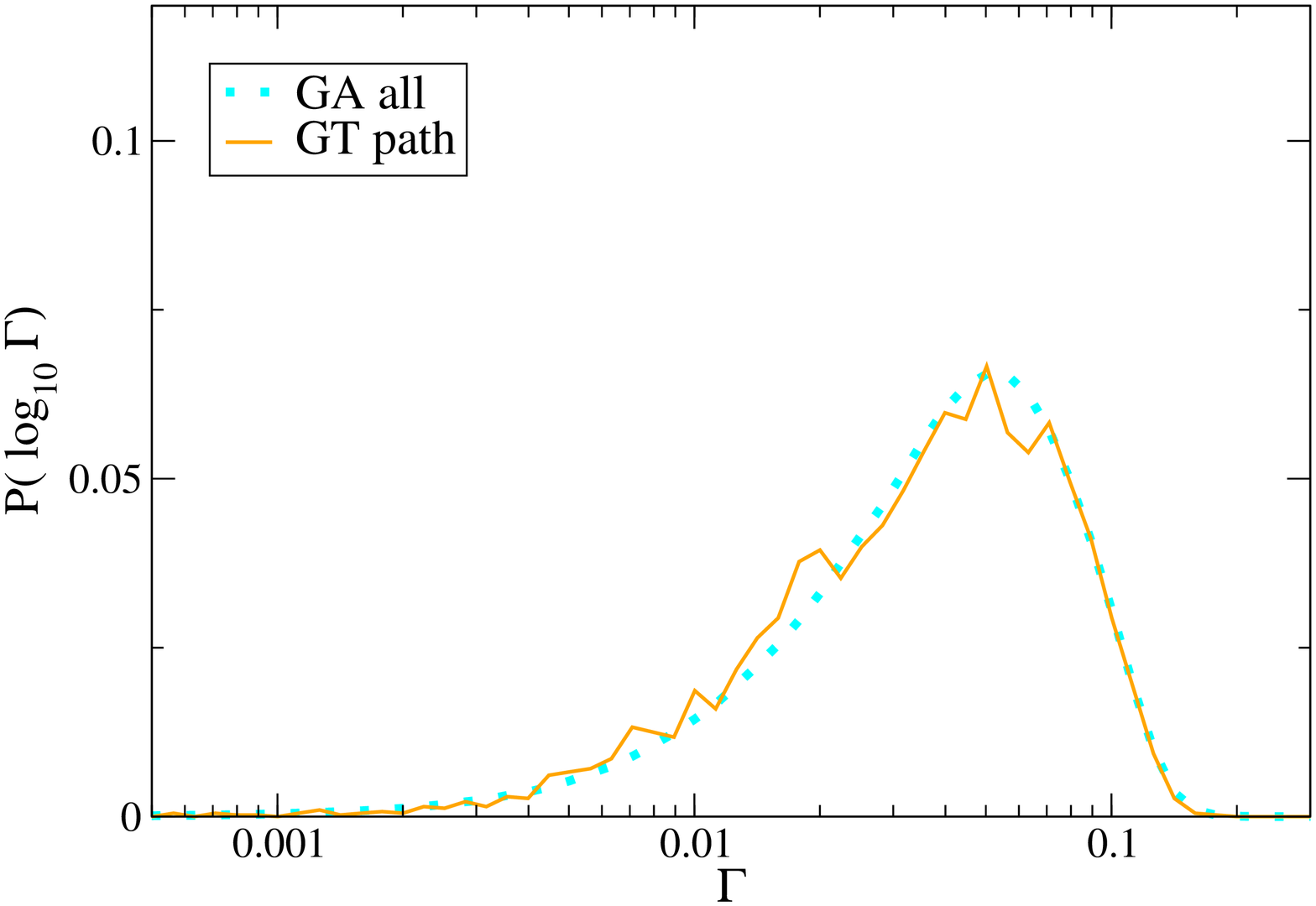}
(w) \includegraphics[width=0.2\columnwidth]{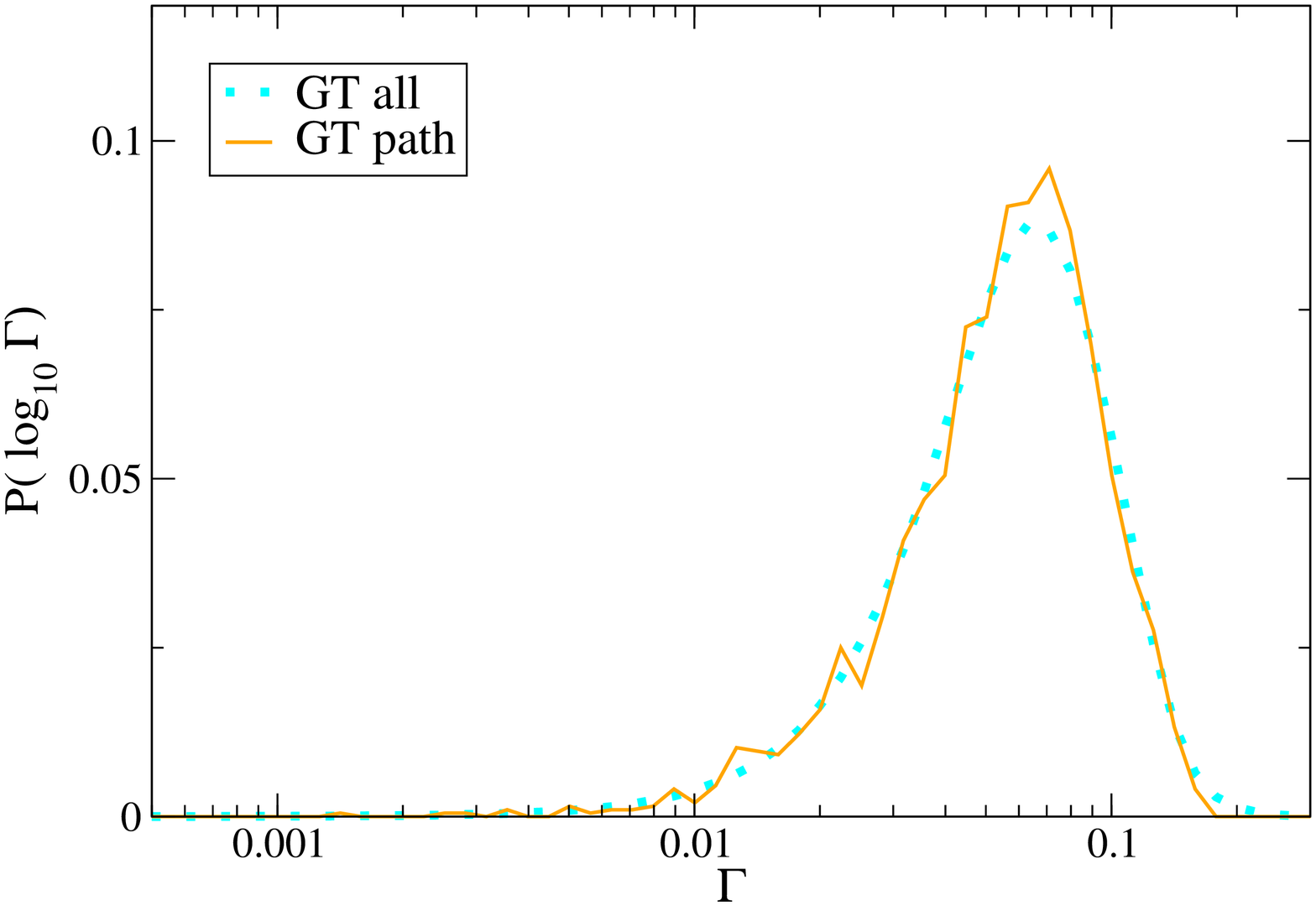}
(x) \includegraphics[width=0.2\columnwidth]{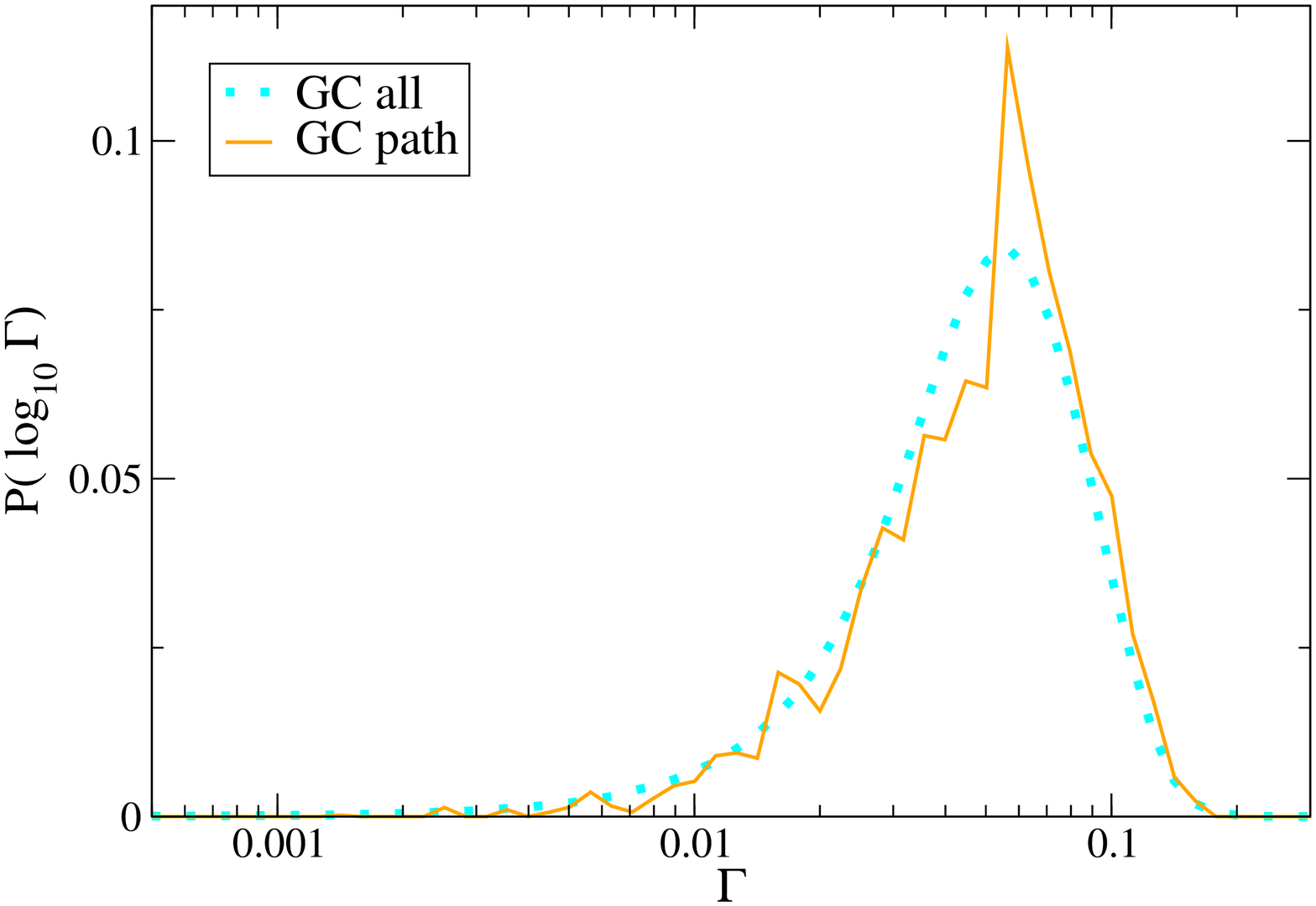}

\caption{}
\end{figure}
\clearpage
\begin{figure}
\centering
(a)\includegraphics[width=0.45\columnwidth]{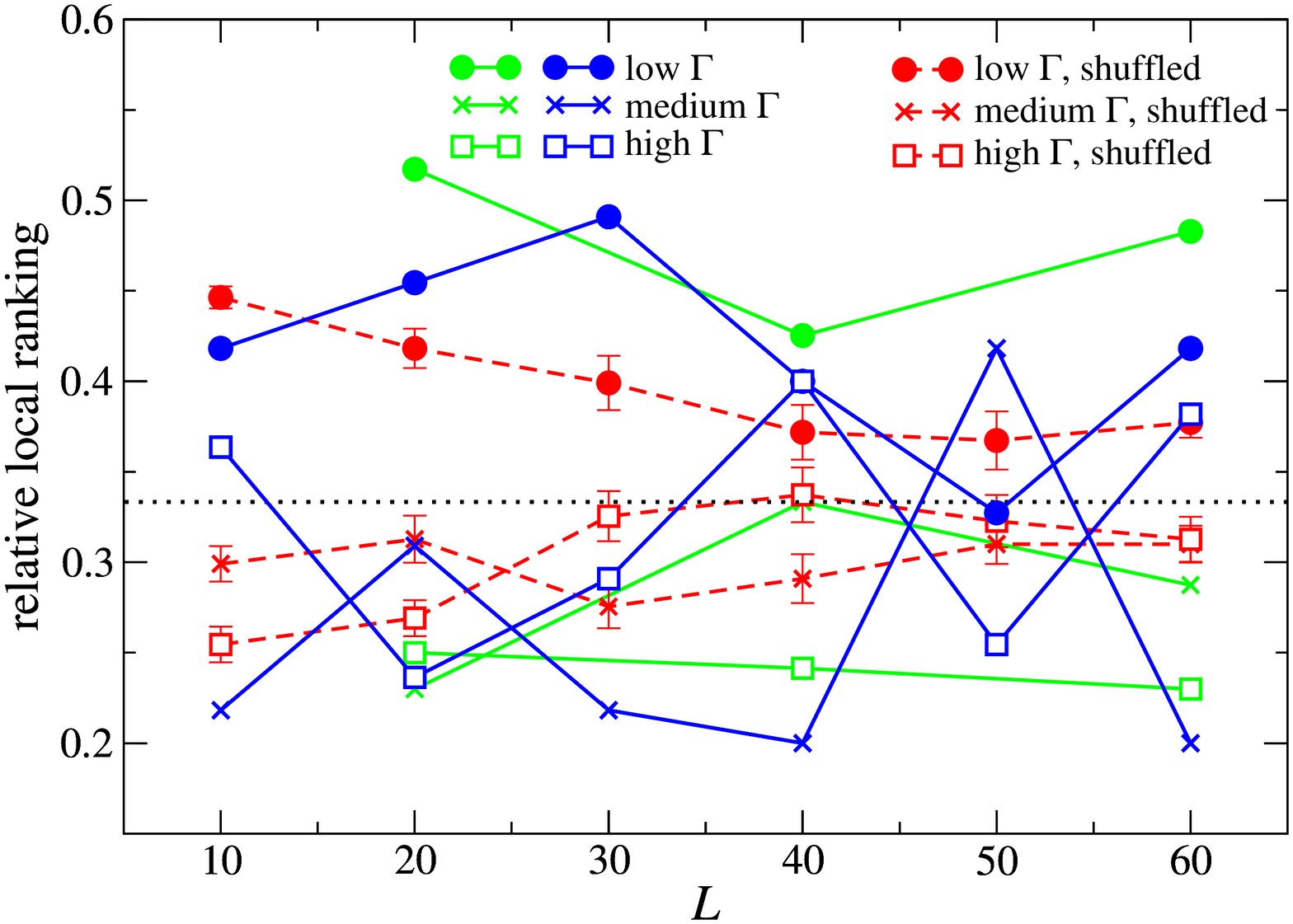}
(b)\includegraphics[width=0.45\columnwidth]{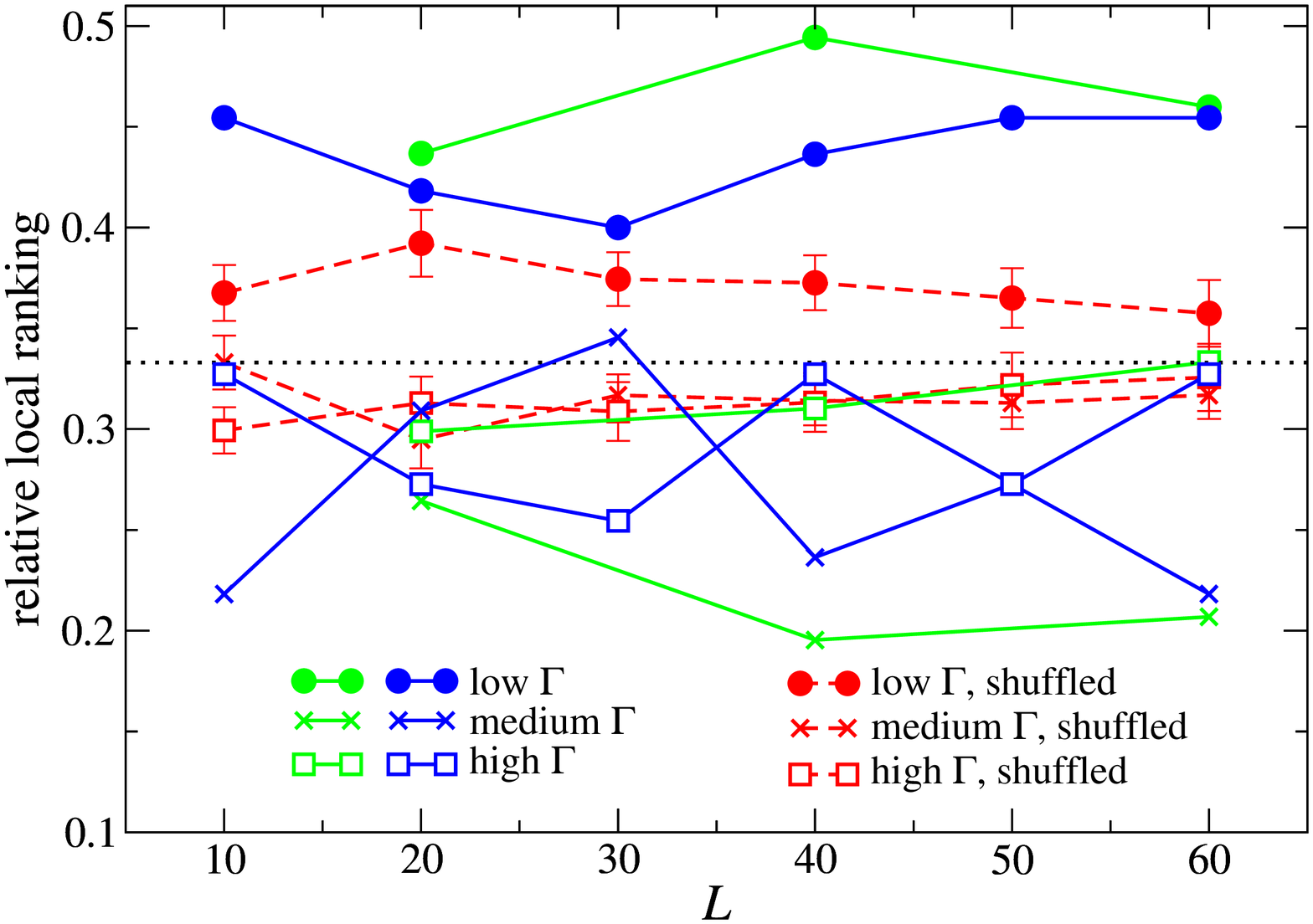}
(c)\includegraphics[width=0.45\columnwidth]{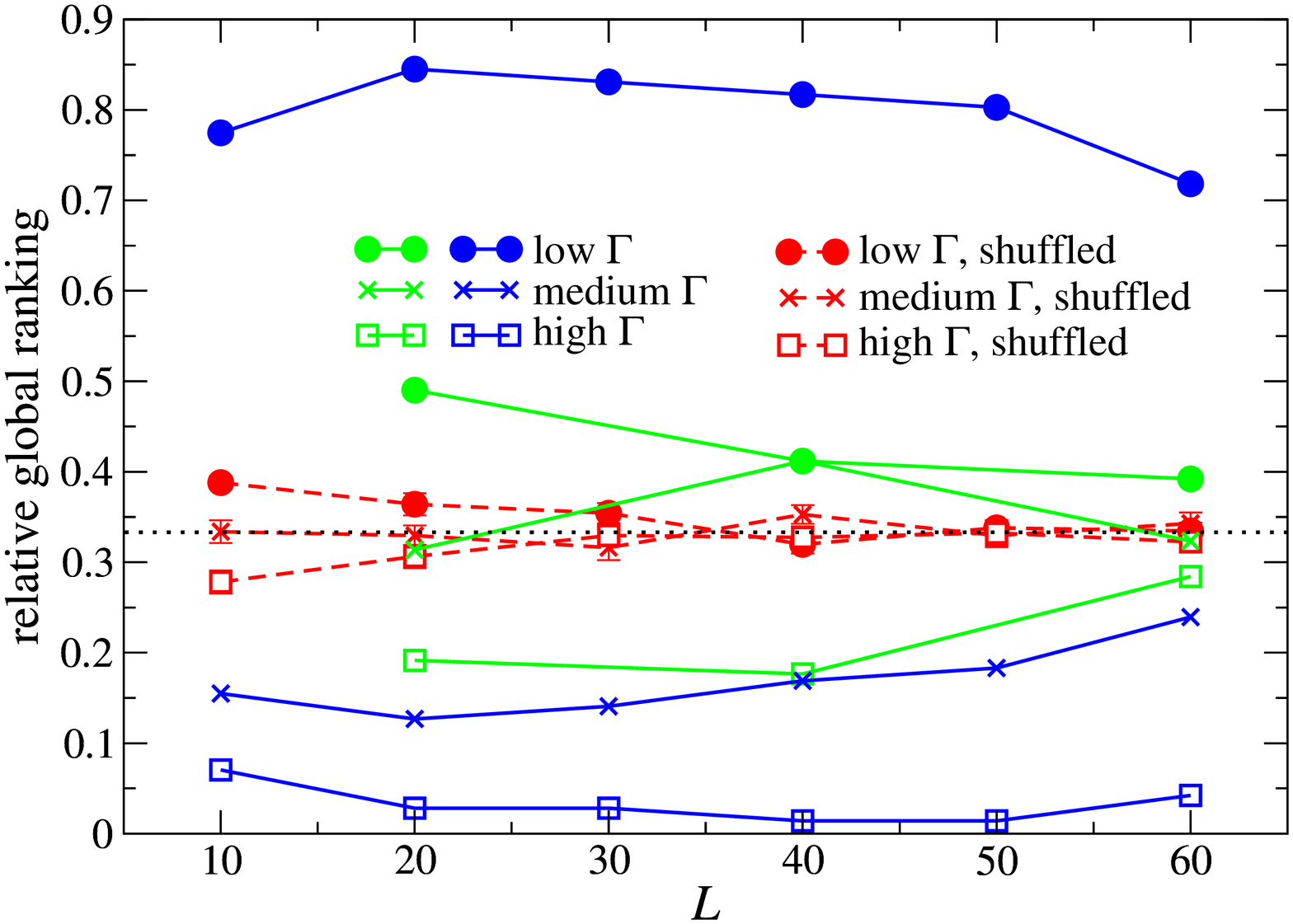}
(d)\includegraphics[width=0.45\columnwidth]{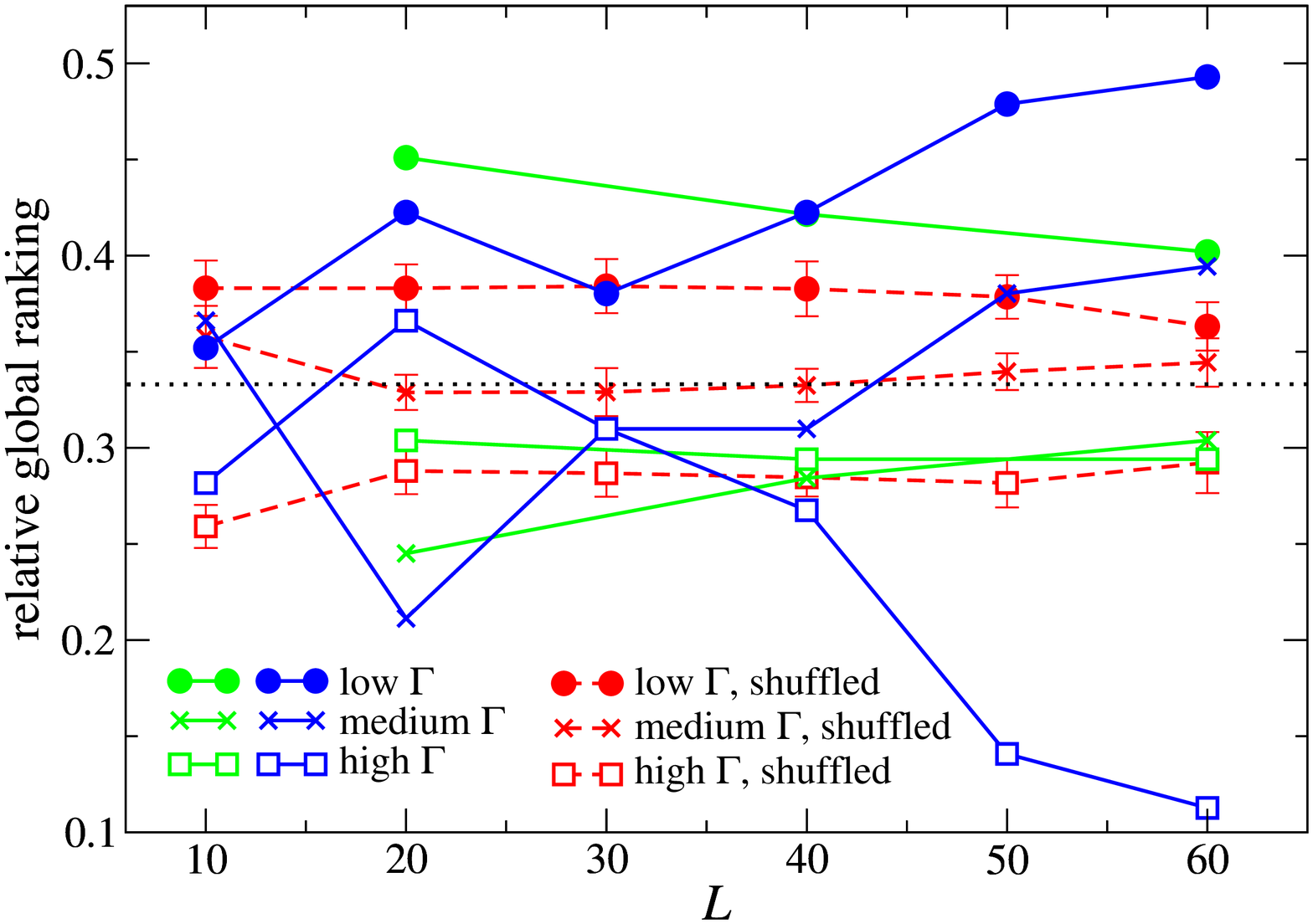}
\caption{}
\end{figure}
\clearpage
\begin{figure}
\centering
(a)\includegraphics[width=0.45\columnwidth]{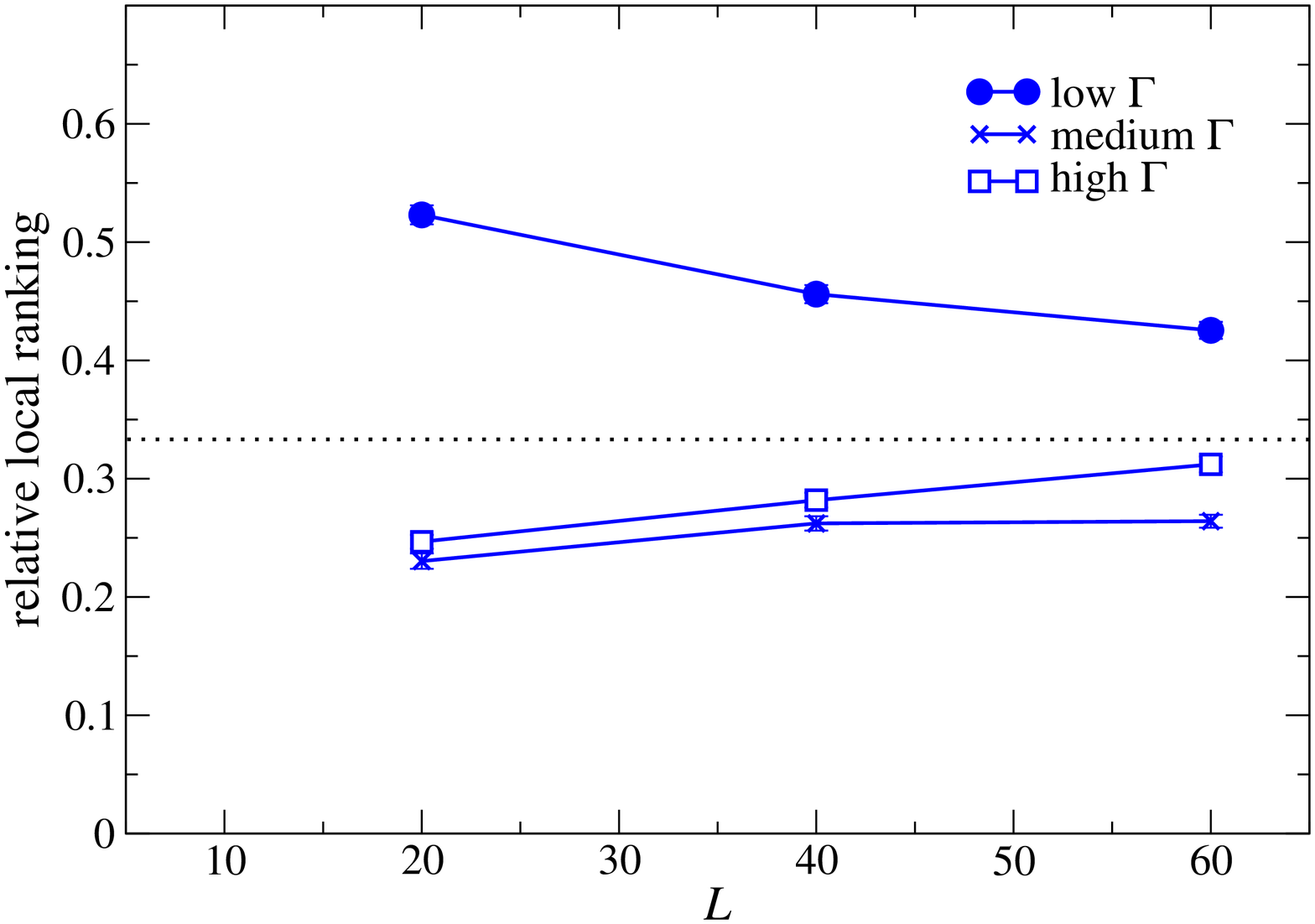}
(b)\includegraphics[width=0.45\columnwidth]{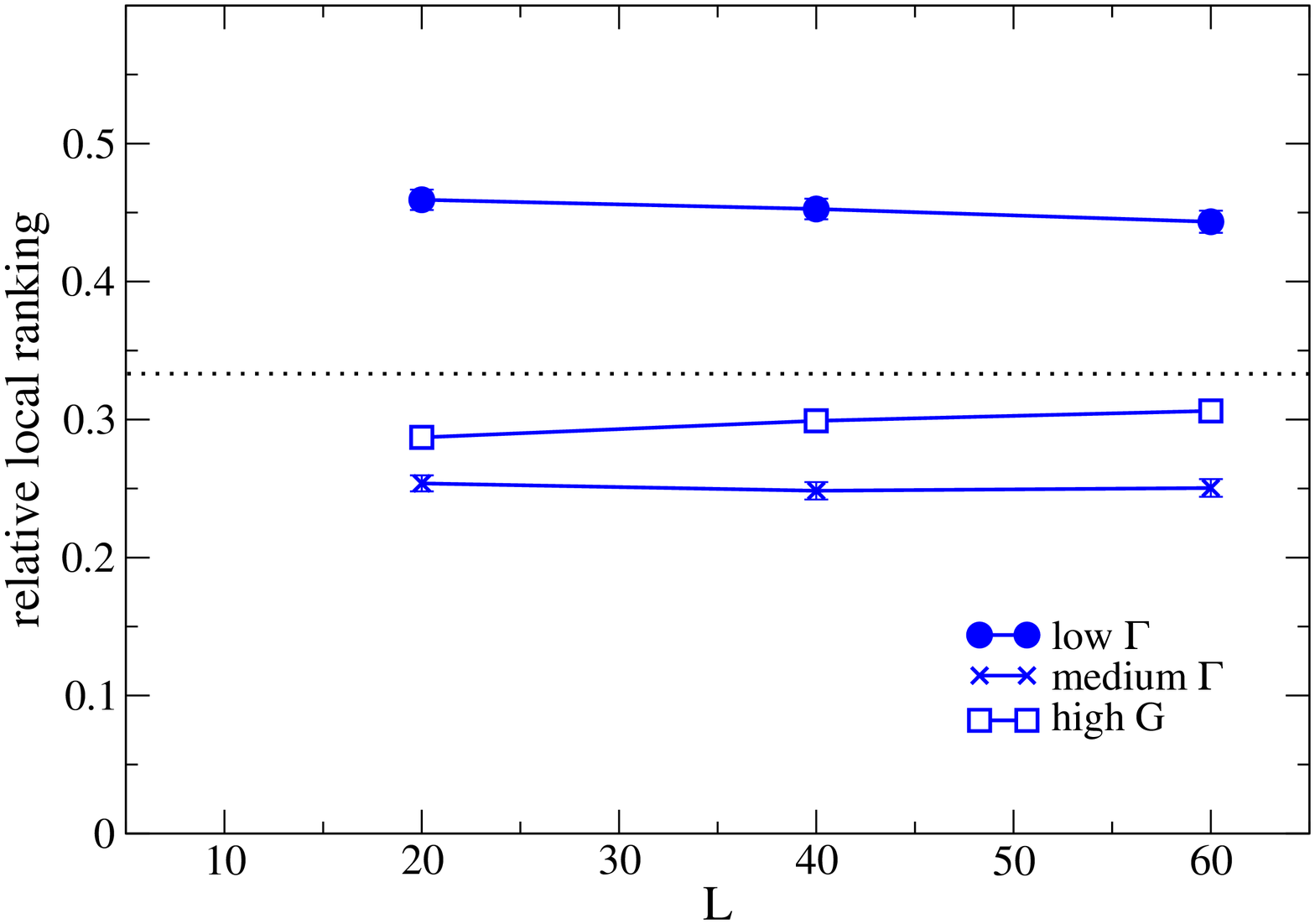}
(c)\includegraphics[width=0.45\columnwidth]{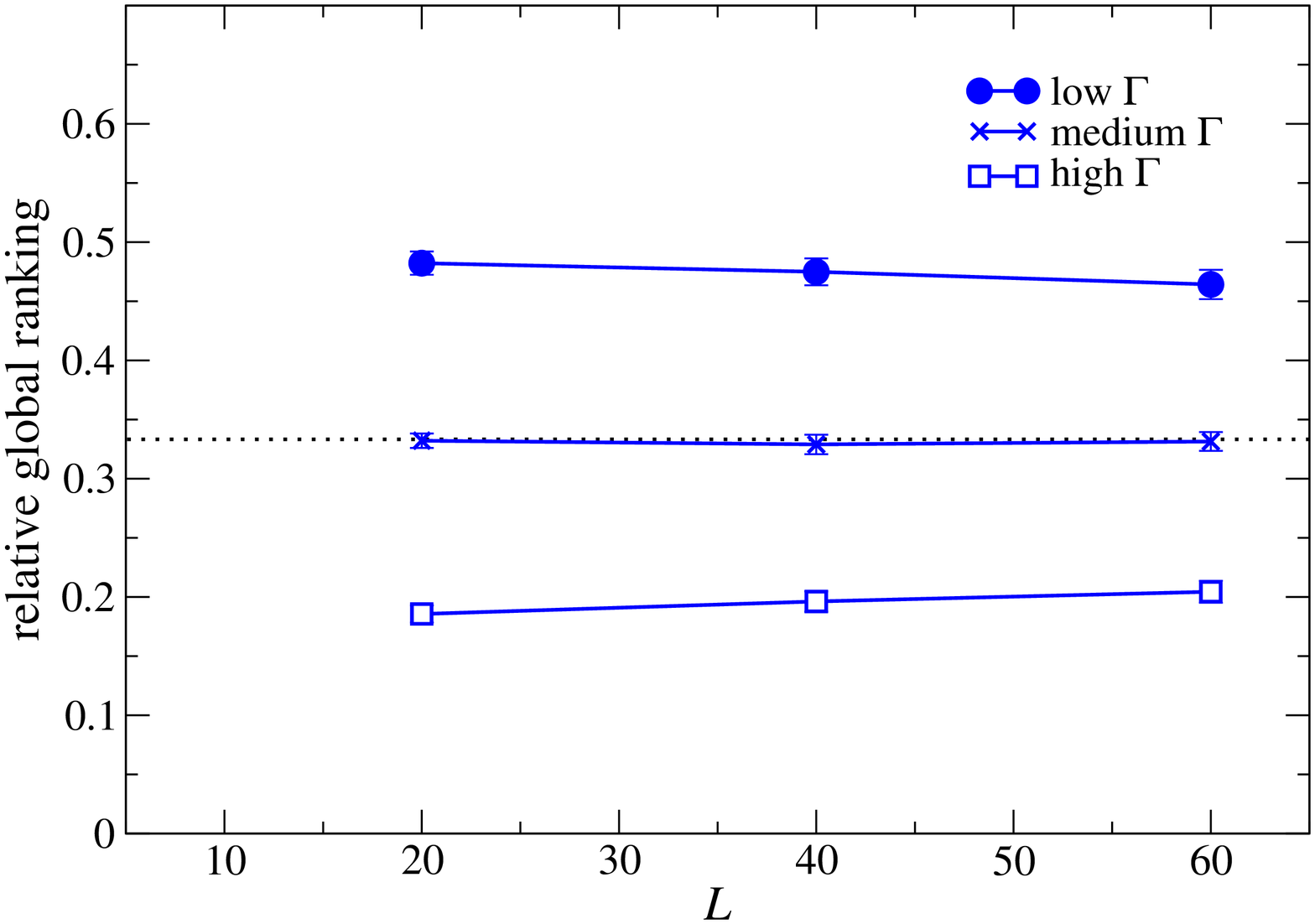}
(d)\includegraphics[width=0.45\columnwidth]{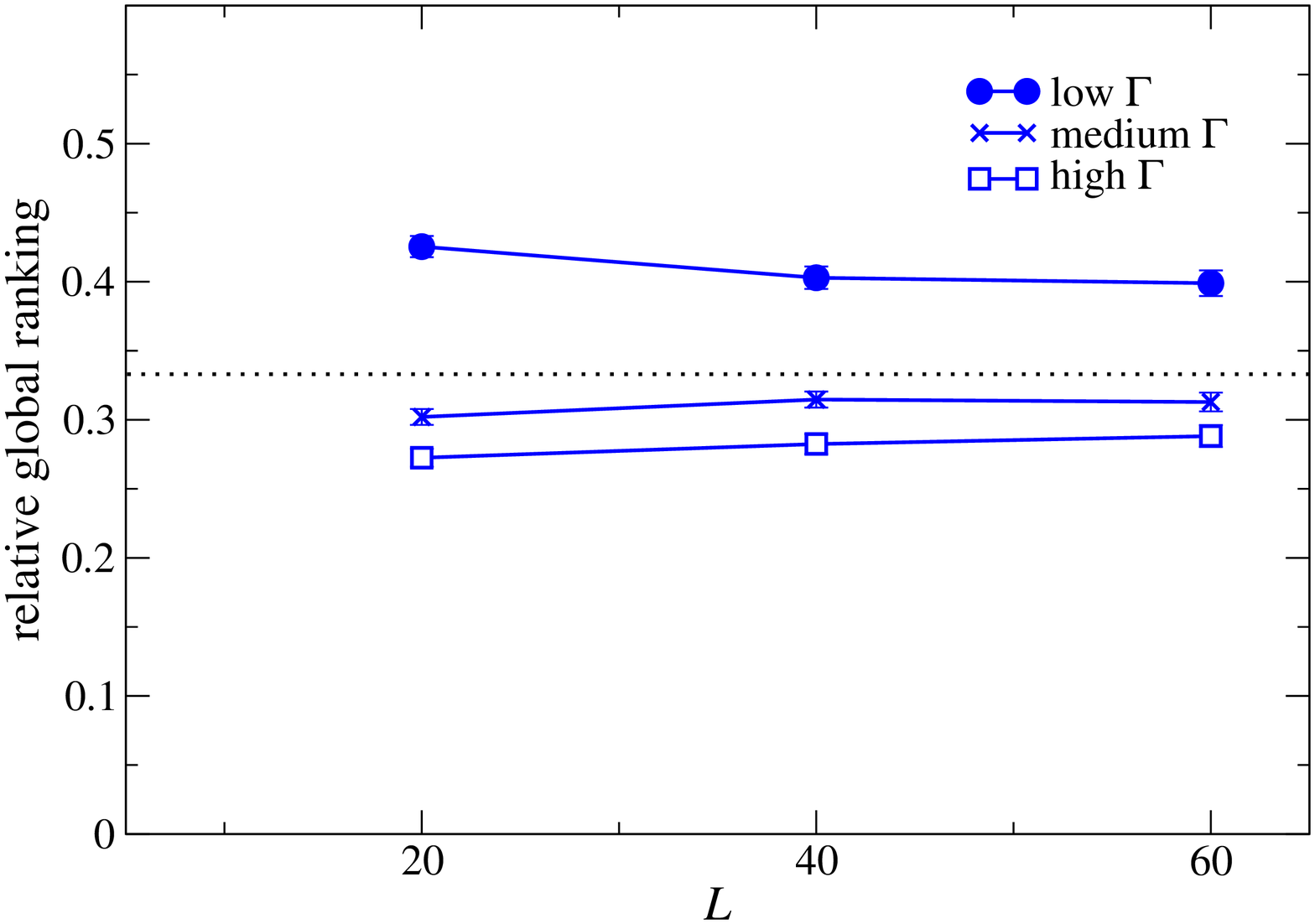}
\caption{}
\end{figure}
\clearpage
\begin{figure}
\centering
\includegraphics[width=0.99\columnwidth,angle=-270,scale=0.41]{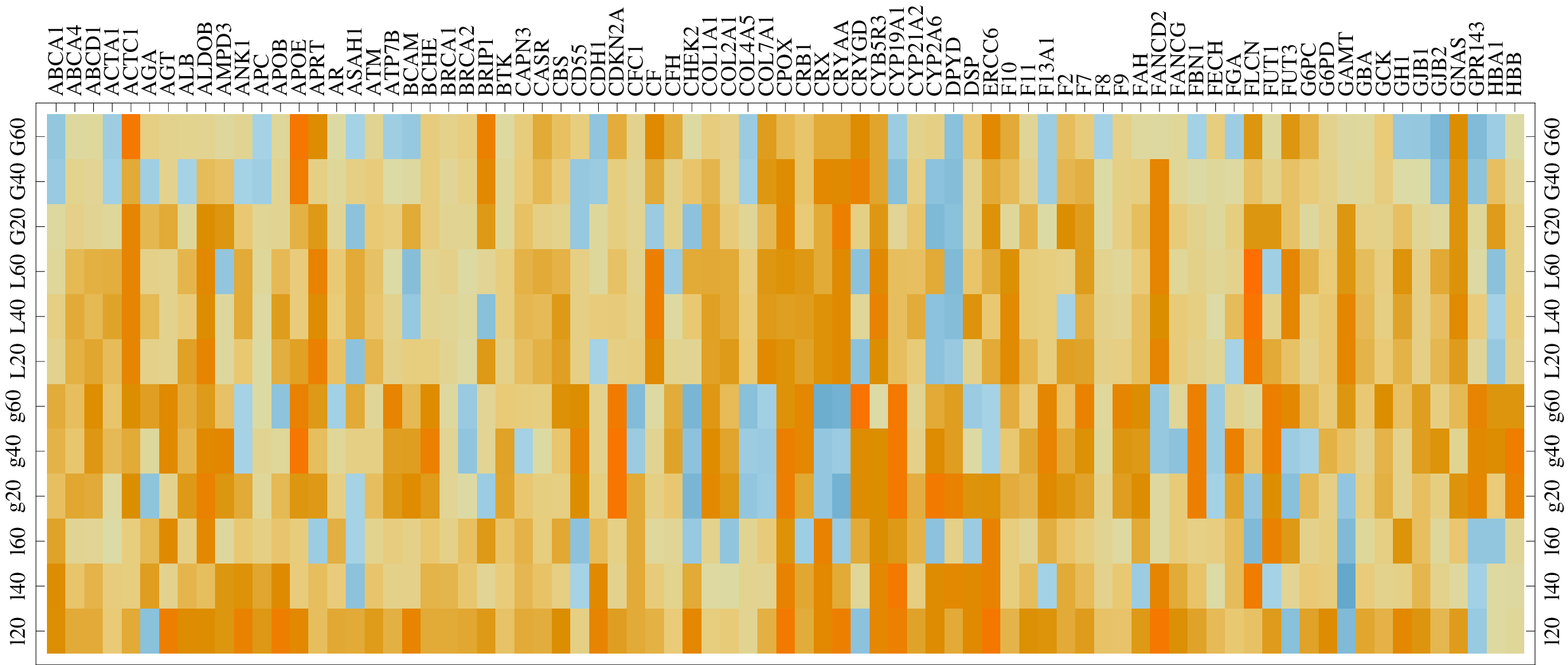}
\includegraphics[width=0.99\columnwidth,angle=-270,scale=0.406]{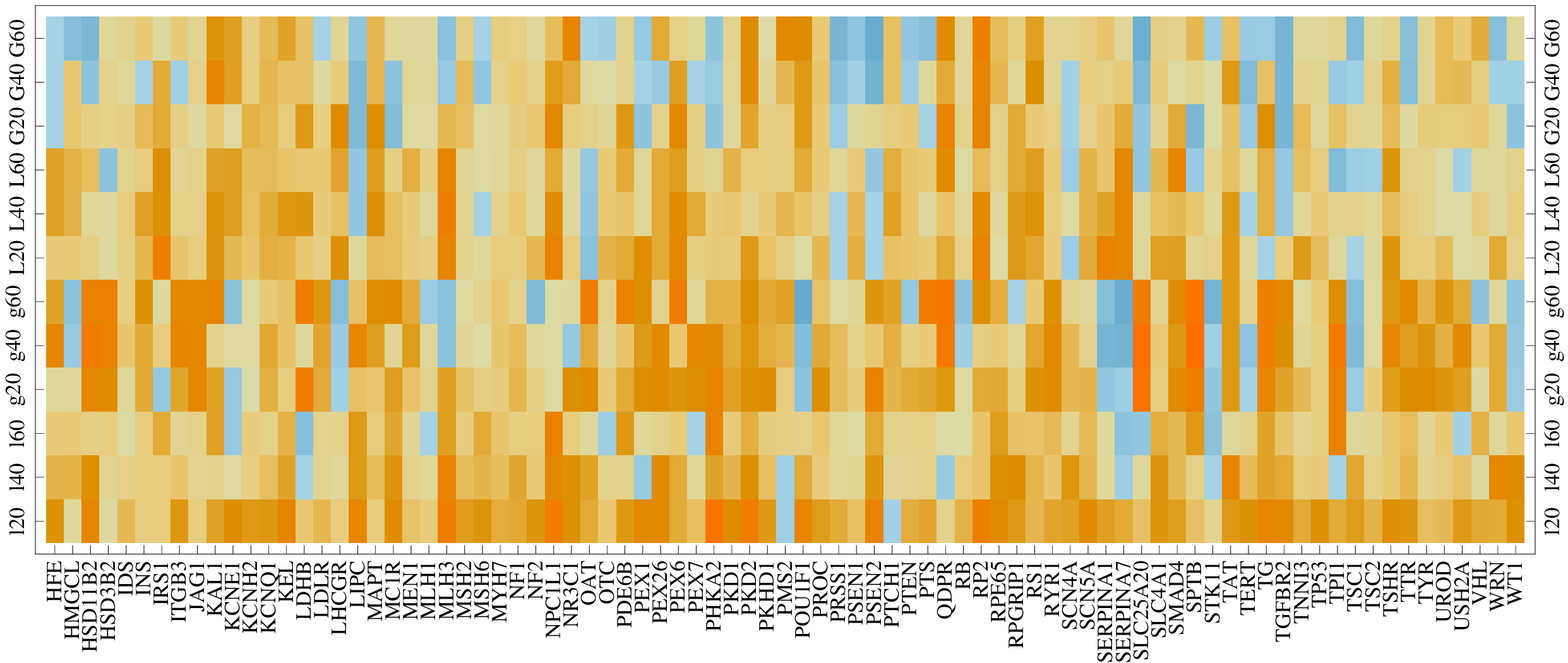}
\caption{
}
\end{figure}
\clearpage
\begin{figure}
\centering
\includegraphics[width=0.99\columnwidth,angle=-270,scale=0.35]{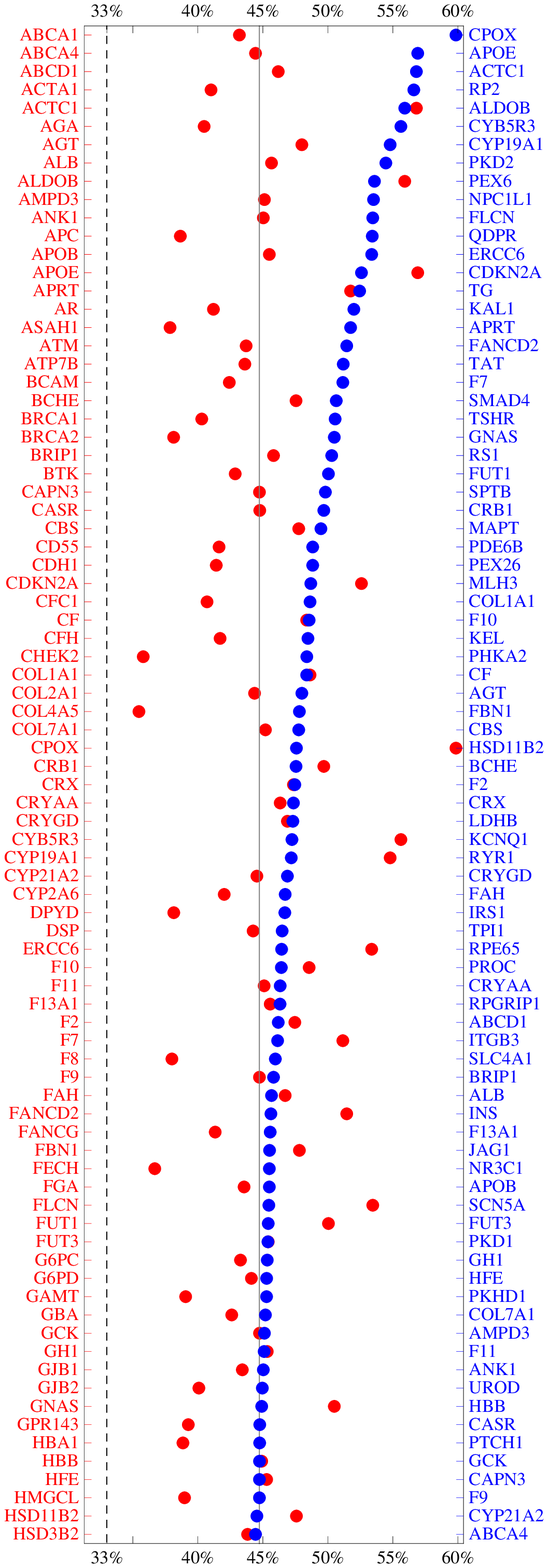}
\includegraphics[width=0.99\columnwidth,angle=-270,scale=0.355]{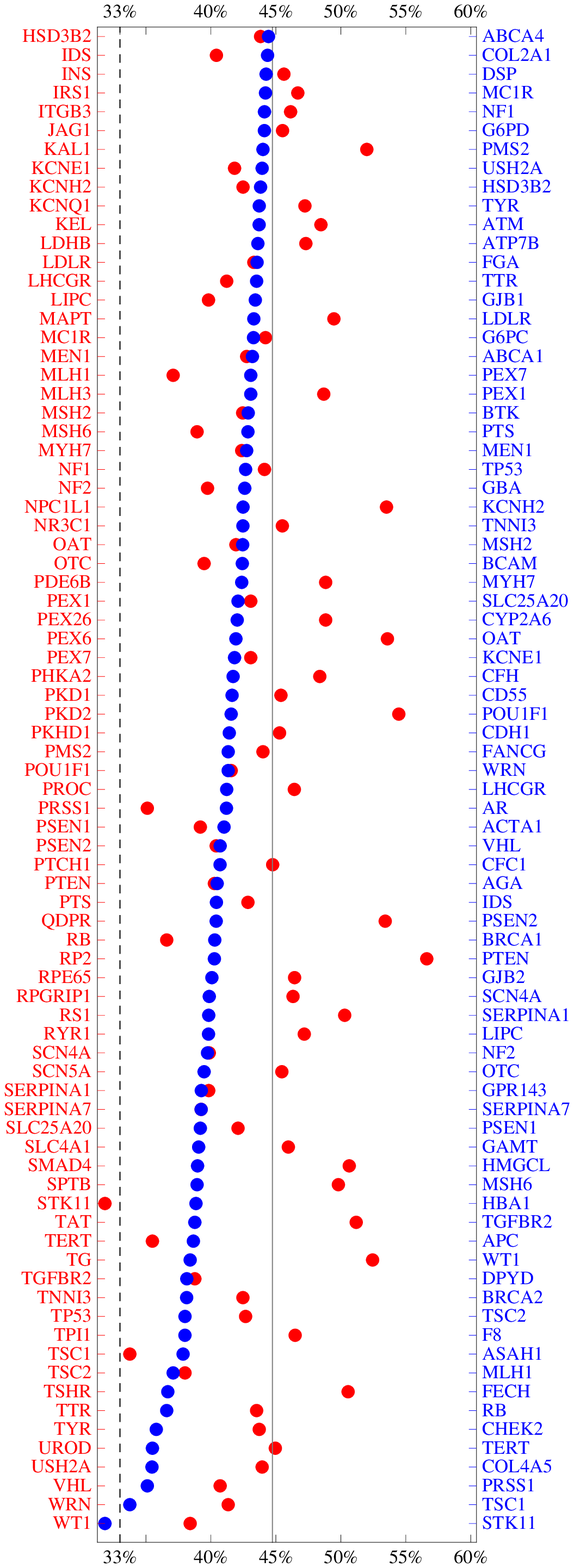}
\caption{}
\end{figure}

\clearpage
\begin{figure}
\centering
(a)\includegraphics[width=0.45\columnwidth]{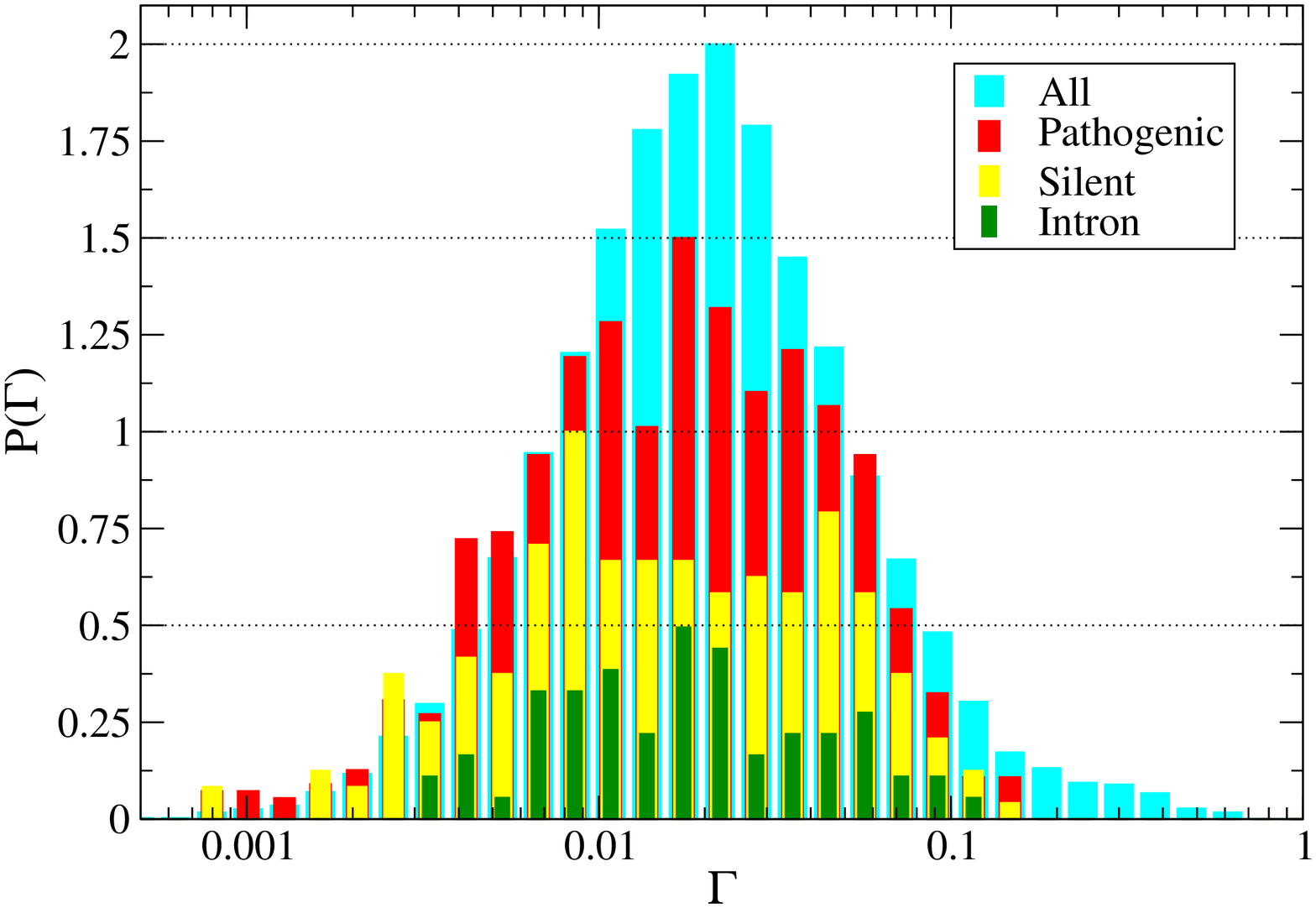}
(b)\includegraphics[width=0.45\columnwidth]{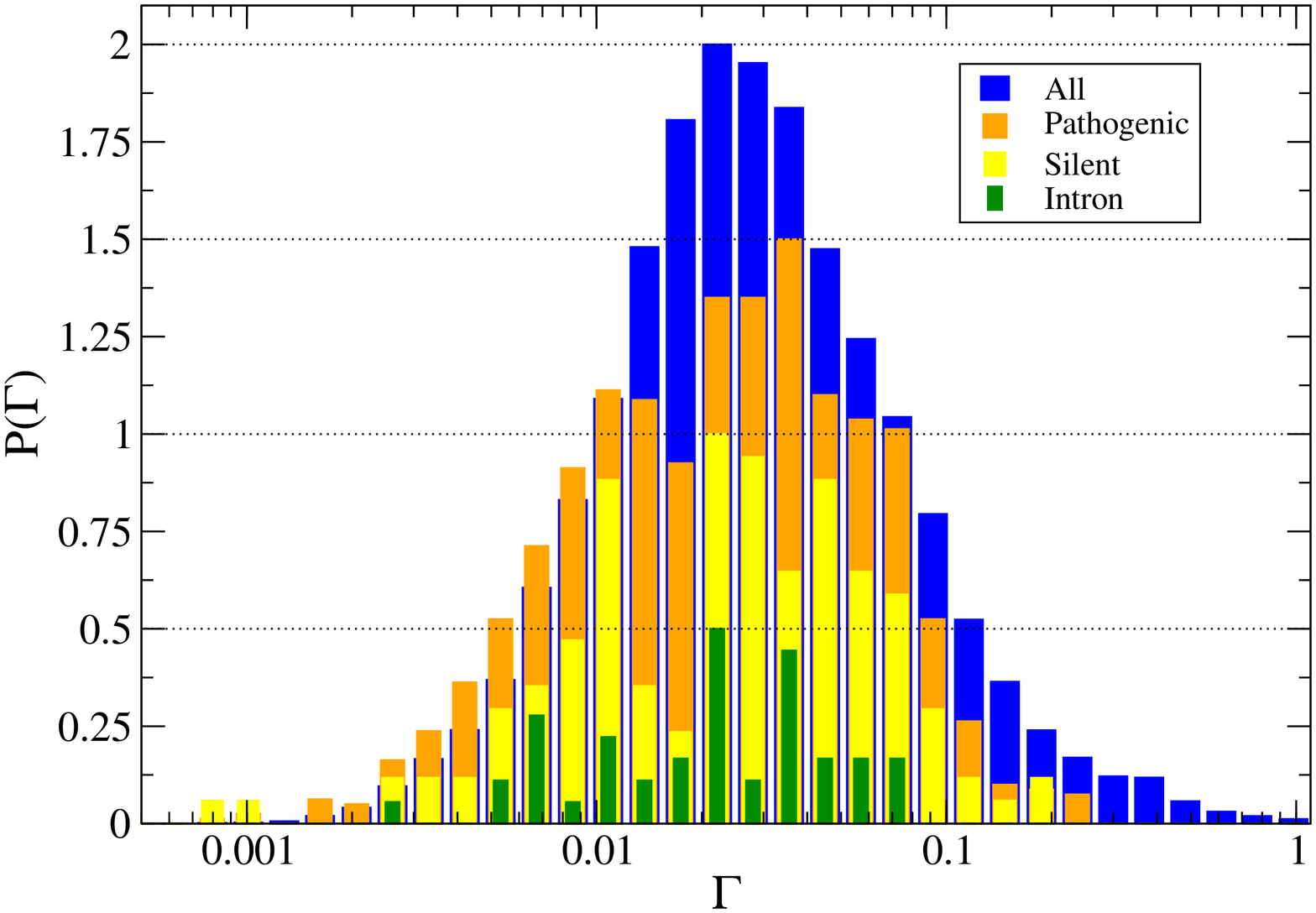}
\caption{}
\end{figure}

\fi\end{document}